\documentclass[letterpaper]{article} % DO NOT CHANGE THIS
\usepackage{aaai2026}  % DO NOT CHANGE THIS
\usepackage{times}  % DO NOT CHANGE THIS
\usepackage{helvet}  % DO NOT CHANGE THIS
\usepackage{courier}  % DO NOT CHANGE THIS
\usepackage[hyphens]{url}  % DO NOT CHANGE THIS
\usepackage{graphicx} % DO NOT CHANGE THIS
\usepackage{natbib}  % DO NOT CHANGE THIS AND DO NOT ADD ANY OPTIONS TO IT
\usepackage{caption} % DO NOT CHANGE THIS AND DO NOT ADD ANY OPTIONS TO IT
\usepackage{url}            % simple URL typesetting
\usepackage{booktabs}       % professional-quality tables
\usepackage{amsfonts}       % blackboard math symbols
\usepackage{nicefrac}       % compact symbols for 1/2, etc.
\usepackage{microtype}      % microtypography
\usepackage{xcolor}         % colors

\usepackage{soul}

\usepackage{afterpage}
\usepackage{graphicx}
\usepackage{makeidx}
\usepackage{amsmath,amssymb,amsfonts,amsthm,mathtools}
\usepackage{textcomp}
\usepackage{xcolor}
\usepackage{array}
\usepackage{times}
\usepackage{url}
\usepackage{caption}

\usepackage{booktabs}
\usepackage{multirow}
\usepackage{color}
\usepackage{comment}
\usepackage{graphics}
\usepackage{rotating}
\usepackage{epstopdf}
\usepackage{lscape}
\usepackage{thm-restate}
\usepackage{natbib}
\usepackage{amsmath}
\usepackage{bm}
\usepackage{amssymb}
\usepackage{amsthm}
\usepackage{mathtools}
\usepackage{multirow}
\usepackage{multicol}
\usepackage{subcaption}
\usepackage{tabularx}
\usepackage{xcolor}
\usepackage{listings}
\usepackage{chngcntr}
\usepackage{amsfonts}
\usepackage{blkarray}
\usepackage[capitalise]{cleveref}
\usepackage{adjustbox}
\usepackage{xspace}

\usepackage{booktabs}

\theoremstyle{definition}

\theoremstyle{remark}

\newcommand{\bsub}[1]{\vspace{1px}\noindent \textbf{#1}}

\newcommand{\mylink}[2]{\textbf{#1:}  {\textcolor{blue}{#2}} \\}

\newcommand{\vitfeat}{x_{\text{vit}}}

\newcommand{\vaefeatsrc}{x_{\text{src}}}

\newcommand{\vaefeattarget}{x_{\text{tgt}}}

\newcommand{\textfeat}{x_{\text{text}}}

\newcommand{\noisesrc}{\epsilon_{\text{src}}}

\newcommand{\noisetarget}{\epsilon_{\text{tgt}}}

\newcommand{\bbR}{\mathbb{R}}

\newcommand{\tpack}{p}

\newcommand{\mmname}{\text{LGCC}\xspace}
\newcommand{\vaedecoder}{\text{Dec}}

\usepackage{algorithm}
\usepackage{algorithmic}

\usepackage{newfloat}
\usepackage{listings}
\DeclareCaptionStyle{ruled}{labelfont=normalfont,labelsep=colon,strut=off} % DO NOT CHANGE THIS
\floatstyle{ruled}
\newfloat{listing}{tb}{lst}{}
\floatname{listing}{Listing}
\nocopyright
\title{LGCC: Enhancing Flow Matching Based Text-Guided Image Editing with Local Gaussian Coupling and Context Consistency}

\author{
    Fangbing Liu,\textsuperscript{\rm 1}
    Pengfei Duan,\textsuperscript{\rm 1}
    Wen Li,\textsuperscript{\rm 1}
    Yi He\textsuperscript{\rm 2}
}
\affiliations{
    \textsuperscript{\rm 1}ZanChat Inc., \textsuperscript{\rm 2}Wuhan University\\
}

\usepackage{bibentry}
\begin{document}

\maketitle

\begin{abstract}

Recent advancements have demonstrated the great potential of flow matching-based Multimodal Large Language Models (MLLMs) in image editing. 
However, state-of-the-art works like BAGEL face limitations, including detail degradation, content inconsistency, and inefficiency due to their reliance on random noise initialization. 
To address these issues, we propose LGCC, a novel framework with two key components: Local Gaussian Noise Coupling (LGNC) and Content Consistency Loss (CCL). LGNC preserves spatial details by modeling target image embeddings and their locally perturbed counterparts as coupled pairs, while CCL ensures semantic alignment between edit instructions and image modifications, preventing unintended content removal.
By integrating LGCC with the BAGEL pre-trained model via curriculum learning, we significantly reduce inference steps, improving local detail scores on I$^2$EBench by 1.60\% and overall scores by 0.53\%. LGCC achieves 3x–5x speedup for lightweight editing and 2x for universal editing, requiring only 40\%–50\% of the inference time of BAGEL or Flux. 
These results demonstrate LGCC's ability to preserve detail, maintain contextual integrity, and enhance inference speed,  offering a cost-efficient solution without compromising editing quality.

\end{abstract}

\begin{links}
\mylink{Website}
{https://zan.chat/fast-bagel}
\mylink{Github}{https://github.com/ZanChat/fast-bagel}
\mylink{Huggingface}{https://huggingface.co/zanchat-ai/fast-bagel}

\end{links}

% \begin{links}
% \link{Website}{{https://zan.chat/fast-bagel}
% \link{Github}{https://github.com/ZanChat/fast-bagel}
% \link{Huggingface}{https://huggingface.co/zanchat-ai/fast-bagel}
% \end{links}
% \color{blue}

% \begin{links}
% \link{Website}
% {https://zan.chat/fast-bagel
% }
% \link{Github}{https://github.com/ZanChat/fast-bagel
% }
% \link{Huggingface}{https://huggingface.co/zanchat-ai/fast-bagel}
% \end{links}

% Uncomment the following to link to your code, datasets, an extended version or similar.
% You must keep this block between (not within) the abstract and the main body of the paper.
% \begin{links}
%     \link{Code}{https://aaai.org/example/code}
%     \link{Datasets}{https://aaai.org/example/datasets}
%     \link{Extended version}{https://aaai.org/example/extended-version}
% \end{links}

% LLM语义控制， 
% 目前已经出现了BAGEL这样的支持细粒度的语言描述的工作，但是其细节和context有问题，并且推理成本过高。
% 因此我们提出一种轻量级方法来提升细节。

% SOTA Flux BAGEL are two open-source state-of-the-art works. 
% 1. 缺少细节
% 2. 过度编辑
% 并且我们方法的好处是,在保持质量不变的情况下极大的增加了图片质量

% 分别体现了两种技术线路，BAGEL在FLUX上支持了细节编辑。

\section{Introduction}
The emerging diffusion models and flow-matching techniques has revolutionized and fundamentally reshaped the field of image generation and editing, offering unprecedented control over the creation and manipulation of visual content~\citep{ho2020denoising, song2021score, lipman2022flow}. 
Recent advancements in multimodal large language models have further pushed the boundaries of text-guided image editing. 
State-of-the-art (SOTA) models such as FLUX~\citep{labs2025flux} and Bagel~\citep{bagel} enable intuitive manipulation of images through natural language instructions~\citep{saharia2022photorealistic, team2024chameleon, liu2025step1x}. 
Notably, approaches leveraging flow-matching techniques have demonstrated significant promise in generative tasks, showcasing their potential to redefine the capabilities of image editing and synthesis.

% Meanwhile, flow matching based approaches significantly improve the 

% \hy{add a figure to show how flux and bagel missing details.}

\begin{figure}[t]
    \centering
    \includegraphics[width=\linewidth]{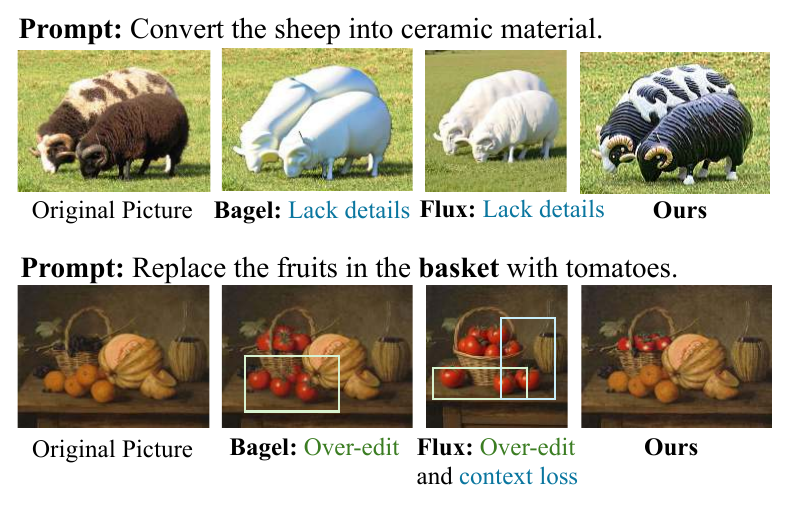}
    \vspace{-15px}
    \caption{Existing text-guided image editing often lacks detail, misses context, and sometimes over-edits images.}
    \label{fig:limtation-sample}
    \vspace{-15px}
\end{figure}

State-of-the-art image editing systems, such as Flux and Bagel~\citep{labs2025flux, bagel}, however, still face critical limitations in understanding image context and generating flawless outputs.
First, these systems often \textit{degrade the details} of target objects during the editing process. For example, object boundaries frequently become blurred, making it challenging to blend them seamlessly with the background (see the sheep generated by Bagel and Flux in Figure~\ref{fig:limtation-sample}). Additionally, material textures are often overlooked or poorly rendered, resulting in visible flaws that significantly undermine the overall visual quality and realism.

Another significant limitation is that existing systems often suffer from \textit{context loss}, where the editing process becomes overly aggressive and unintentionally modifies more elements than desired. For instance, irrelevant background regions may be mistakenly altered, or important foreground objects may be incorrectly removed~\cite{hertz2022prompt, tumanyan2023plug}. This issue arises from the global nature of current flow-matching approaches, which fail to accurately preserve the contextual relationships between regions that should remain unchanged and those requiring editing.

To mitigate these limitations, we introduce $\mmname$, a novel flow matching framework for multimodal large language models.
$\mmname$ incorporates two key innovations: Local Gaussian Noise Coupling (LGNC) and Content Consistency Loss (CCL).
This approach is motivated by the observation that existing flow matching methods treat image editing as a global transformation problem, often overlooking the importance of local spatial structures and semantic relationships within images.

To preserve object details during editing, we introduce the Local Gaussian Noise Coupling (LGNC) method, which represents a paradigm shift in flow matching by treating target image embeddings and their locally noise-perturbed counterparts as coupling pairs.
LGNC adopts a local coupling strategy that ensures the flow matching process respects the spatial coherence of the original image, particularly around critical object boundaries where detail preservation is most challenging.
Consequently, LGNC enables fine-grained spatial control while maintaining essential visual details.

To address context loss, we propose the Content Consistency Loss (CCL) approach, which provides semantic regularization by measuring the distance between text embeddings and the difference between generated and source image embeddings in latent space.
This ensures that the editing process remains semantically consistent with the input instructions while preventing the unintended removal of objects that should be preserved.

% Third, computational inefficiency poses a significant barrier to real-world applications. Current diffusion-based methods for image editing typically require dozens to hundreds of denoising steps to gradually refine an embedding into an image~\citep{zhang2025nexus, zhang2023adding}. Flow matching based methods also require 
% multiple neural function estimations (NFEs) during inference~\citep{labs2025flux,bagel}. Too many denoising steps and NFEs resulting in prohibitively long processing times that limit their scalability and availability for online service. 

% The CCL acts as a semantic anchor, maintaining the integrity of the original content while allowing for targeted modifications.

% Furthermore, \fb{Our method can be incorporated with any flow matching based methods.} 
% Moreover, 

% 用原始图片作为噪音，在一开始的时候更接近缘原图，在求解空间里面里最优解更近，因此能step更少达到更好的效果
% Benefiting from LGCC's new coupling approach, which starts from the original image as local noises, LGCC achieves better results in fewer steps.

% Benefiting from these new approaches, LGCC achieves better results with fewer steps.

Integrating LGCC into existing flow-matching workflows can effectively address issues such as detail degradation and context loss, while also accelerating inference by reducing the steps needed to achieve satisfactory results. This improvement is driven by LGCC's use of local Gaussian methods, which initiate from the original image as noise and constrain flows to smoother, shorter paths near the source image, thereby reducing the required steps. Additionally, CCL also enhances the process by aligning edits with text instructions, minimizing unnecessary changes. 

% To further facilitate the application of $\mmname$, we propose a curriculum learning strategy that enables seamless integration with pre-trained models through fine-tuning.
% This strategy allows $\mmname$ to leverage the architectures and capabilities of existing models without disrupting their learned representations~\citep{bengio2009curriculum, wang2021survey}.

% Experimental results show that LGNC and CCL effectively address issues such as detail degradation and context collapse while improving inference efficiency by reducing the required Neural Function Evaluations (NFEs). The LGNC approach constrains flows to smoother and shorter paths around the source image, significantly reducing NFEs. Similarly, the CCL approach simplifies flows by aligning edits with text instructions, minimizing unnecessary modifications. Both approaches successfully lower NFEs and accelerate the inference process.

% The LGNC approach constrains flows to smoother and shorter paths around the source image, while CCL aligns edits with text instructions, simplifying flows and minimizing unnecessary modifications in the editing results. 

We apply the \mmname method to the Bagel pre-trained model~\cite{bagel} via curriculum learning.
Experimental results demonstrate that LGNC and CCL effectively address issues such as detail degradation and context collapse while significantly improving inference efficiency by reducing the required inference steps. 
Our methods improve local detail scores of the I$^2$EBench by 1.60\% and overall scores by 0.53\%. 
Furthermore, \mmname only requires 40\% to 50\% of the inference time of Bagel or Flux, accelerating lightweight editing by 3x–5x and universal editing by 2x.
Thereby showcasing its ability to preserve detail, maintain contextual integrity, and enhance inference speed.

% \hy{update results.}

The contributions of this work are summarized as follows:
\begin{itemize}
\item We introduce LGNC, a novel approach that leverages locally perturbed image embeddings as coupling pairs, effectively preserving details during flow-matching-based image editing.
\item We propose CCL, a semantic regularization method that ensures consistency between text instructions and image modifications while preventing unwanted changes to objects.
\item We implement LGNC and CCL as the $\mmname$ framework, which can be integrated with existing flow-matching based pre-trained models such as Bagel. We thoroughly evaluate its performance, demonstrating superior local detail editing effects while requiring fewer inference steps.
\end{itemize}

\section{Related Work}

% \textbf{Vision-Language Models (VLMs).}
% VLMs form the foundation of many multimodal reasoning and image editing systems. CLIP\citep{clip} pioneers large-scale contrastive learning of image-text pairs, enabling strong zero-shot alignment across modalities. Building on this idea, SigLIP and its improved variant SigLIP-2\citep{siglip,siglip2} replace the softmax-based loss in CLIP with a sigmoid loss, resulting in more robust text-image representations and better transfer performance. Flamingo~\citep{flamingo} further extends VLMs by integrating frozen language models with cross-attention layers, enabling few-shot multimodal reasoning and dynamic conditioning on both visual and textual inputs. 
% These models are widely adopted as backbones or feature extractors for unified multimodal large language models, providing the core alignment needed to interpret editing instructions and manipulate visual content effectively.

\subsubsection{Text-Guided Image Editing.}
Text-guided image editing frameworks such as InstructPix2Pix~\citep{brooks2023instructpix2pix}, MagicBrush~\citep{zhang2023magicbrush} and HQ-Edit~\citep{hui2024hq} use conditional diffusion models to translate user intents into detailed edit operations. MGIE~\citep{fu2023guiding} uses MLLMs to derive expressive instructions from terse prompts, improving controllability while maintaining efficiency. Other frameworks like SmartEdit and GoT integrate spatial and semantic reasoning to generate precise object‑level edits guided by MLLMs~\citep{huang2024smartedit, fang2025got}. These methods improve adherence to complex instructions and object consistency, but often require explicit detailed text instructions.

% mask supervision or multi‑step reasoning pipelines.

\subsubsection{Unified Multimodal Models.}
Unifed multimodal models adopt unified architectures that support text‑to‑image generation, in‑context editing, and reasoning in a single architecture. Chameleon uses decoder-only transformer architecture and is trained on interleaved text and image tokens~\citep{team2024chameleon}. It treats vision-language generation as next-token prediction across modalities. Janus~\citep{wu2025janus} and Janus-Pro~\citep{chen2025januspro} address limitations of Chameleon that rely on a single visual encoder for both tasks by decoupling visual encoding into separate pathways while utilizing a single, unified transformer architecture for processing. SEED-X~\citep{ge2024seed-x}, Nexus-Gen~\citep{zhang2025nexus} and Step1X-Edit~\citep{liu2025step1x} both use multimodal large language models for text prediction, which is followed by pretrained diffusion models or flow matching models for image generation. BAGEL and FLUX.1 Kontext adopt MLLM under flow matching framework in latent space and the generated embeddings are put into an image decoder for final image generation~\citep{bagel, labs2025flux}.

% FLUX.1 Kontext excels at iterative local editing with strong character consistency, while BAGEL demonstrates general-purpose image and video modeling but suffers from detail degradation and over‑editing especially around object boundaries \cite{fluxkontext2025, bagelreview2025}.

\subsubsection{Flow Matching.}
Similar to Stable Diffusion~\citep{croitoru2023diffusion}, flow matching has emerged as a powerful generative modeling framework.
It is versatile and applicable to various modalities, including images, videos, and more~\citep{lipman2022flow}.
Recent systems like Rectified Flow~\citep{liu2022flow} and Conditional Flow Matching~\citep{pooladian2023multisample} further optimize flow matching by iteratively directing trajectory or approximating optimal transport plan, resulting in faster inference with maintained visual fidelity. Transformer‑based latent flow models such enables scalable and fine‑grained manipulation of latent space, though fine-detail preservation remains a challenge \cite{hu2024latent}. 

\subsubsection{Our Work.}
We propose $\mmname$, introducing two novel approaches into the flow-matching workflow: the Local Gaussian Noise Coupling (LGNC) method to address detail degradation, and the Content Consistency Loss (CCL) approach to mitigate context loss during image editing.
Meanwhile, $\mmname$ achieves high-quality results in fewer steps, offering a faster and cost-efficient solution.

\section{Background and Preliminaries}
We provide a brief background and preliminaries to introduce the essential knowledge required for our approach.

\subsection{Optimal Transport}
For two distributions $p_0,p_1\in \mathcal{P}_{2,ac}(\bbR^D)$ and a cost function $c: \bbR^D \times \bbR^D \rightarrow \bbR $, Monge's optimal transport formulation~\citep{villani2021topics} is:
\begin{equation}\label{eq_bagel_def_monge_ot}
    \inf_{T\# p_0=p_1}\int_{\bbR^D} c(x_0, T(x_0)) p_0(x_0) d_{x_0},
\end{equation}
where $T:\bbR^D \rightarrow \bbR^D$ is a transport map, which is a function satisfying the mass preservation constrain $T\#p_0=p_1$. The transport map achieving the minimum is called the optimal transport map, denoted as $T^*$.

% For the quadratic cost function $c(x_0,x_1)=\frac{\|x_0-x_1\|^2}{2}$, the square root of the optimal value of~\cref{eq_bagel_def_monge_ot} is called the Wasserstein-2 distance~\citep{villani2021topics}:
% \begin{equation}\label{eq_bagel_def_w2_dis}
%     \mathbb{W}_2^2(p_0,p_1):= \min_{T\#p_0=p_1} \int_{\bbR^D} \frac{\|x_0-T(x_0)\|^2}{2}p_0(x_0) d_{x_0}.
% \end{equation}

% It is known that problem in~\cref{eq_bagel_def_w2_dis} has an equivalent dual form~\citep{villani2021topics}:
% \begin{align*}
%     \mathbb{W}_2^2(p_0,p_1) &= Const(p_0,p_1) - \min_{convex\quad \Phi} \mathcal{L}_{OT}(\Phi)  \\
%     \mathcal{L}_{OT}(\Phi) &= \int_{\bbR^D} \Phi(x_0)p_0(x_0) d_{x_0} \\
%     &\quad + \int_{\bbR^D} [\sup_{x_0\in \bbR^D} [<x_0,x_1>-\Phi(x_0)] ] p_1(x_1) d_{x_1},
% \end{align*}
% where $Const(p_0,p_1)$ does not depend on the convex function $\Phi$. Minimizing the Wasserstein-2 distance in~\cref{eq_bagel_def_w2_dis} is equivalent to minimizing $\mathcal{L}_{OT}(\Phi)$. According to~\citep{villani2021topics}, the optimal transport map $T^*$ is 
% \begin{equation}\label{eq_bagel_solution_ot}
%     T^*=\nabla \Phi^*
% \end{equation}

% Further, it is pointed that above  has an 
\subsection{Dynamic Optimal Transport and Flow Matching}
When the cost $c$ in~\cref{eq_bagel_def_monge_ot} is the the quadratic cost $c(x_0,x_1)=\frac{\|x_0-x_1\|^2}{2}$, the calculation of optimal transport map $T^*$ can be formulated in a dynamic form~\citep{benamou2000computational}.
% which operates with a vector field defining time dependent mass transport.
For a time interval $[0,1]$, let $u_t(\cdot): [0,1]\times \bbR^D \rightarrow \bbR^D $ be a vector field and $\{\{z_t\}_{t\in[0,1]}\}$ be a set of trajectories that each trajectory $\{z_t\}_{t\in[0,1]}$ starts with $z_0$ sampled from $p_0$ and all $z_t$ satisfies the differential equation:
\begin{equation}\label{eq_bagel_ode}
    d_{z_t} = u_t(z_t) d_t,\quad z_0\sim p_0.
\end{equation}

% The flow map $\phi_t^u(\cdot):[0,1]\times \bbR^D \rightarrow \bbR^D $ is a function that maps the initial point $z_0$ to its position at time $t$ according to~\cref{eq_bagel_ode}:
% \begin{equation}
%     d_{\phi_t^u(z_0)} = u_t(\phi_t^u(z_0)), \quad \phi_0^u(z_0)=z_0.
% \end{equation}

% As the initial point $z_0$ is sampled from $p_0$, then~\cref{eq_bagel_ode} defines a distribution $p_t$ of $z_t$ at time $t$, which is expressed by the push-forward operator $p_t^u:=\phi_t^u\# p_0$.

% The dynamic OT is the following minimization problem:
% \begin{equation}\label{eq_bagel_def_dynamic_ot}
%     \begin{split}
%          \mathbb{W}_2^2(p_0,p_1) &= \inf_u \int_0^1 \int_{\bbR^D} \frac{ \|u_t(x_t)\|_2^2 }{2} \phi_t^u\#p_0(x_t) d_{x_t} d_t,\\
%     &s.t. \quad \phi_1^u\#p_0=p_1
%     \end{split}
% \end{equation}
% where vector field $u$ defines the flows starting at $p_0$ and end at $p_1$ and the optimal $u^*$ minimizes the kinetic energy over entire time interval.

% A connection between the solution $T^*=\nabla \Phi^*$ in~\cref{eq_bagel_solution_ot} to the OT problem and the solution $u^*$ to the dynamic OT problem in~\cref{eq_bagel_def_dynamic_ot} is that: for every initial point $z_0$, $u^*$ defines a linear trajectory $\{z_t\}_{t\in[0,1]}$, i.e.,
% \begin{equation*}
%     z_t=t\nabla \Phi^*(z_0) + (1-t)z_0,\quad \forall t\in [0,1].
% \end{equation*}

% \subsection{Flow Matching}
In flow matching, for a sample point $x_0,x_1$ from transport plan $p_0\times p_1$ $\pi \in \prod(p_0,p_1)$, the vector field $u$ is encouraged to follow the direction $x_1-x_0$ so that for any time $t\in[0,1]$, we have $x_t=(1-t)x_0 + t x_1$. Therefore, the objective function of flow matching is defined as:
\begin{equation}\label{eq_bagel_obj_flow_matching}
\small
     \mathcal{L}_{FM}^{\pi}(u)= \int_0^1 \int_{\bbR^D\times \bbR^D} \|u_t(x_t)-(x_1-x_0)\|_2^2 \pi(x_0,x_1) d_{x_0} d_{x_1}  d_t,
\end{equation}
where the linear interpolation $x_t=(1-t)x_0 + t x_1$ is an intuitive way to move $p_0$ to $p_1$.
% , but it requires us to know $x_1$ in advance. 
With a known $x_1$, one can fit $u$ to the direction $x_1-x_0$ to minimize the objective loss.
% $\mathcal{L}_{FM}^{\pi}(u)$.

Certain initial plans $\pi$ can lead to more straight path rather than the independent plan $p_0\times p_1$, such as the Rectified Flow~\citep{liu2022flow} and Optimal Transport Conditional Flow Matching~\citep{pooladian2023multisample}. 

\subsubsection{Rectified Flow.}
RF uses an iterative method to make trajectories more straight within each iteration. At each iteration $k$, the velocity field $u_t^k$ is updated as
\begin{equation*}\label{eq_bagel_obj_flow_matching}
\small
\begin{split}
     u_t^{k+1}&= \min_{u} \\
     &\int_0^1 \int_{\bbR^D\times \bbR^D} \|u_t(x_t)-(x_1-x_0)\|_2^2 \pi^k(x_0,x_1) d_{x_0} d_{x_1} d_t,\\
    % &\quad x_t=(1-t)x_0 + t x_1   
\end{split}
\end{equation*}
Then we compute the flow map $\phi_{k+1}^u$ from the velocity field $u_t^{k+1}$ that $x_1^{k+1}=\phi^u_{k+1}(x_0)$. Further, the transport plan $\pi^{k+1}= [I(\cdot),\phi^u_k] \# p_0$, where $I(\cdot)$ denotes the identity function. 
% \fb{change eq format}

% RF optimizes for trajectory straightness while optimal transport aims to minimize a cost function. They are different geometric aims. Further, the random coupling for the initial transport plan in RF may create non-optimal pairing between points. 
The iteration process in RF straightens trajectories locally and the global optimal coupling are never solved and thus RF is biased toward the initial transport plan.

\subsubsection{Optimal Transport Conditional Flow Matching.}
% When using the optimal transport plan $\pi^*$ as the initial plan $\pi^0$, then we got the optimal velocity field $u^*$ of \cref{eq_bagel_obj_flow_matching}, which generates exactly straight trajectories. Thus, 
Making the initial transport plan $\pi^0$ close to the optimal $\pi^*$ may leads to the learned trajectories close to the ideal straight trajectories. Instead of random pairing, OT-CFM approximates the optimal transport plan $\pi^*$ locally using the minibatch optimal transport. They sample small batches of samples:
\begin{equation*}
    \{x_0^i\}_{i=1}^B\sim p_0,\quad \{x_1^i\}_{i=1}^B\sim p_1
\end{equation*}
where $B$ is the batch size. 
Then they solve the discrete optimal transport problem between them:
\begin{equation*}
    \pi^0_{ij}= \arg\min_{\pi\in \prod} \sum_{i,j} c(x_0^i, x_1^j) \pi_{ij}
\end{equation*}
where $c$ is the cost function and $\prod$ is the set of coupling pairs.

Over multiple mini batches, the local approximation provides an initial point $\pi^0$ that is much closer to $\pi^*$ than random pairing.

\subsubsection{Optimal Flow Matching.}
According to~\citep{villani2021topics}, the optimal transport map $T^*$ is the gradient of a convex function, i.e.,
\begin{equation}\label{eq_bagel_solution_ot}
    T^*=\nabla \Phi^*
\end{equation}
OFM restricts the search space to optimal vector fields induced by a convex potential and guarantees in theory that the learned trajectories are straight by construction through the convex function parameterization: \begin{equation*}
    u_t^{\Phi}(x)=\nabla \Phi(x_0) - z_0
\end{equation*}
where $z_0$ solves the inverse map:
\begin{equation*}
    x_t=(1-t)z_0 + t\nabla\Phi(z_0)
\end{equation*}

However, OFM introduces substantial computational overhead through the need to solve a convex optimization subproblem at each training step to find initial points $z_0$. Its restriction to convex potential functions may also limit its expressiveness compared to unrestricted vector field learning, potentially struggling with complex transport scenarios where the optimal map cannot be well-approximated by gradients of convex functions.

% Approach: 第一段前面
% 1. 图片编辑场景和生成场景不一样，需要在已有的图片上改。在分布上距离上是很近的。编辑的去噪过程用local gaussian比较有优势。更加能保持原图的语义和结构。local gaussion去更符合图片编辑的场景
% 2. local gaussian假设下的去燥过程，更能够突出图片的小区域细节
% Problem Formulation 
% Local Guassian 
% 加更多东西顺下来

\section{Approach}

\subsection{The LGCC Framework}

In image editing tasks, the primary goal is to preserve the content of the original image while making precise modifications based on the provided instructions. This approach stands in sharp contrast to image generation tasks, which focus on creating entirely new images from scratch. Denoising methods grounded in the local Gaussian assumption offer significant advantages in this context: they excel at highlighting fine details within small regions while effectively preserving the semantics and structure of the original image. As a result, local Gaussian methods are particularly well-suited for scenarios where maintaining the integrity of the original content is crucial.
Moreover, aligning the image latent modification with the latent states of the instructions can help maintain consistency with the instructions and accurately follow the image context.

\begin{figure}[t]
    \centering
    \includegraphics[width=\linewidth]{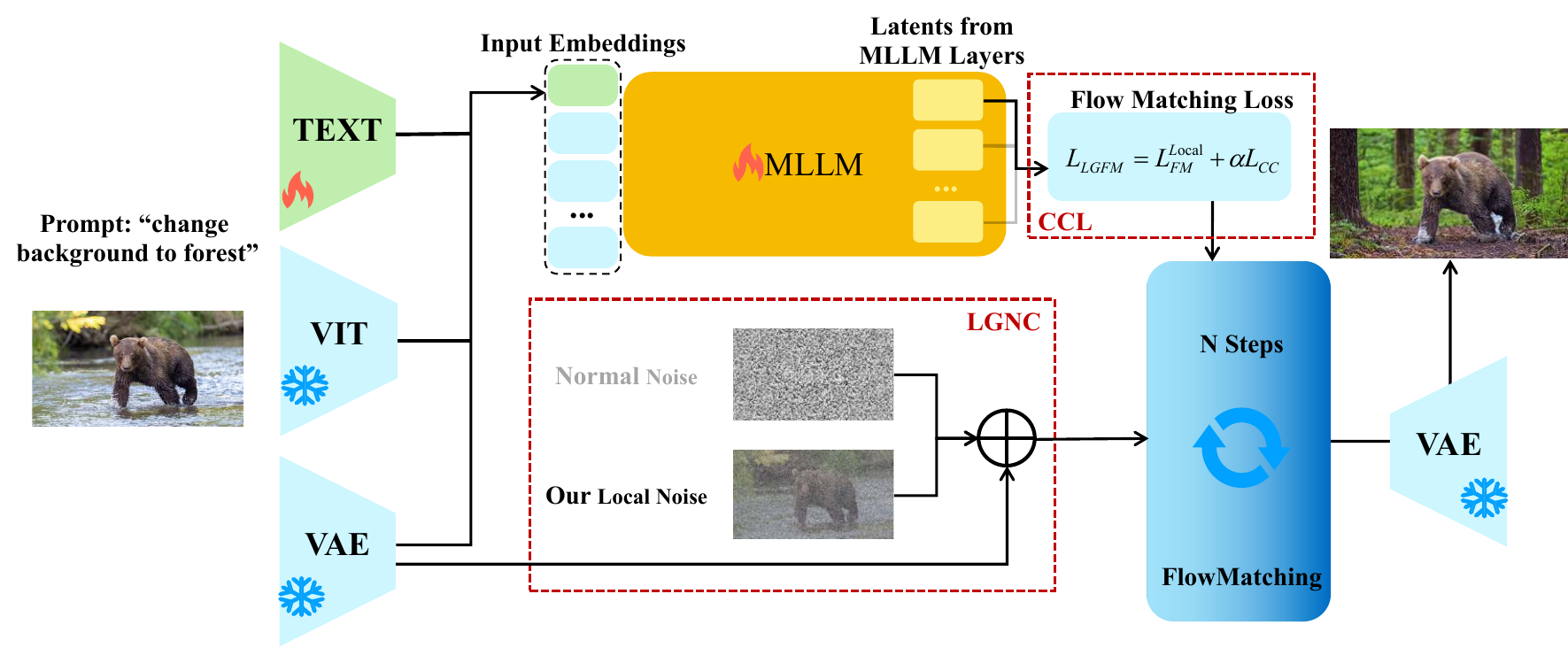}
    % \vspace{-15px}
    \caption{The workflow of LGCC. LGCC consists of two key modifications in the flow matching approach, including Local Gaussian Noise Coupling (LGNC) and Context Consistency Loss (CCL), which are in red boxes.}
    \label{fig:workflow}
\end{figure}

Based on these insights, we propose the LGCC framework, which operates in latent space and combines local Gaussian methods with context-preserving training to achieve efficient and high-quality image editing.
As illustrated in Figure~\ref{fig:workflow}, $\mmname$ adopts a flow-matching-based architecture with three complementary encoders for capturing multimodal features: a text encoder for processing editing instructions, SigLIP-2 \citep{siglip2} for extracting high-level visual features from source images, and the VAE encoder from Flux for obtaining detailed latent representations of source and target images.
Specifically, $\mmname$ introduces two novel approaches into the flow-matching workflow. 
The core innovation is conducting flow matching within an enriched latent space, where the network takes as input the concatenation of text features, SigLIP-2 visual features, and VAE features of source and target images at time $t$. The network's output for the target image is decoded through Flux’s VAE decoder to produce the final edited image. 
This architecture allows us to leverage existing pre-trained models while enhancing performance in preserving fine-grained details, maintaining semantic consistency, and improving inference speed.

\subsection{LGCC's Enhancements on Flow Matching}

\subsubsection{Flow Matching Problem Formulation.}
Given a source image $X_s \in \mathbb{R}^{H \times W \times 3}$ and a text instruction $M$ describing the desired edit, the goal of text-guided image editing is to generate a target image $X_t$ that incorporates the specified modifications while preserving unrelated content from the source. The text-guided image editing task is formulated as a flow matching problem in the latent space. 
The text instruction $M$ is encoded into a sequence of semantic vector $\textfeat = \text{TextEncoder}(M) \in \bbR^{|M|\times d_t}$, where $\text{TextEncoder}$ is a linear function and $|M|$ is the word number in sentence. The source image $X_s$ is encoded via a SigLIP-2 encoder to extract structure-aware features $\vitfeat = \text{ViT}(X_s) \in \bbR^{|P|\times d_v}$. The Flux VAE encoder is used to obtain both features from source image $\vaefeatsrc = \text{VAE}(X_s) \in \bbR^{|P|\times d_s}$ and target image $\vaefeattarget = \text{VAE}(X_t) \in \bbR^{|P|\times d_s}$ where $P$ is the patch number of images. $d_t$, $d_v$,$d_s$ are different feature dimensions in latent space.

We formulate the problem as learning a flow matching process that operates on multimodal latent representations. Our neural network $u_t(\cdot;\Theta)$ takes the latent noisy at time $t$ as input where text feature $\textfeat$ and Vit feature $\vitfeat$ appear as global context and outputs a velocity prediction that is subsequently decoded via $\vaedecoder$ to produce the final edited image $\hat{X}_t$.

\subsubsection{Our Enhancements on Flow Matching.}

In conventional flow matching, the initial state $x_{\tpack}^0$ consists of pure Gaussian noise $\mathcal{N}(0, I)$, which bears no structural relationship to the target embedding $x_{\tpack} = [\vaefeatsrc, \vaefeattarget]$. The defined trajectory $x_{\tpack}^t = (1-t) x_{\tpack}^0 + t x_{\tpack}$ require to traverse from completely uncorrelated noise to highly structured image representations in latent space, inevitably losing fine-grained spatial details during this drastic transformation.

To operate in latent space, we introduce the LGNC method, which employs a local Gaussian noise coupling approach to fundamentally alter this dynamic by constructing the initial state as follows:
\begin{equation*}
    x_{\tpack}^0 = [\vaefeatsrc^{(0)}, \vaefeattarget^{(0)}],
\end{equation*}
where
\begin{align*}
    &\vaefeatsrc^{(0)} = \vaefeatsrc + \noisesrc, \quad \vaefeattarget^{(0)} = \vaefeattarget + \noisetarget\\
    & \noisesrc, \noisetarget \sim \mathcal{N}(0, \sigma^2 I), \quad \sigma = \text{Std}(\vaefeatsrc).
\end{align*}
$\text{Std}(\cdot)$ is the standard deviation of variable. This design ensures that the initial state retains the underlying spatial structure of the true embeddings while introducing controlled perturbations at an appropriate scale. The trajectory now becomes a smooth interpolation between two structurally similar representations rather than a generation from pure noise.

Mathematically, the structural similarity can be quantified through the signal-to-noise ratio: $\text{SNR} = \frac{||\vaefeattarget||^2}{\sigma^2 \cdot d_z}$, where $d_z$ is the latent dimension. When the VAE features of the source and target image in the latent space have same variance that $\text{Var}(\vaefeatsrc)\approx \text{Var}(\vaefeattarget)$, we we have $\text{SNR} \approx \frac{\text{Var}(\vaefeattarget)}{\text{Std}(\vaefeatsrc)^2} = 1$, indicating that signal and noise contributions are balanced. This balanced perturbation preserves the spatial correlation structure inherent in $\vaefeattarget$ while providing sufficient stochasticity for effective flow matching. Consequently, the learned velocity field operates on representations that maintain spatial and semantic coherence throughout the entire flow trajectory in the latent space, leading to superior preservation of edges, textures, and fine-grained details in the final decoded images.
We illustrate the LGNC method in \S~\ref{sec:lgnc}

Previous flow-matching works, such as Bagel, often suffer from context loss and over-editing, which deviate from the intended purpose of the instructions. 
To address this issue, we introduce CCL to ensure that the context consistently aligns with the user instructions during training, which is detailed in \S~\ref{sec:ccl}.

\subsection{Local Gaussian  Noise Coupling (LGNC)}
\label{sec:lgnc}
% Flow Matching Loss
The LGNC methods can be formulated as the flow matching loss during training.
We employ Qwen2 as the neural network backbone to parameterize the velocity field $u_\theta(\cdot, t)$ in our rectified flow framework. The network learns to predict the instantaneous change direction by modeling cross-modal interactions between text, visual, and latent representations:
\[
    u_{\Theta}(x_{\tpack}^t, t)=\text{Qwen2}(x_{\tpack}^t,t|\textfeat,\vitfeat;\Theta)
\]
where $\textfeat,\vitfeat$ serve as conditioning contexts derived from text and SigLIP-2 features respectively, and $\Theta$ denotes the learnable parameters.

The key innovation of our approach lies in the coupling transport map between the noisy initialization $x_0$ and the clean target $x_1$. Unlike traditional rectified flow that uses random Gaussian noise, our LGC defines the transport map as:
\begin{equation*}
\pi(x_{\tpack}^0, x_{\tpack}) = p(x_{\tpack}) \mathcal{N}(x_{\tpack}^0| x_{\tpack}, \sigma^2 I)
\end{equation*}
This coupling ensures that $x_{\tpack}^0$ is a structured perturbation of $x_{\tpack}$ rather than independent noise. The optimal transport between coupled pairs follows straight-line trajectories in latent space, where the flow direction is simply:
\begin{equation*}
    x_{\tpack}-x_{\tpack}^0 = -\epsilon_{\tpack} = -[\noisesrc;\noisetarget].
\end{equation*}
with $\epsilon_{\text{src}}, \epsilon_{\text{tgt}} \sim \mathcal{N}(0, \sigma^2 I)$. This formulation enables the velocity field to learn from structured noise-to-clean mappings, preserving spatial coherence while maintaining the computational advantages of rectified flow matching.

Under our LGC framework, we derive the flow matching objective by substituting the structured transport map into the standard flow matching loss in~\cref{eq_bagel_obj_flow_matching}. The theoretical objective function for our local Gaussian flow matching (LGFM) is expressed as:
\begin{equation*}
\small
    \begin{split}
        \mathcal{L}_{FM}^{local}(u) &= \int_0^1 \int_{\bbR^d\times \bbR^d } \| u_t(x_{\tpack}^t) - (x_{\tpack}-x_{\tpack}^0) \|_2^2 \pi(x_{\tpack}^0, x_{\tpack}) d_{x_{\tpack}^0} d_{x_{\tpack}} d_t,\\
    \end{split}
\end{equation*}
where $\quad x_{\tpack}^t = (1-t) x_{\tpack}^0 + t x_{\tpack}$ represents the linear interpolation between the coupled noisy and clean embeddings at time $t \in [0,1]$, and $d = 2d_s$ accounts for the concatenated source and target VAE features. The key distinction from traditional flow matching lies in the target velocity $x_{\tpack}-x_{\tpack}^0$, which corresponds to the negative noise direction $-\epsilon_p$ due to our coupling design.

For practical implementation with gradient-based optimization, we reformulate this as a tractable expectation over the data distribution and our structured noise coupling:
\begin{equation}\label{eq_bagel_lgc_loss}
\small
        \mathcal{L}_{FM}^{local}(u) = \mathbb{E}_{x_{\tpack}\sim p_{data}, \epsilon_{\tpack}\sim \mathcal{N}(0,\sigma^2 I_d), t\sim U[0,1] } \|u_t(x_{\tpack}^t)-(-\epsilon_{\tpack}) \|_2^2
\end{equation}
This formulation enables the network to learn the mapping from structured noisy representations back to their clean counterparts by predicting the negative noise direction $-\epsilon_p$. Unlike conventional approaches that learn to denoise from random Gaussian noise, our objective leverages the preserved spatial structure in the coupled initialization, facilitating more accurate detail recovery during the generation process.

\subsection{Content Consistency Loss (CCL)} \label{sec:ccl}
To ensure that the generated image adheres closely to the provided text instruction without deviating unnecessarily from the source image, we introduce a context consistency loss that explicitly aligns the semantic change in the image with the intended modification described by the text.

The generated features from the predicted flow corresponding to the source and target image are obtained respectively:
\begin{equation*}
    \small
    z_{\text{src}}=u_{\Theta}(x_{\tpack}^t,t)[-2|P|:-|P|,:],\; z_{\text{tgt}}=u_{\Theta}(x_{\tpack}^t,t)[-|P|:,:]
\end{equation*}
where $[\cdot]$ denotes the slicing operation of tensor. We pool the representations using a mean operation over their respective dimensions (tokens for text, patches for images):
\[
    \bar{x}_{\text{text}} = mean(\textfeat),\; \bar{z}_{\text{src}}=mean(z_{\text{src}}),\; \bar{z}_{\text{tgt}} = mean(z_{\text{tgt}})
\]
The semantic delta vector in the image space is defined as the difference:
\[
    \Delta z_{img}=\bar{z}_{\text{tgt}} - \bar{z}_{\text{src}}
\]
We normalize both the pooled text vector and the image delta vector:
\[
    \hat{x}_{\text{text}}=\frac{\bar{x}_{\text{text}}}{\|\bar{x}_{\text{text}}\|_2},\quad \Delta \hat{z}_{img} = \frac{\Delta z_{img}}{\|\Delta z_{img}\|_2}
\]
The context consistency loss is then defined as the squared distance between these normalized vectors:
\begin{equation}
    \mathcal{L}_{CC} = \|\hat{x}_{\text{text}} - \Delta \hat{z}_{img}\|_2^2
\end{equation}
This encourages the semantic shift in image space to align with the intended change encoded in the text instruction, while discouraging unrelated or excessive modifications. Instead of comparing entire representations of the source and target images directly, we measure their difference in latent space and ensure this change vector corresponds semantically to the instruction.

We incorporate this regularization term into the overall loss function as follows:
\begin{equation*}
    \mathcal{L}_{LGFM} = \mathcal{L}_{FM}^{local} + \alpha \mathcal{L}_{CC}
\end{equation*}
where $\mathcal{L}_{FM}^{local}$ is the flow matching loss under local Gaussian noise coupling framework in~\cref{eq_bagel_lgc_loss} and  $\alpha>0$ is a hyperparameter that controls the trade-off between flow generation and semantic preservation.
A higher $\alpha$ enforces stronger alignment to the text semantics, potentially at the cost of visual quality or diversity; a lower $\alpha$ prioritizes realism but may allow semantic drift.

Mathematically, the gradient of $\mathcal{L}_{CC}$ with respect to the network parameters provides a corrective signal that steers the prediction away from directions that would cause unwanted semantic changes. This creates an implicit semantic prior that preserves object identity and spatial relationships unless explicitly contradicted by the text instruction, thereby preventing the context collapse phenomenon observed in traditional approaches.

% The content consistency loss prevents over-editing through two key theoretical properties. First, by constraining $\Delta \hat{z}_{img}$ to align with $\hat{x}_{\text{text}}$, the regularization creates a semantic bottleneck that limits the scope of modifications to those consistent with the text instruction. Second, the normalization ensures that the constraint focuses on the direction of change rather than its magnitude, allowing for natural variations in edit strength while maintaining semantic consistency.

\subsection{Achieve Superior Edit Results with Fewer Steps via LGCC}
% \textbf{Fine-Tuning via Curriculum Learning.}

To integrate \mmname into existing flow matching approaches, rather than retraining the model, \mmname supports fine-tuning on pretrained models.
This is achieved through a curriculum learning strategy, which gradually transitions from random noise to local Gaussian noise. 
This approach can prevent catastrophic forgetting, ensures stable gradients, and preserves pretrained knowledge while adapting to structured noise coupling

One benefit of adopting \mmname is its ability to achieve appropriate results with fewer inference steps. This is made possible by the LGCC approach, which reduces the number of function evaluations (NFEs). 
Unlike traditional flow matching that relies on global random noise and results in long trajectories, LGCC restricts flows to smoother and shorter paths near the source image. This significantly decreases NFEs while preserving both semantic and content consistency.

\section{Experiments}

We conduct experiments to answer the following research questions: \\
% 细节对比，指标和消融实验。图片质量
\noindent \textbf{Q1}: How does LGCC perform in text-guided image editing, especially in preserving fine details? (\S~\ref{sec:eva_perf}) \\
% over-editing
\noindent \textbf{Q2}: Can LGCC effectively address over-editing issues in real-world cases? (\S~\ref{sec:eva_quality})  \\
% step and 推理速度 , less steps with better quality
\noindent \textbf{Q3}: How does LGCC improve inference speed while ensuring output quality? (\S~\ref{sec:eva_spend})

% \noindent \textbf{Q3}: How each component of LGCC contribute to the performance? 

% \fb{fine tune based on bagel.}

\subsection{Experiments Setup}
\label{sec:setup}

\bsub{Datasets.} 
We evaluate \mmname on two widely used benchmark datasets: I$^2$EBench~\citep{ma2024i2ebench} and GEdit-Bench~\citep{liu2025step1x}.
I$^2$EBench contains over 2,000 images and is designed to assess the quality of edited images across multiple dimensions.
GEdit-Bench is commonly used by existing open-source MLLM models to evaluate performance on image editing tasks~\citep{liu2025step1x,bagel}.

% GEdit-Bench~\citep{liu2025step1x}, SmartEdit~\citep{huang2024smartedit} and RISE~\citep{zhao2025envisioning}. They are widely used in evaluations for image editing tasks~\cite{bagel}.

\bsub{Baselines.} We use several baselines for text-guided image editing, including InstructPix2Pix~\citep{brooks2023instructpix2pix}, MagicBrush~\citep{zhang2023magicbrush}, HQ-Edit~\citep{hui2024hq}, {Step1X-Edit~\citep{liu2025step1x}, FLUX.1 Kontext~\citep{labs2025flux} and BAGEL~\citep{bagel}}.

\bsub{LGCC Setup.}
We fine-tune Bagel with LGCC using a curriculum learning approach.
Specifically, we apply a warm-up phase of 300 steps with only 25\% local Gaussian noise. Afterward, we use a combination of 50\% local Gaussian noise and 50\% normal noise for an additional 1,000 steps or fewer.
Each step processes 30–40 packed examples.

% HIVE~\citep{zhang2024hive}, InstructDiffusion~\citep{geng2024instructdiffusion}, MGIE~\citep{fu2023guiding}, 

\bsub{Evaluation Metrics.}
Following VIEScore~\citep{ku2024viescore}, we use two primary metrics: Semantic Consistency (SC) and Perceptual Quality (PQ). SC evaluates how well the edited image aligns with the text instruction, while PQ assesses the visual quality of the image.
To better evaluate the edited area, we introduce Local Semantic Consistency (LSC) and Local Perceptual Quality (LPQ), which focus on the cropped edited region.
Using the source image, text instruction, and edited output, Qwen2.5-VL-72B~\citep{bai2025qwen2.5vl} scores each metric on a scale of $[0,10]$. The Local Score (LScore) is the geometric mean of LSC and LPQ, while the overall score is calculated as the geometric mean of SC, PQ, LSC, and LPQ.

\subsection{Performance of LGCC}
\label{sec:eva_perf}

% We compare \mmname with the baseline methods (described in \S~\ref{sec:setup}) for the text-guided image editing tasks on the I$^2$EBench and GEdit-Bench datasets.

% \bsub{Performance on I$^2$EBench dataset.}

We evaluate $\mmname$'s image detail editing quality through benchmarks and case studies.
As shown in Table~\cref{tab:bagel_score_i2ebench}, $\mmname$ demonstrates competitive performance across a variety of text-guided image editing tasks on the I$^2$EBench dataset. Specifically, LGCC consistently excels in most categories, particularly in RegionAccuracy, DirectionPerception, and Replacement (see Appendix), where it outperforms or matches state-of-the-art baselines such as Bagel and Flux.
improving the overall scores by 0.53\%. Furthermore, LGCC demonstrates balanced improvements in both the SC (6.019) and PQ (6.101) metrics, highlighting its effectiveness in maintaining structural consistency and perceptual quality simultaneously.

% Our methods improve local detail scores of the I$^2$EBench by 1.60\% and overall scores by 0.53\%. 

\begin{figure}[t]
    \centering
    \includegraphics[width=\linewidth]{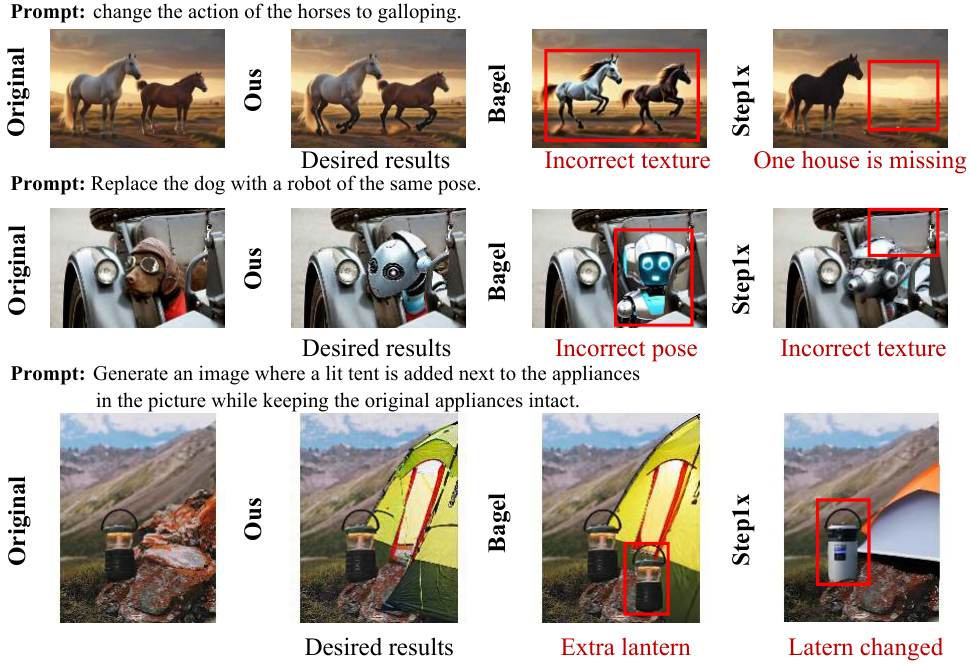}
    \vspace{-15px}
    \caption{Exploring image details across different works.}
    \label{fig:eva_detail}
\end{figure}

One of $\mmname$'s key advantages is its ability to preserve the details and context of the edited areas. As shown in \cref{fig:eva_detail}, $\mmname$ adheres to instructions to accurately generate the desired results while keeping non-target objects unchanged. This advantage is also evident in the benchmark scores, particularly in the LSC and LPQ metrics in \cref{tab:bagel_score_i2ebench}, where LGCC improves the local detail scores of the I$^2$EBench by 1.60\%.
Another advantage of $\mmname$ is its efficiency, requiring only 40\% of the inference time of Bagel and 50\% of the inference time of Flux. As analyzed in \S~\ref{sec:eva_spend}, $\mmname$ can achieve satisfactory results within 10 to 20 steps, while Flux and Bagel require significantly more steps to produce similar-quality results. Despite its faster inference, $\mmname$ achieves comparable performance in global image effects and demonstrates slight improvements in local editing areas compared to previous state-of-the-art methods like Bagel.
Although $\mmname$'s SC score is slightly lower than Bagel's, its PQ score is higher, indicating stronger perceptual quality. We provide an in-depth analysis using the GEdit-Bench dataset, as detailed in the Appendix.

\begin{table}[t]
\centering
\resizebox{\linewidth}{!}{%
\begin{tabular}{c|ccccc} 
\toprule
Model & (SC, PQ)$\uparrow$ & (LSC, LPQ)$\uparrow$ & LScore$\uparrow$ & Overall$\uparrow$ \\ 
\midrule
Step1X-Edit~\citep{liu2025step1x} & (5.144, 5.907) & (5.831, 6.080) & 5.160  & 5.735\\
FLUX.1 Kontext~\citep{labs2025flux} & (5.875, 6.379) & (5.418, 6.346) & 4.852 & 5.994\\
BAGEL~\citep{bagel} & (6.255, 5.918) & (6.418, 5.667) & 5.375 &6.058\\
LGCC(ours) & (6.019, 6.101) & (6.359, 5.886) & \textbf{5.461} & \textbf{6.090} \\
% LGCC(ours, lightweight) & (7.337, 7.079) & (6.359, 5.886) & 5.461 &6.090 \\
\bottomrule
\end{tabular}
}
\caption{Performance scores on I$^2$EBench and MagicBrush dataset. For each score, the high the better. All scores are evaluated by Qwen2.5-VL-72B.}
\label{tab:bagel_score_i2ebench}
\end{table}

% \begin{table}[t]
%     \centering
%     \resizebox{0.5\textwidth}{!}{%
%     \begin{tabular}{c|c c c}
%      \toprule
%      Model    &  SC & PQ & Overall \\
%      \midrule
%      InstructPix2Pix~\citep{brooks2023instructpix2pix} & 4.746 & 6.913 & 4.578 \\
%      MagicBrush~\citep{zhang2023magicbrush} & 5.752 & 7.069 & 5.558\\
%      % AnyEdit & 3.713 & 6.730 & 3.635 \\
%      % OmniGen & 6.845 & 6.700 & 6.352 \\
%      \midrule
%      Gemini 2.0~\citep{Gemini2.0} & 7.274 & 7.327 & 6.971\\
%      Doubao~\citep{doubao} & 7.353 & 7.651 & 7.230 \\
%      GPT-4o~\citep{gpt-4o} & 7.847 & 7.705 & 7.692\\
%      \midrule
%      % Step1X-Edit~\citep{liu2025step1x} & 7.388 & 7.279 & 7.067\\
%      Step1X-Edit~\citep{liu2025step1x} & 6.950 & 7.260 & 6.672\\
%      FLUX.1 Kontext~\citep{labs2025flux} & 7.025 & 7.000 & 6.672\\
%      BAGEL~\citep{bagel} & 7.454 & 7.044 & 7.021\\
%      LGCC(ours) & 7.319 & 7.159 & 6.948\\
%      % \midrule
%      % Fast-BAGEL 1500 & 7.097 & 6.766 & 6.577\\
%      \bottomrule
%     \end{tabular}
%     }
%     \caption{Performance scores on GEdit-Bench dataset. For each score, the high the better. All scores are evaluated by Qwen2.5-VL-72B.\fb{include private model gemini,gpt?}}
%     \label{table_bagel_score_gedit}
% \end{table}

% \textbf{Component wise ablation study }
% LGCC (OFM without context-preserving )
% LGCC (Full)
% A few sets of experiments:
% \begin{itemize}
%     \item performance score on GEdit-Bench by QWen2.5, comparative;
%     \item 40\% inference time of Bagel; 
%     \item 50\% inference time of FLUX.1-Kontext;
% \end{itemize}

% 1. 放猫的图，
\subsection{Qualitative Analysis of Context Preserving}
\label{sec:eva_quality}
% Detail Preservation and Context Consistency

Over-editing issues are common with MLLM based image editing tools (e.g., Bagel, Step1x), as they often modify more areas than expected to make the picture appear logical or coherent. Unfortunately, judging such over-editing is challenging with existing benchmarks, as these cases are subjective. 
It is unclear whether the focus should be on preserving more of the original image or aligning more closely with the prompt. Over-editing becomes obvious only when noticeable inconsistencies or errors arise in the edited image.

\begin{figure}[t]
    \centering
    \includegraphics[width=\linewidth]{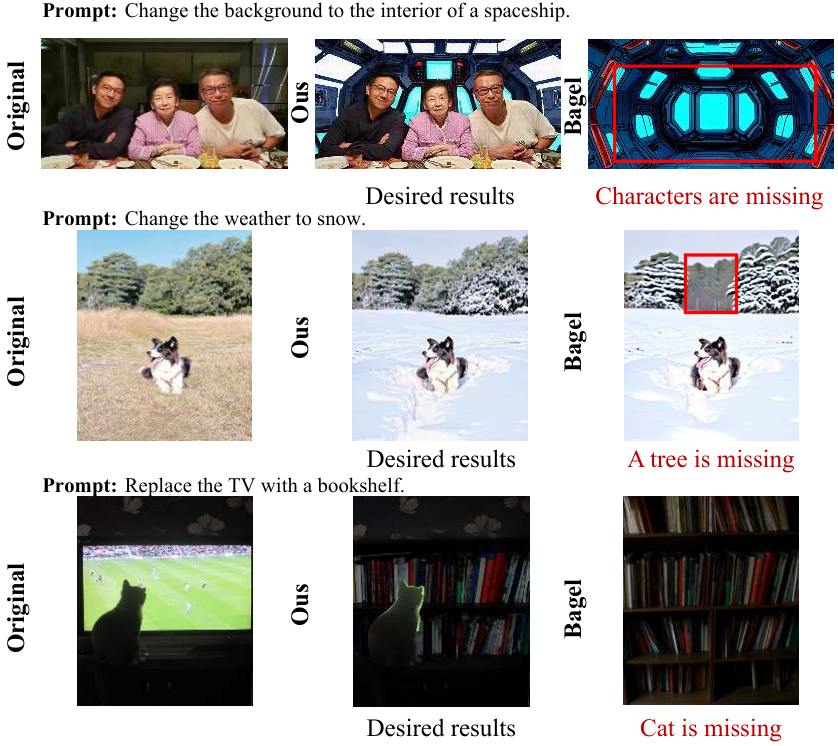}
    \vspace{-15px}
    \caption{Typical over-edit cases by exiting works.}
    \label{fig:over-edit}
\end{figure}

We study real-world examples to qualitatively analyze why existing tools often fail to preserve image context and instead produce over-edited results. 
For these tools, the editing area (as shown in \cref{fig:limtation-sample}) is difficult to constrain to the target object, leading to unintended alterations of nearby elements. 
As illustrated in \cref{fig:over-edit}, this issue becomes even more pronounced when the target object blends with the background.
The root cause of this issue is that these  tools are overfitted to the target label's image context, which is a common flaw in existing MLLM-based approaches.
In contrast, stable diffusion-based methods, such as Flux, are less prone to this problem since they lack the capability for fine-grained editing of individual elements, allowing them to preserve the original context most of the time.
Our method, \mmname, addresses this problem by retaining the context of both the original image and the provided instructions through flow-matching training loss. As a result, \mmname can effectively balance precise editing with context preservation, mitigating the over-editing issue.

% These tools 
% Especially, when the target object is fusion with the background, making it challenging to precisely edit the target area while maintaining the surrounding context. This issue becomes even more pronounced in complex scenarios (as).

% We study the real-world samples to qualitatively analyze how existing tootls failed at preserve the image context and produce over-edit results.
% As the example in \cref{fig:limtation-sample}, the editing area are difficult to be limited in the basets but also alter the nearby fruits.
% The root cause is that  when the target object is fusing in the background, it is difficult to precise edit the target area while keep other context.
% This phenomenon are also confirmed by more complex cases (as shown in \cref{fig:over-edit}).
% \mmname can efficiently mitigate this problem by keep the context from both the original image and the instructions during flowing matching's training loss.

\begin{figure}[t]
    \centering
    \includegraphics[width=\linewidth]{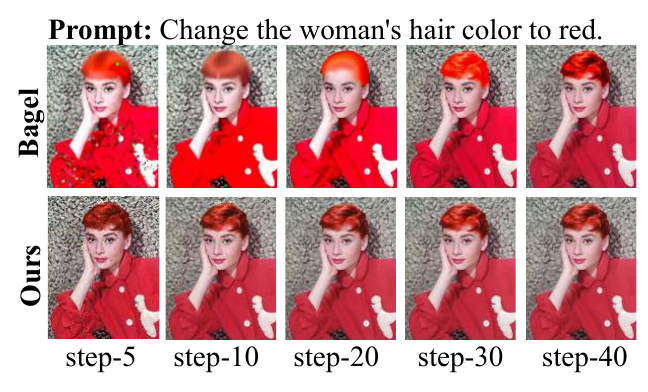}
    \vspace{-25px}
    \caption{Bagel vs LGCC's results of various steps.}
    \label{fig:steps}
\end{figure}

\begin{table}[t]
\centering
\resizebox{0.9\linewidth}{!}{%
\begin{tabular}{ccccccc} 
\toprule
Steps & 5 & 10 & 20 & 30 & 50 & 70 \\ 
\midrule
BAGEL & 36.50 & 32.93 & 29.70 & 29.68 & \textbf{29.16} & 30.24 \\
LGCC & 30.90 & 27.93 & \textbf{26.65} & 26.76 & 26.83 & - \\
Flux & 37.54 & 37.20 & \textbf{36.78} & 36.82 & - & - \\
\bottomrule
\end{tabular}
}
\vspace{-5px}
\caption{Brisque Image quality (lower is better) across different steps.}
\label{tab:required_steps}
\end{table}

\begin{table}[ht]
    \centering
    \resizebox{0.9\linewidth}{!}{%
    \begin{tabular}{ccc}
    \toprule
       Models & Latency &  Brisque Quality$\uparrow$ \\
       \midrule
       BAGEL(50 step, demo setup) & 54.46s & 6.058 \\
       Step1X (28 step, demo setup) & 82.34s & 5.735 \\
       Flux Kontext (28 step, demo setup) & 35.42s  &5.994\\
       LGCC (25 step, demo setup) & \textbf{27.12s} & \textbf{6.090} \\
       LGCC (10 step, lightweight edit) & 12.22s & 5.826 \\
    \bottomrule
    \end{tabular}
    }
    \caption{Inference latency on 1 * A100-PCIE-40G GPU with 800x1024 images.}
    \label{tab:step_quality}
\end{table}

% : Less Steps and Faster Inference
\subsection{The Inference Steps Required by LGCC}
\label{sec:eva_spend}
% 25 steps

% 由于初始噪声和降噪强度的差异，不同模型在进行去噪生成的时候也有不同的最佳inference step。在step过多超过了这一最佳step的情况下，模型反而会在最优解附近震荡导致过拟合，使得图像的质量反而下降。
% Different models exhibit distinct optimal inference steps for the denoising generation process, owing to variations in initial noise and denoising strength. Exceeding this optimal step count can cause the model to oscillate around the optimal solution, leading to overfitting and a paradoxical degradation in image quality.
% Each tools needs to set an approaite interfence steps during image editing.
% Note that more steps do not means better qualitys.
% This is because extra steps can lead to the  oscillate around the optimal solution
% Different models require different optimal inference steps during the denoising generation process due to differences in initial noise levels and denoising strength. If the number of steps exceeds this optimal point, the model may begin to oscillate around the optimal solution. This oscillation can result in overfitting, ultimately reducing the quality of the generated image instead of improving it.

Each tool needs to set appropriate inference steps during image editing. Note that more steps do not necessarily lead to better quality. This is because additional steps can result in over-adjustment around the optimal solution, potentially leading to a decline in image quality.
As shown in \cref{tab:required_steps}, we investigate the number of steps required to achieve optimal image quality for different tools. The results indicate that Flux and \mmname require approximately 20 to 30 steps, while Bagel needs 50 steps.

Based on these findings, we empirically choose the most suitable step count for each tool.
As shown in \cref{tab:step_quality}, \mmname's default demonstration is set to 25 steps, whereas Bagel is configured to run 50 steps by default.
Consequently, $\mmname$ requires only 40\% of the inference time of Bagel and 50\% of the inference time of Flux. 
The comparison of image quality between Bagel and $\mmname$ is further illustrated in \cref{fig:steps}, where $\mmname$'s 25-step output is comparable to Bagel's 50-step result.
\mmname is significantly more cost-efficient due to its reduced inference workload.
For real-world deployment, $\mmname$ can be configured to use 10 steps for scenarios with small edit areas or whole-picture style transformations with minimal overall structural changes.

\section{Conclusion}
% In this paper, we present \mmname, which adopts .

We propose the LGCC framework to address the challenges of detail loss and context inconsistency in image editing. Built on the MLLM flow matching framework, LGCC incorporates Local Gaussian Noise Coupling (LGNC) and Content Consistency Loss (CCL) to mitigate detail degradation and over-editing. By fine-tuning BAGEL with a small dataset, LGCC improves the quality of edited areas while maintaining strong overall performance. Case studies reveal significant reductions in detail degradation and over-editing. Additionally, LGCC achieves a 2x speedup in universal editing with improved overall scores and a 3x–5x boost in lightweight editing, demonstrating its efficiency and effectiveness.

\clearpage
\appendix
\begin{figure}[htbp!]
    \centering
    \resizebox{0.4\textwidth}{!}{
    \begin{tabular}{c c}
    Input Image & Ours\\ 

        \includegraphics[width=0.15\textwidth]{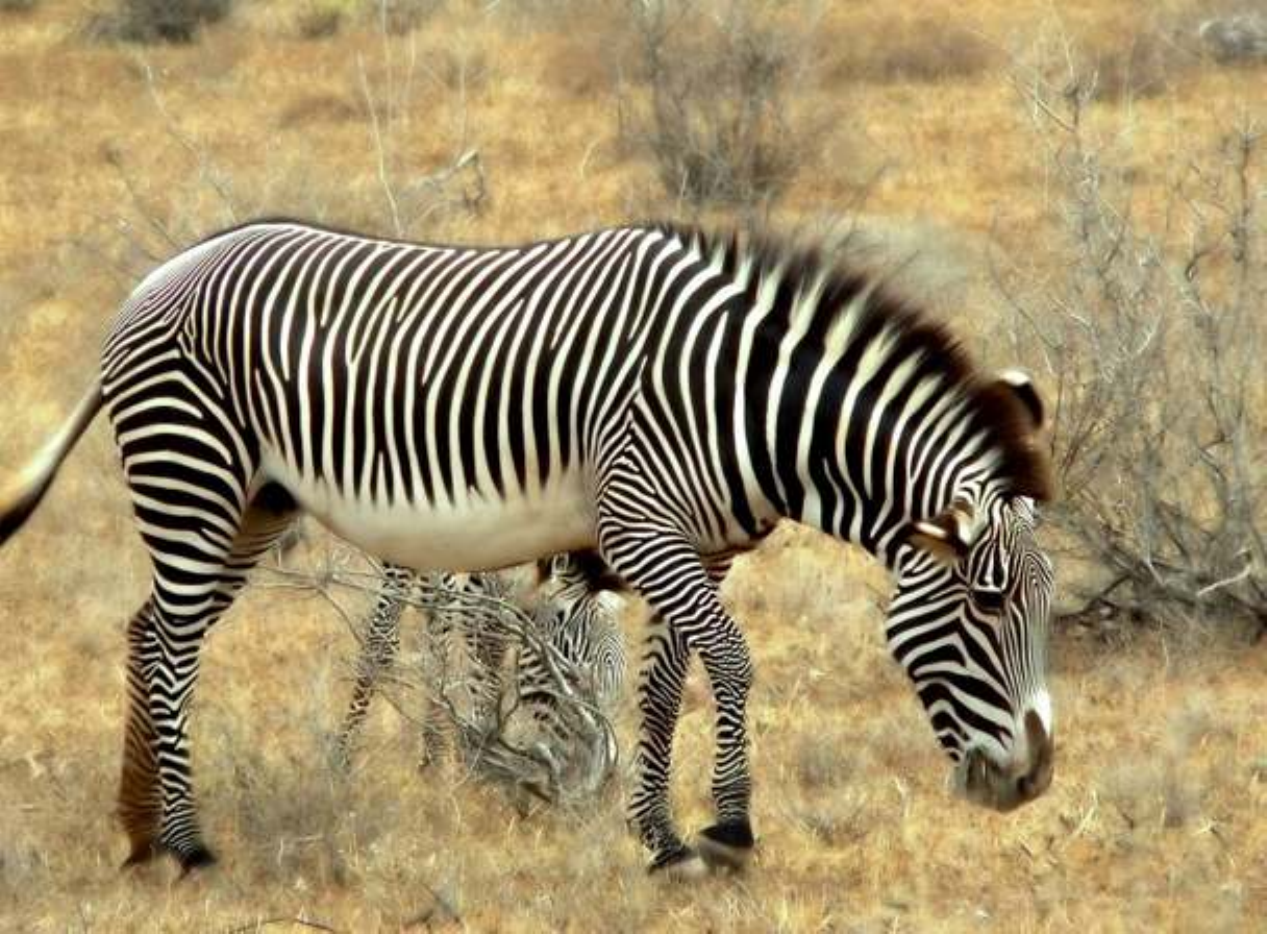} &
        \includegraphics[width=0.15\textwidth]{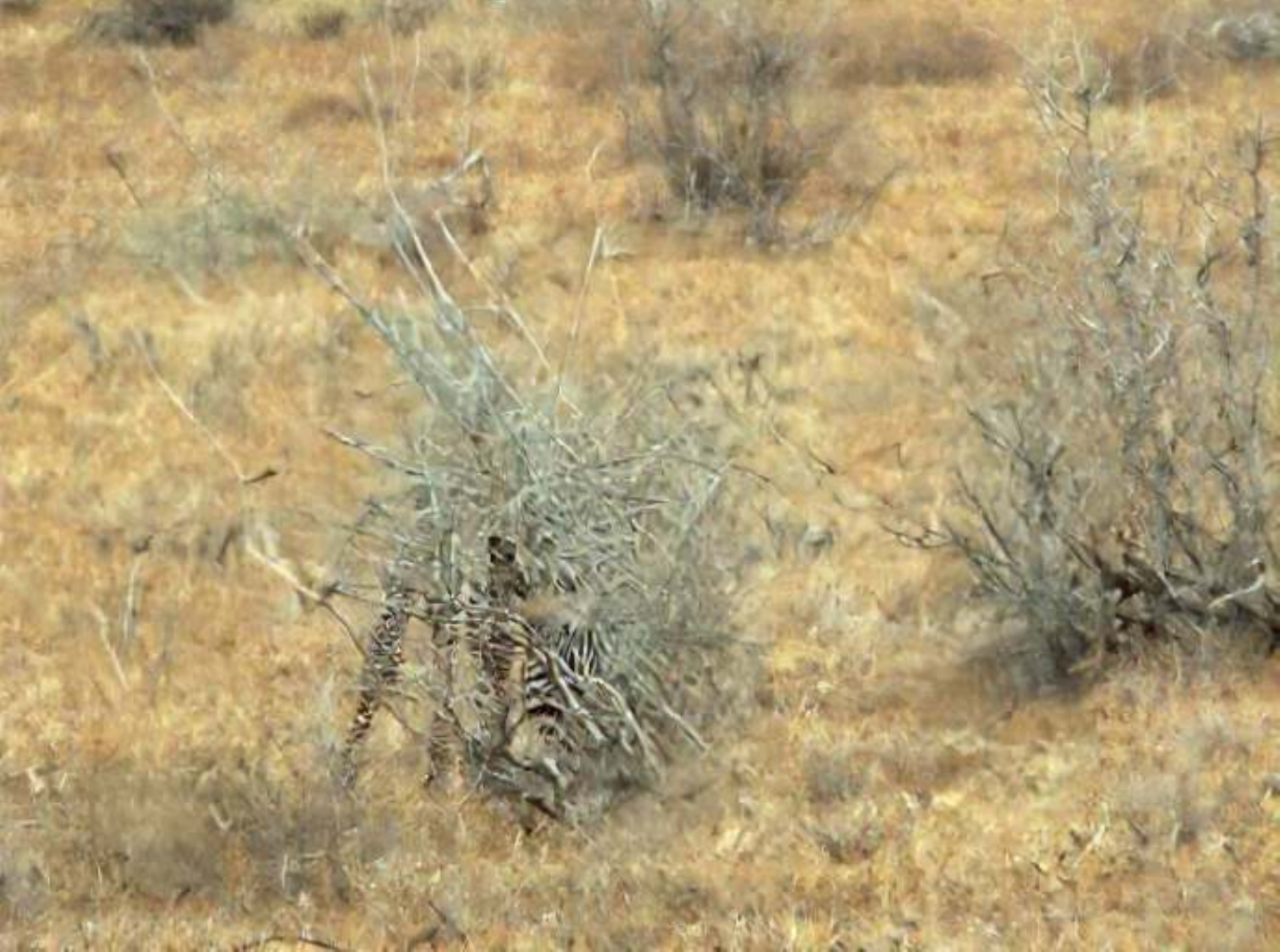} \\
  
        \multicolumn{2}{c}{erase the zebra.}\\
    
        \includegraphics[width=0.15\textwidth]{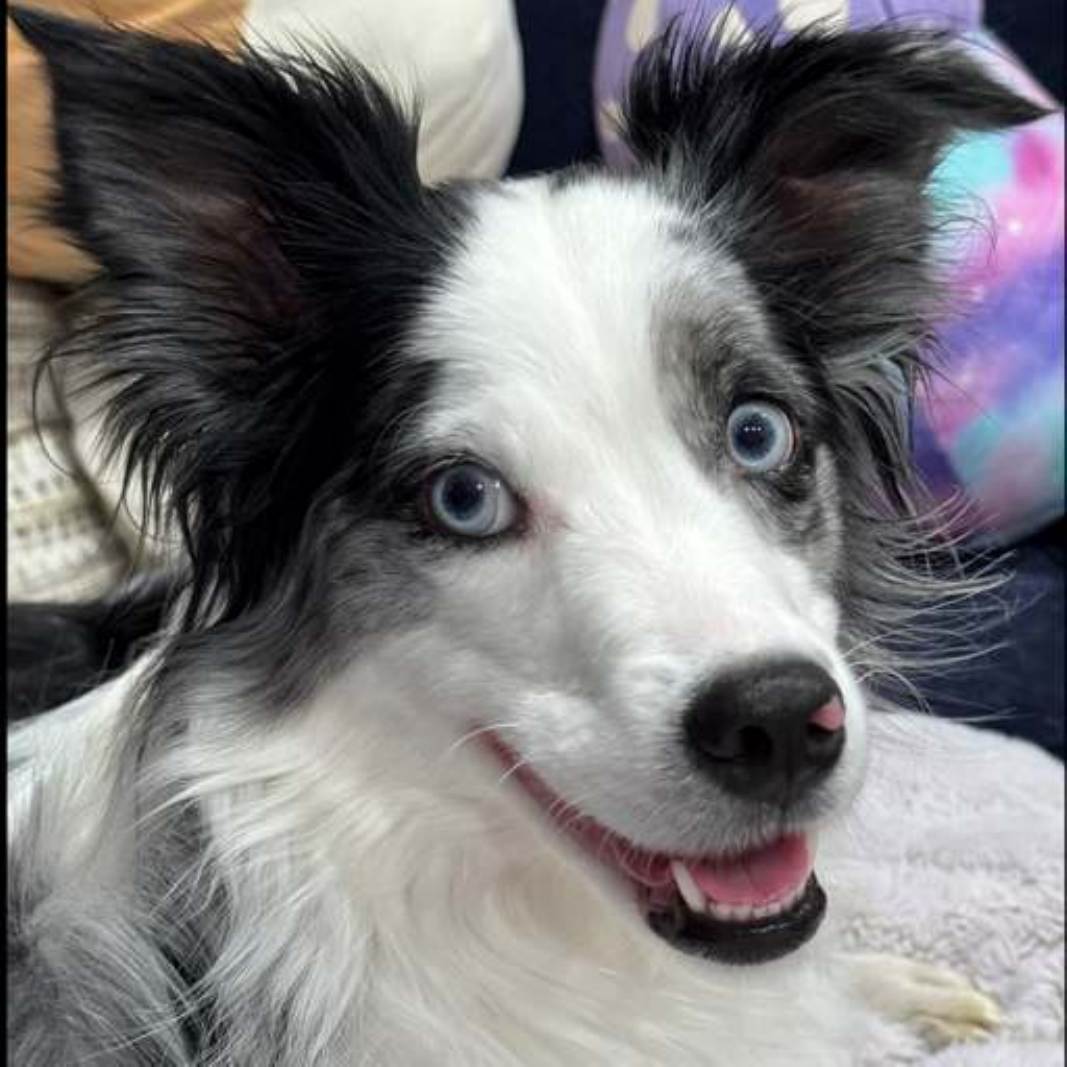} &
        \includegraphics[width=0.15\textwidth]{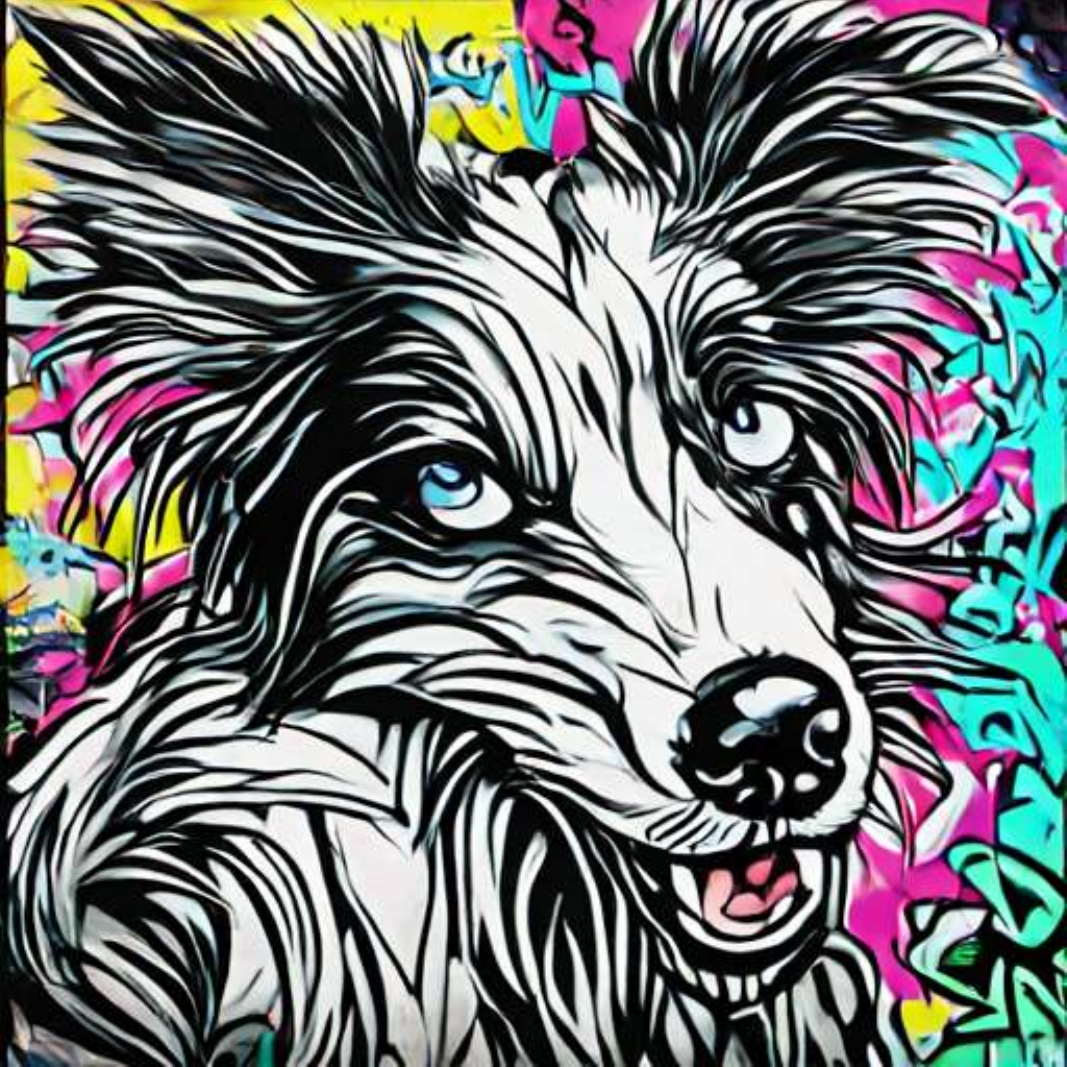} \\

        \multicolumn{2}{c}{Restore and colorize the image.} \\

        \includegraphics[width=0.15\textwidth]{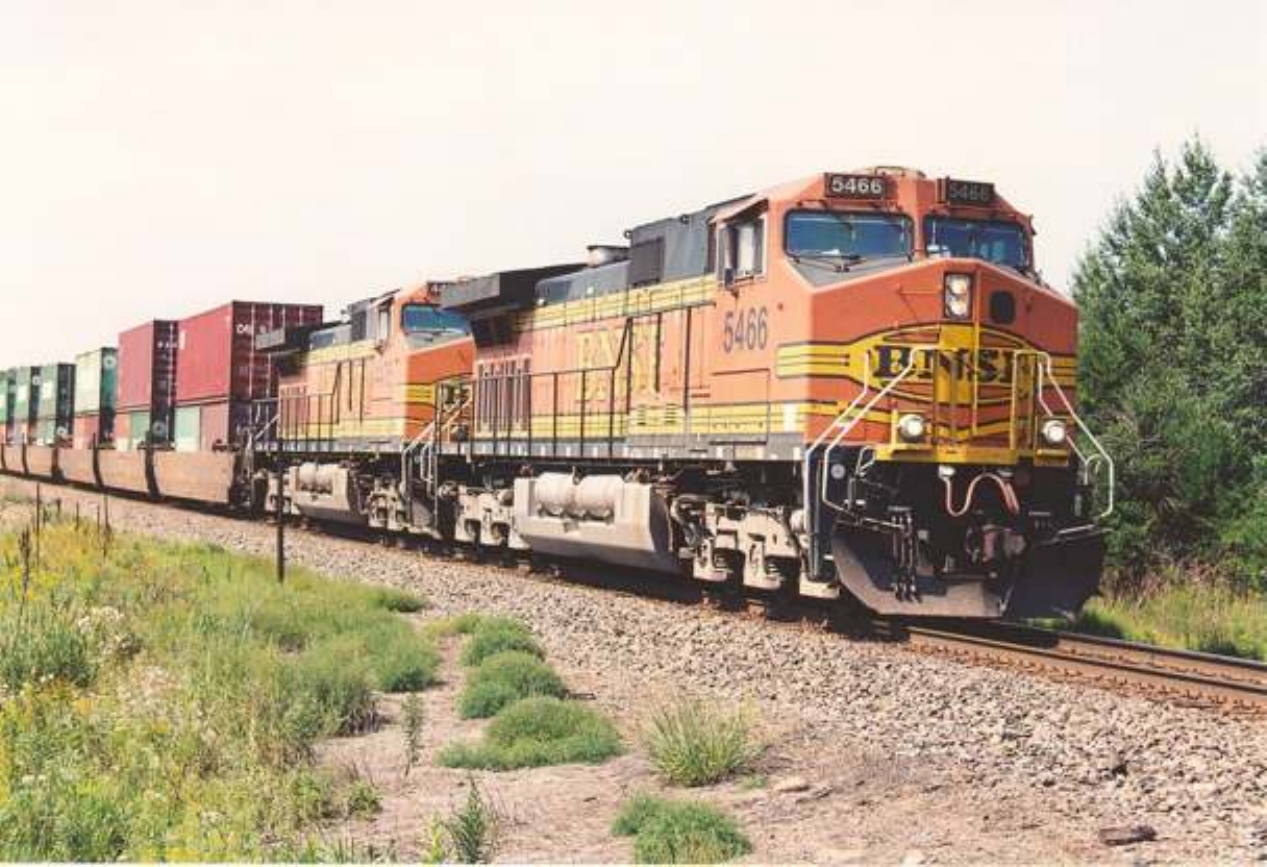} &
        \includegraphics[width=0.15\textwidth]{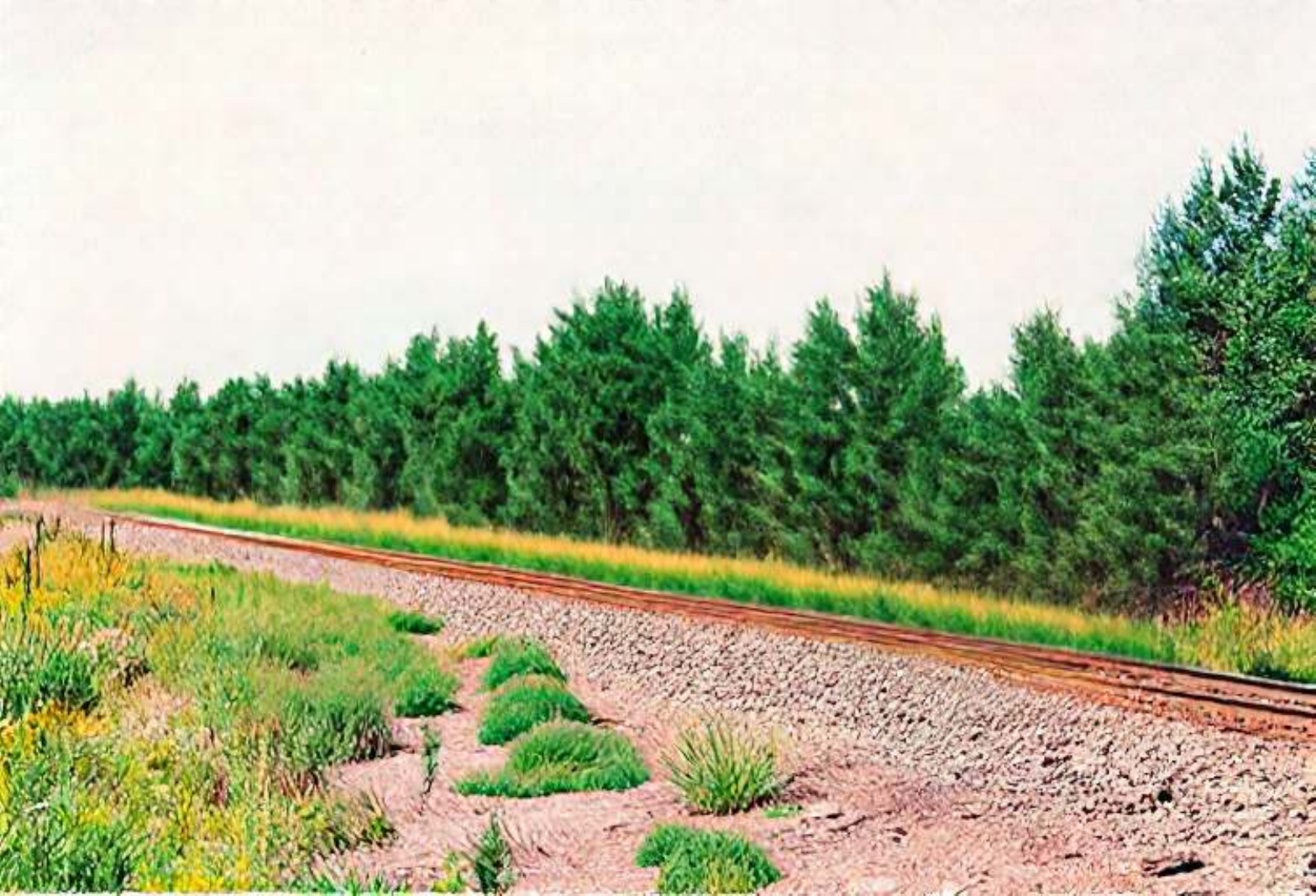} \\

        \multicolumn{2}{c}{Remove the freight train.}\\
 
        \includegraphics[width=0.15\textwidth]{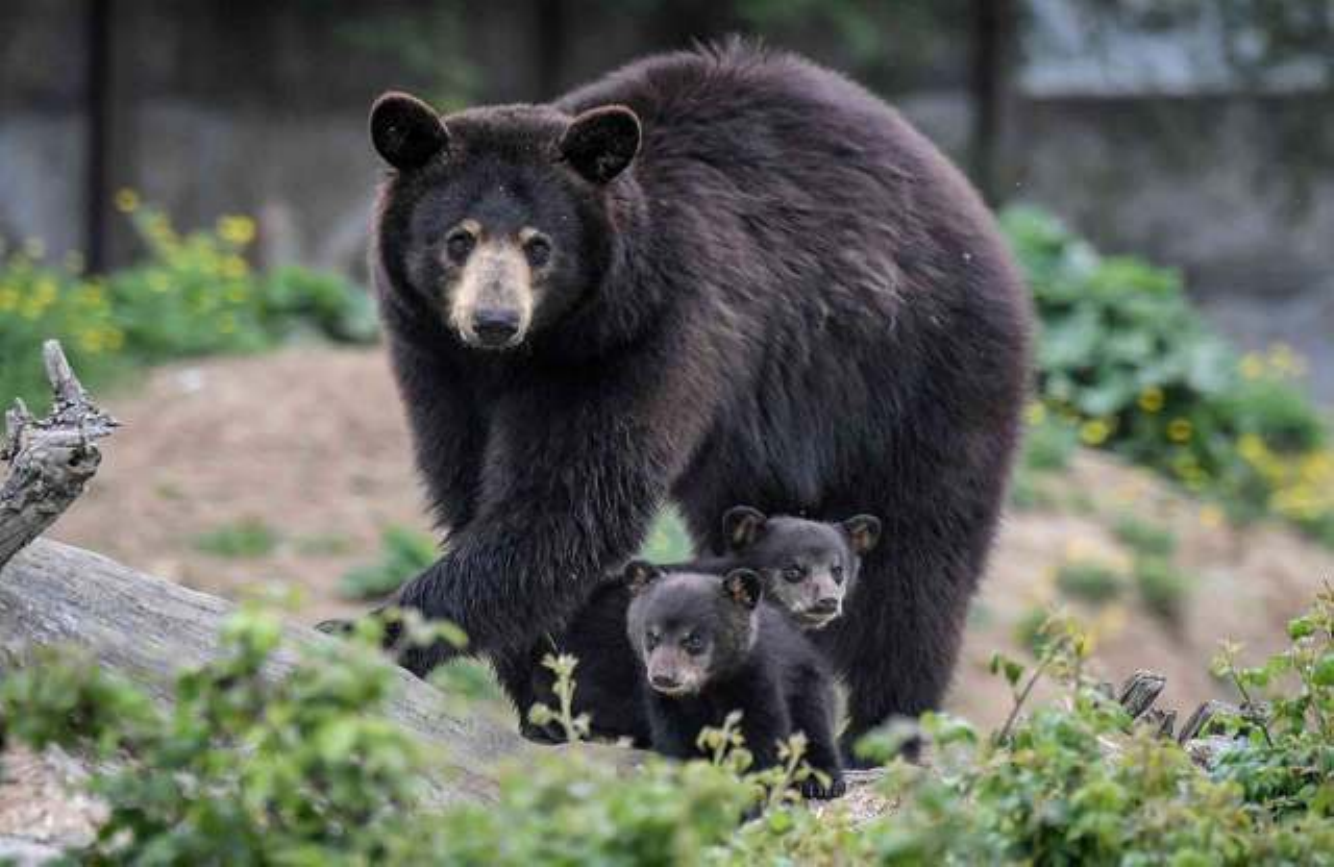} &
        \includegraphics[width=0.15\textwidth]{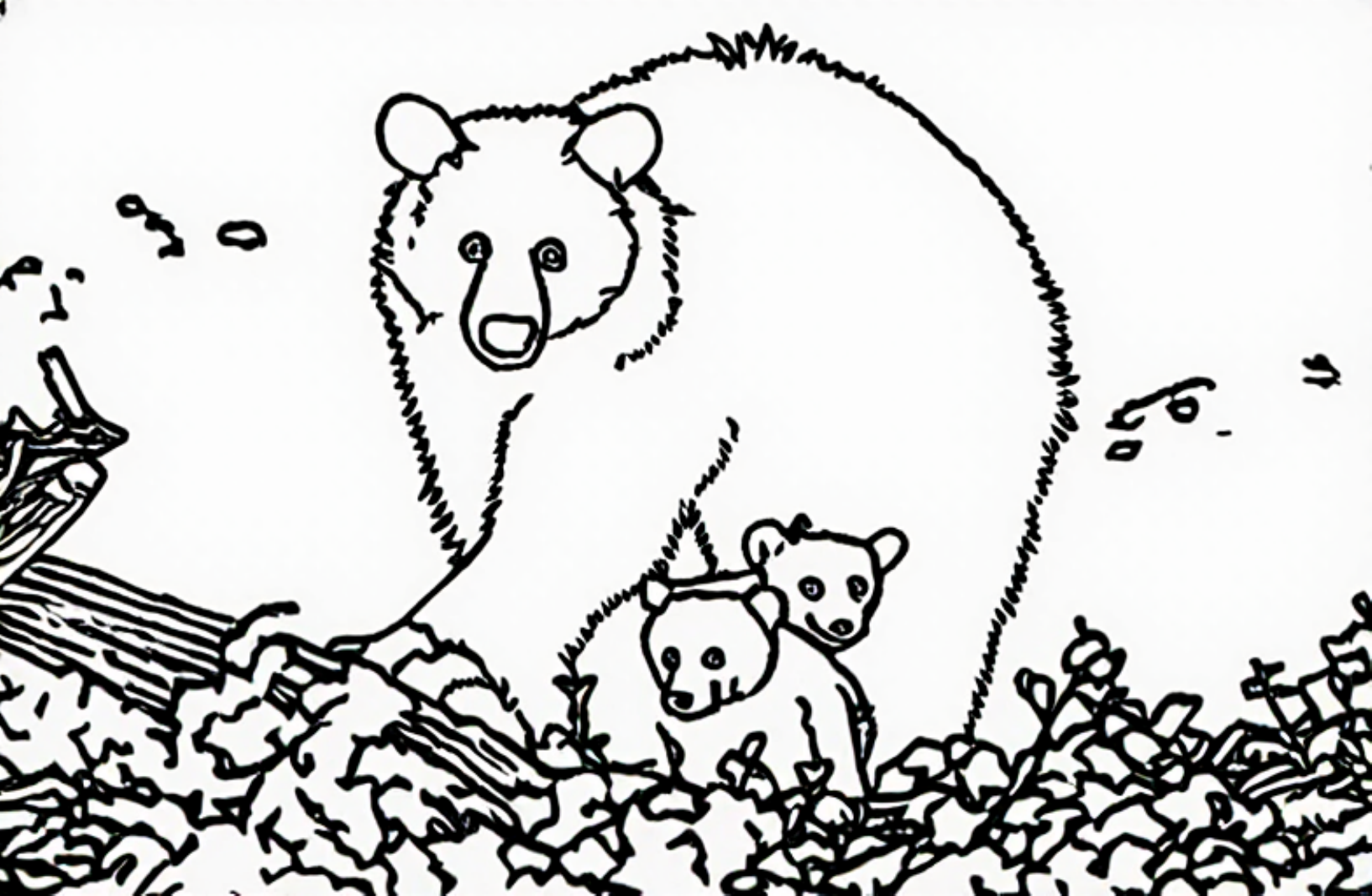} \\
  
        \multicolumn{2}{c}{Modify the image style into line art.}\\

        \includegraphics[width=0.15\textwidth]{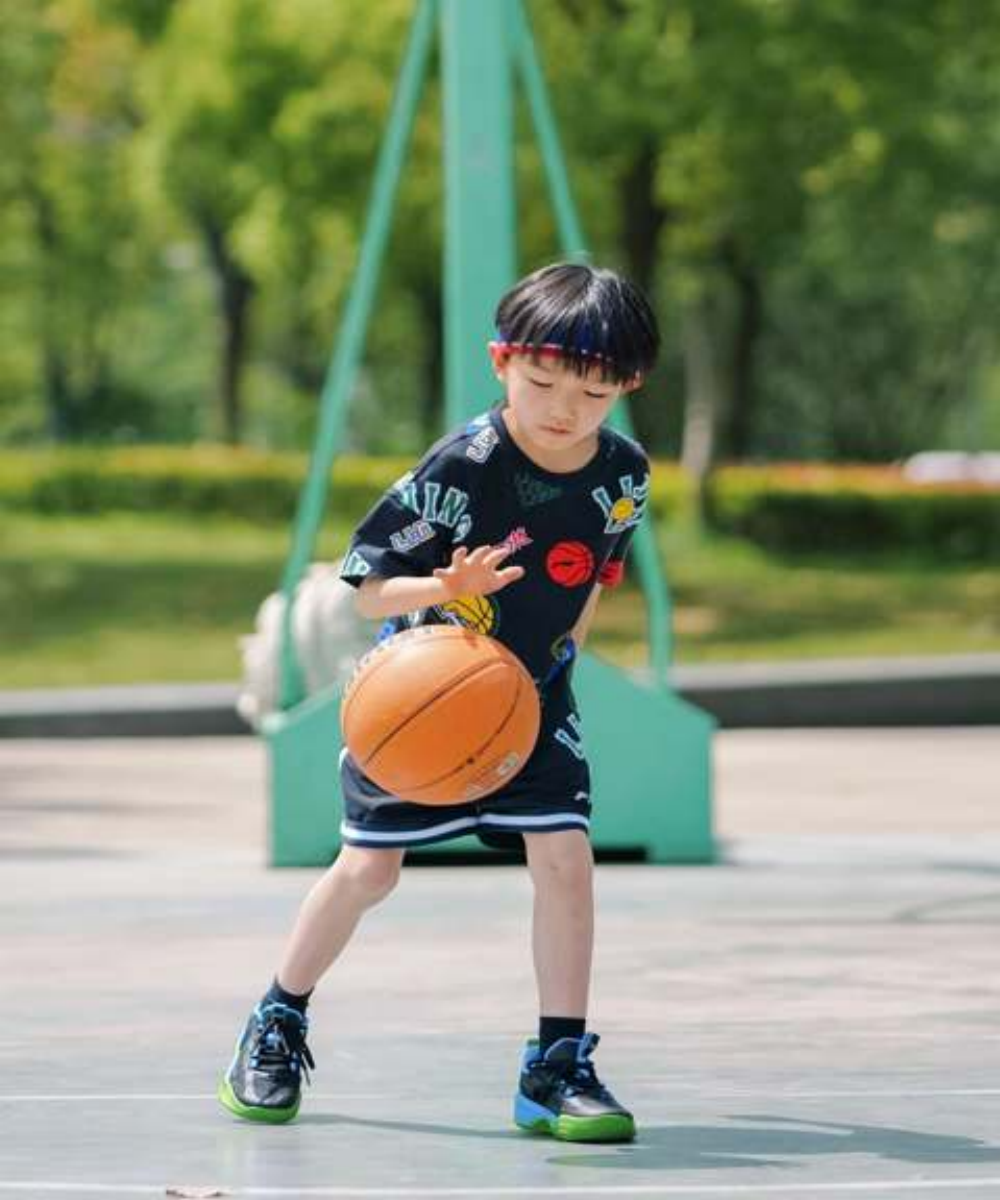} &
        \includegraphics[width=0.15\textwidth]{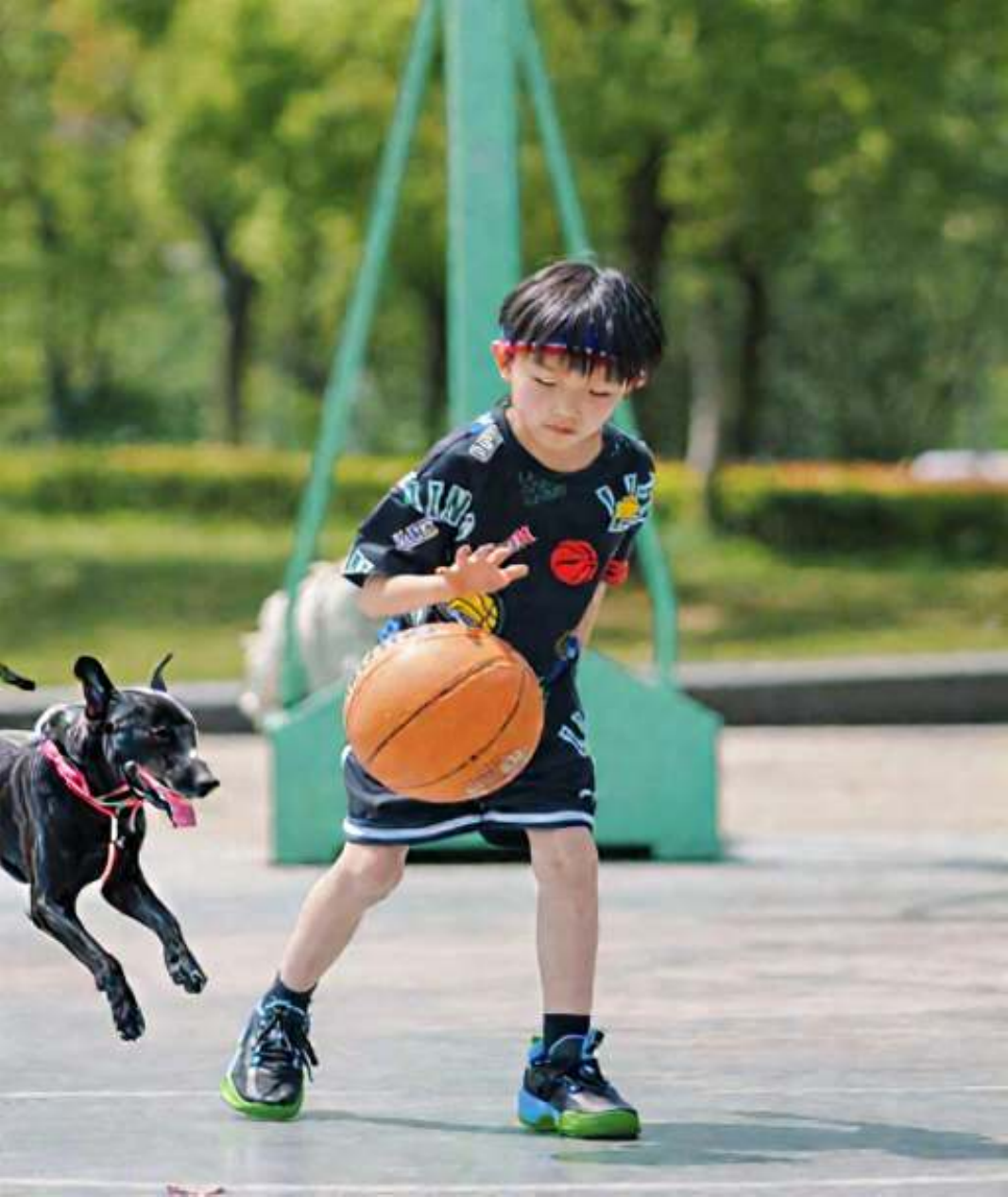}  \\
        \multicolumn{2}{c}{Include a dog running alongside.} \\

        \includegraphics[width=0.15\textwidth]{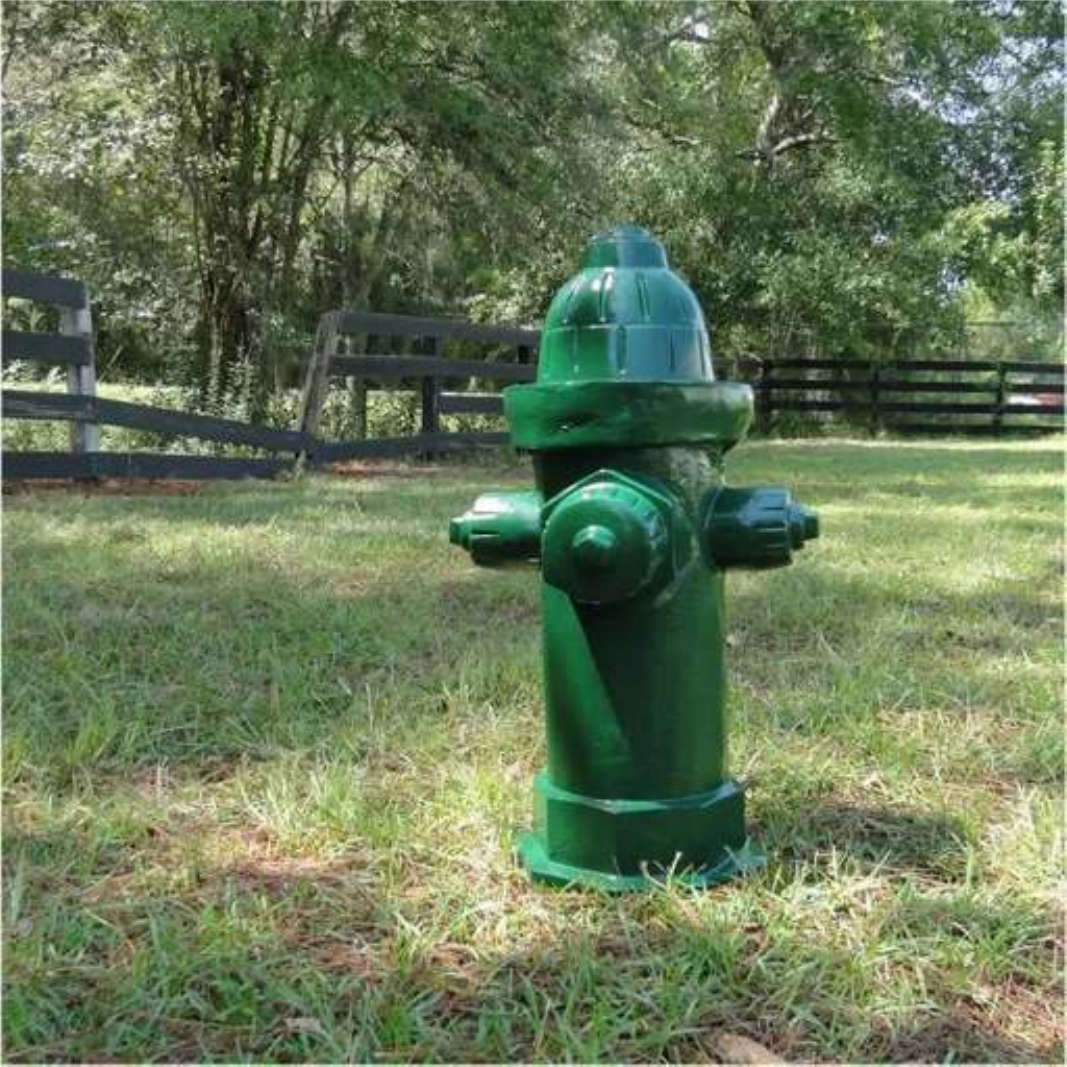} &
        \includegraphics[width=0.15\textwidth]{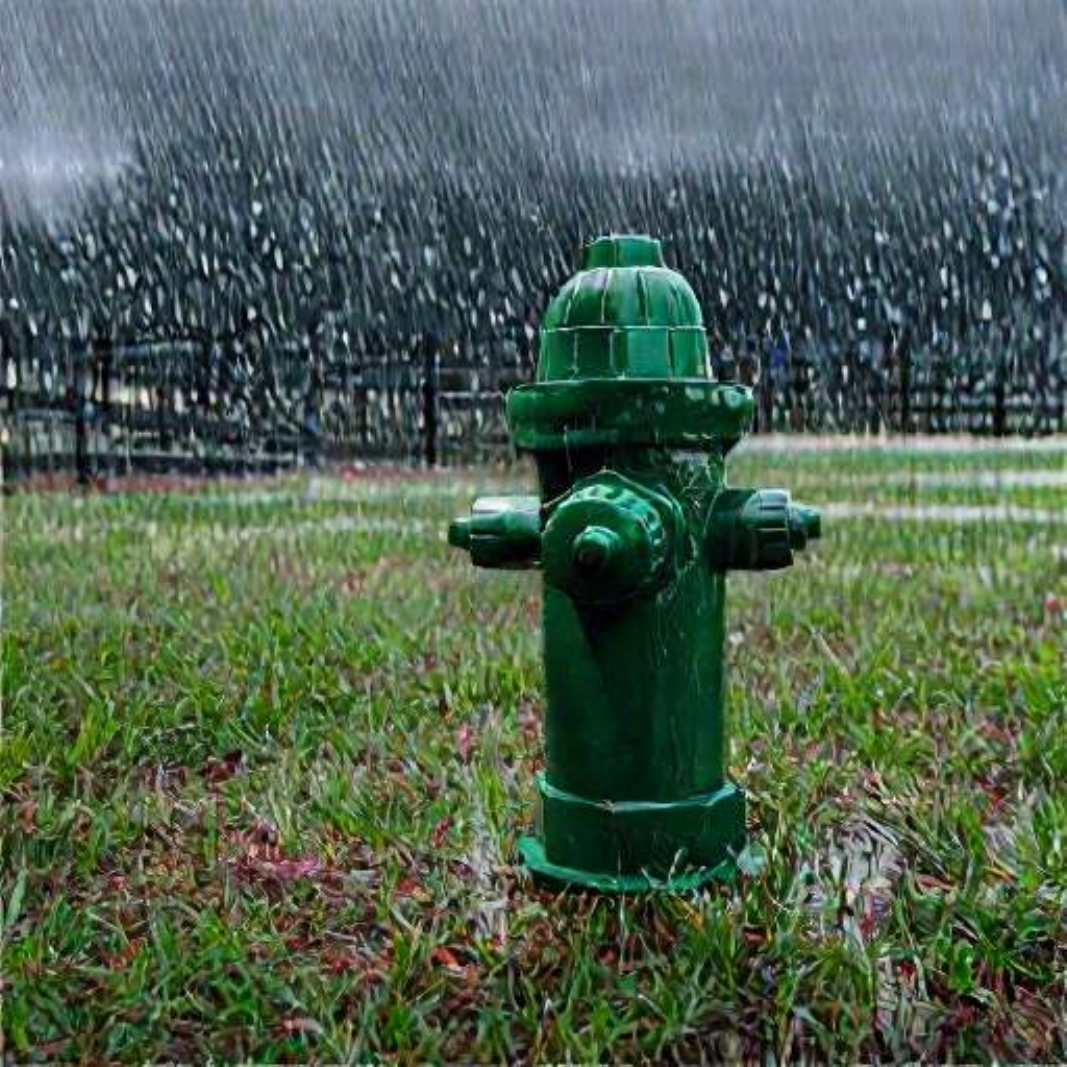} \\
        \multicolumn{2}{c}{Change the weather to heavy rain.} \\
        
    \end{tabular}
    }
    \caption{Demo images.}
    \label{figure-more-demos-1}
\end{figure}

\begin{figure}[htbp!]
    \centering
    \resizebox{0.5\textwidth}{!}{
    \begin{tabular}{c c}
    Input Image & Ours\\

        \includegraphics[width=0.21\textwidth]{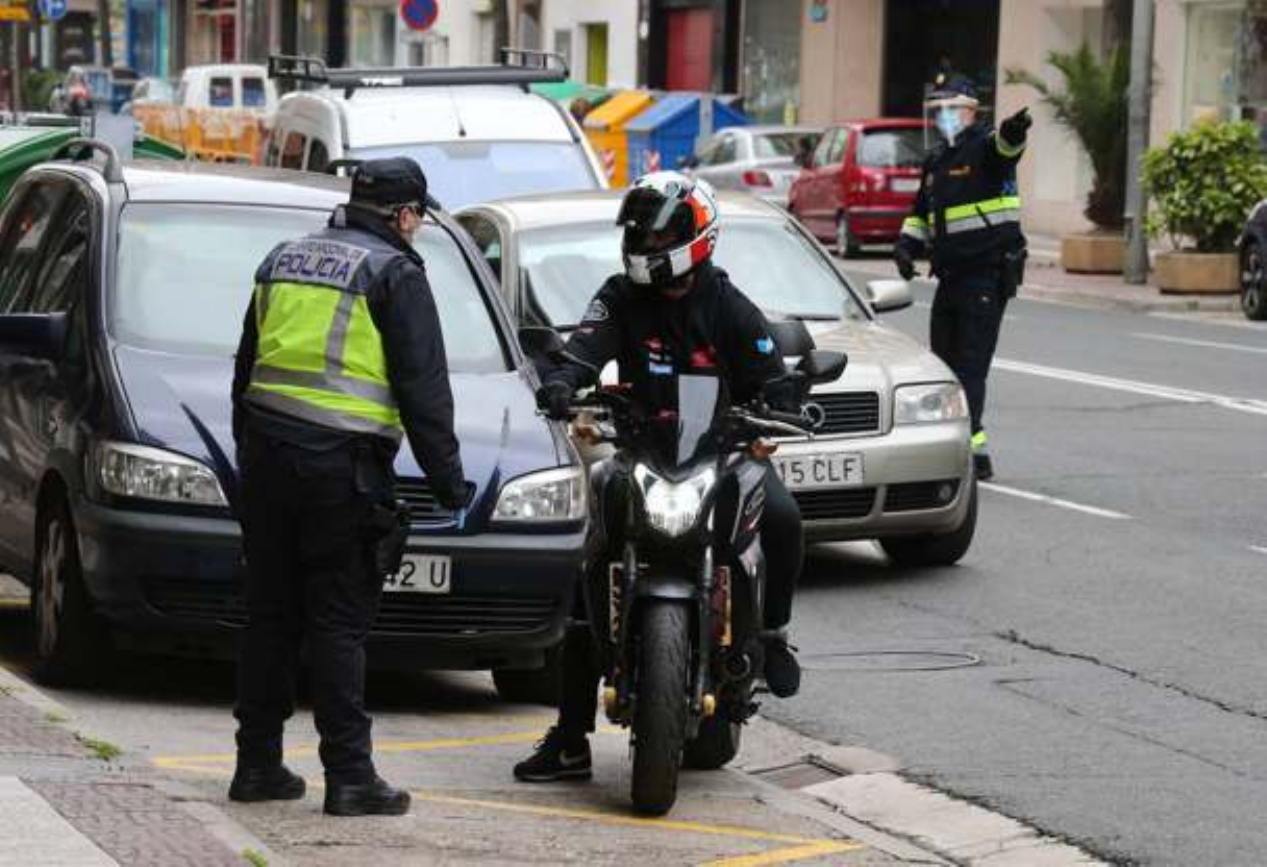} &
        \includegraphics[width=0.21\textwidth]{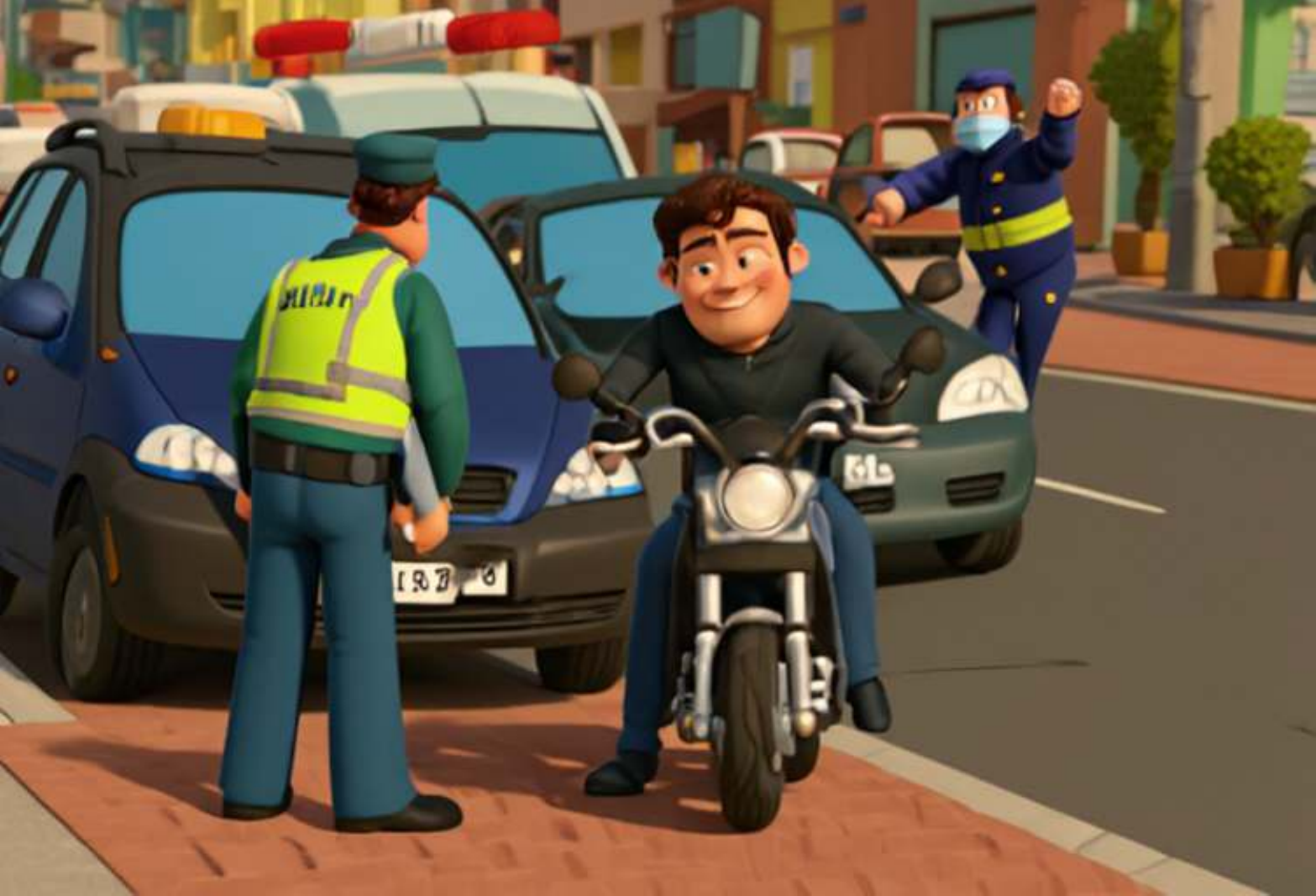} \\

        \multicolumn{2}{c}{Render with Pixar Animation Studios' 3D texture.} \\

        \includegraphics[width=0.21\textwidth]{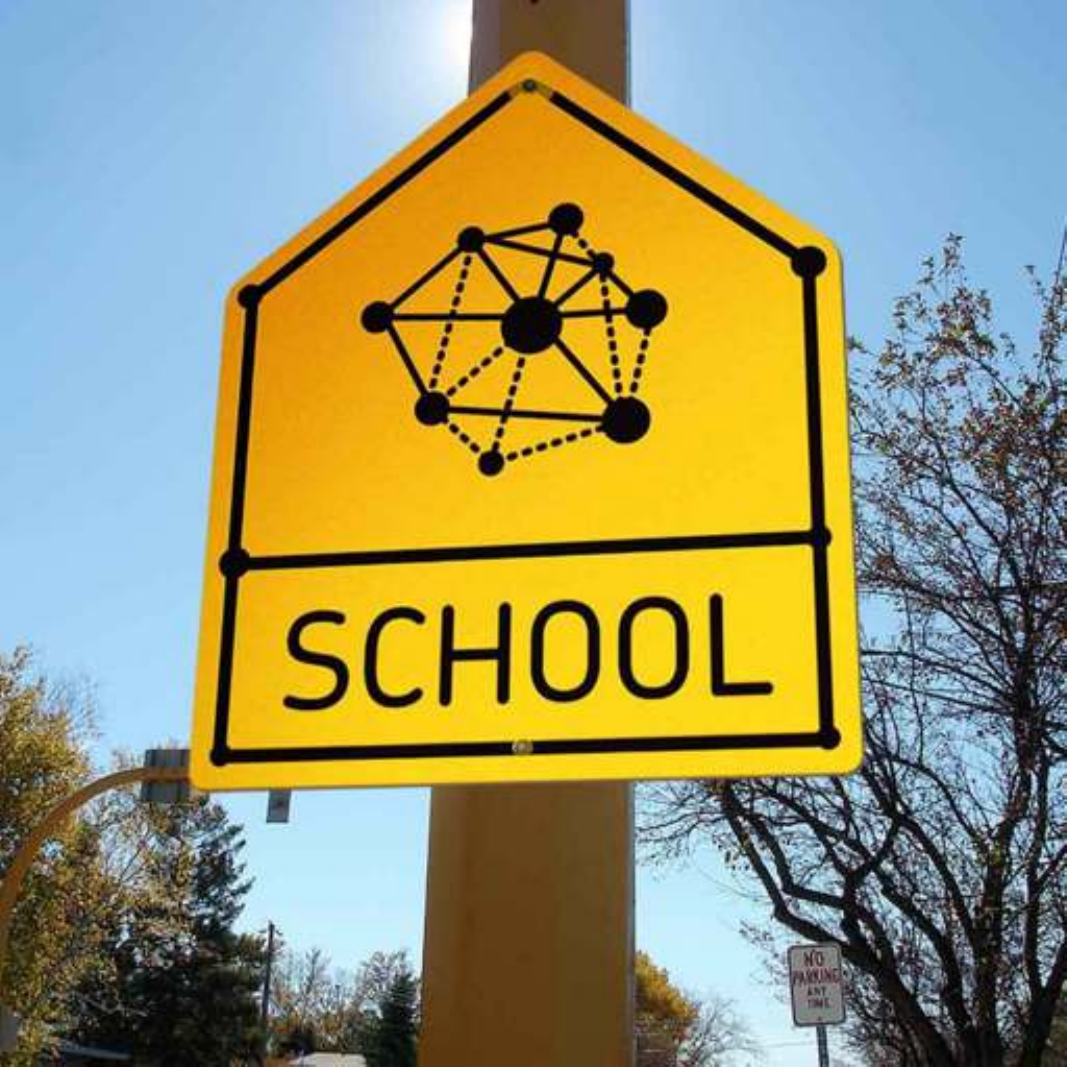} &
        \includegraphics[width=0.21\textwidth]{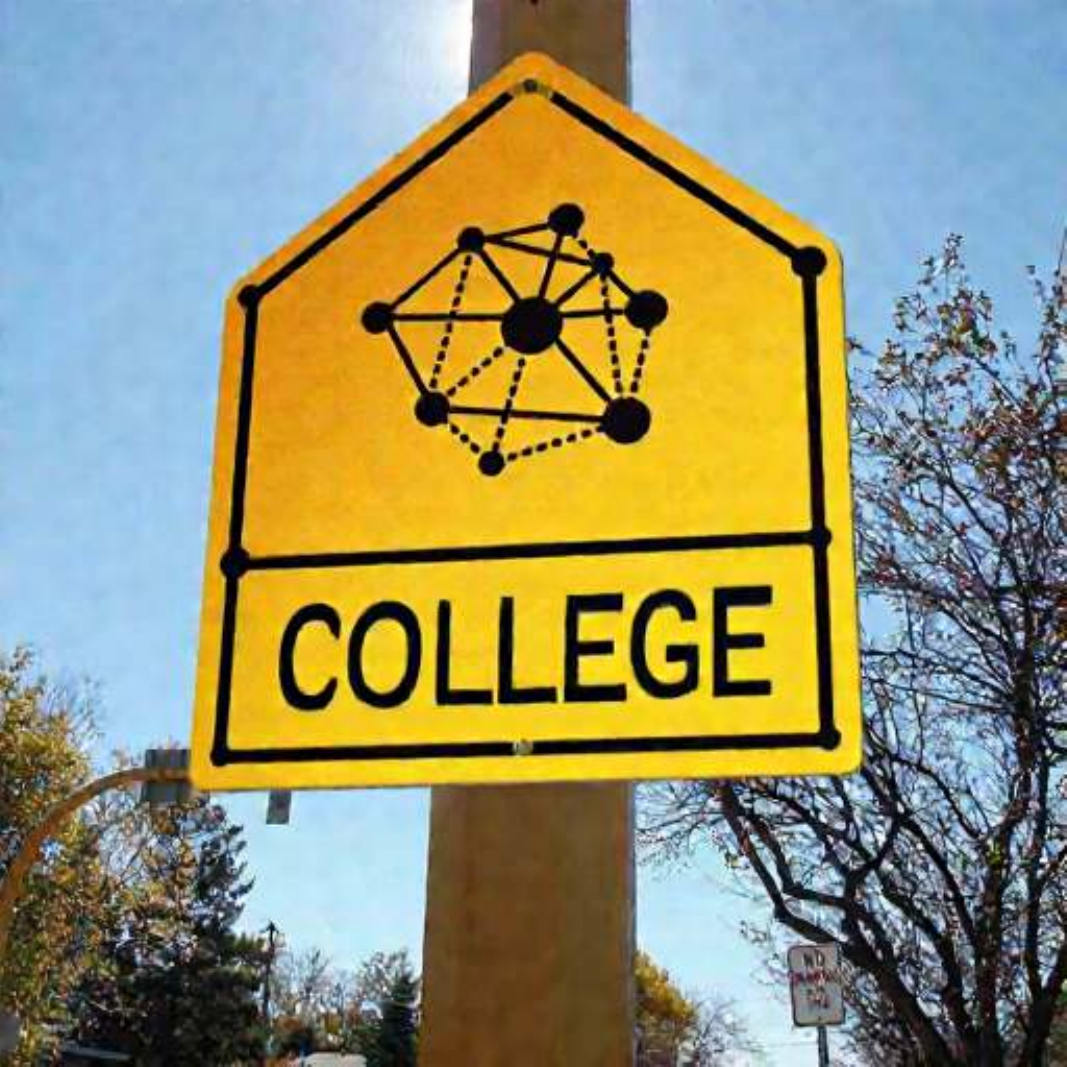} \\
   
        \multicolumn{2}{c}{Change the text school to college.} \\

        \includegraphics[width=0.21\textwidth]{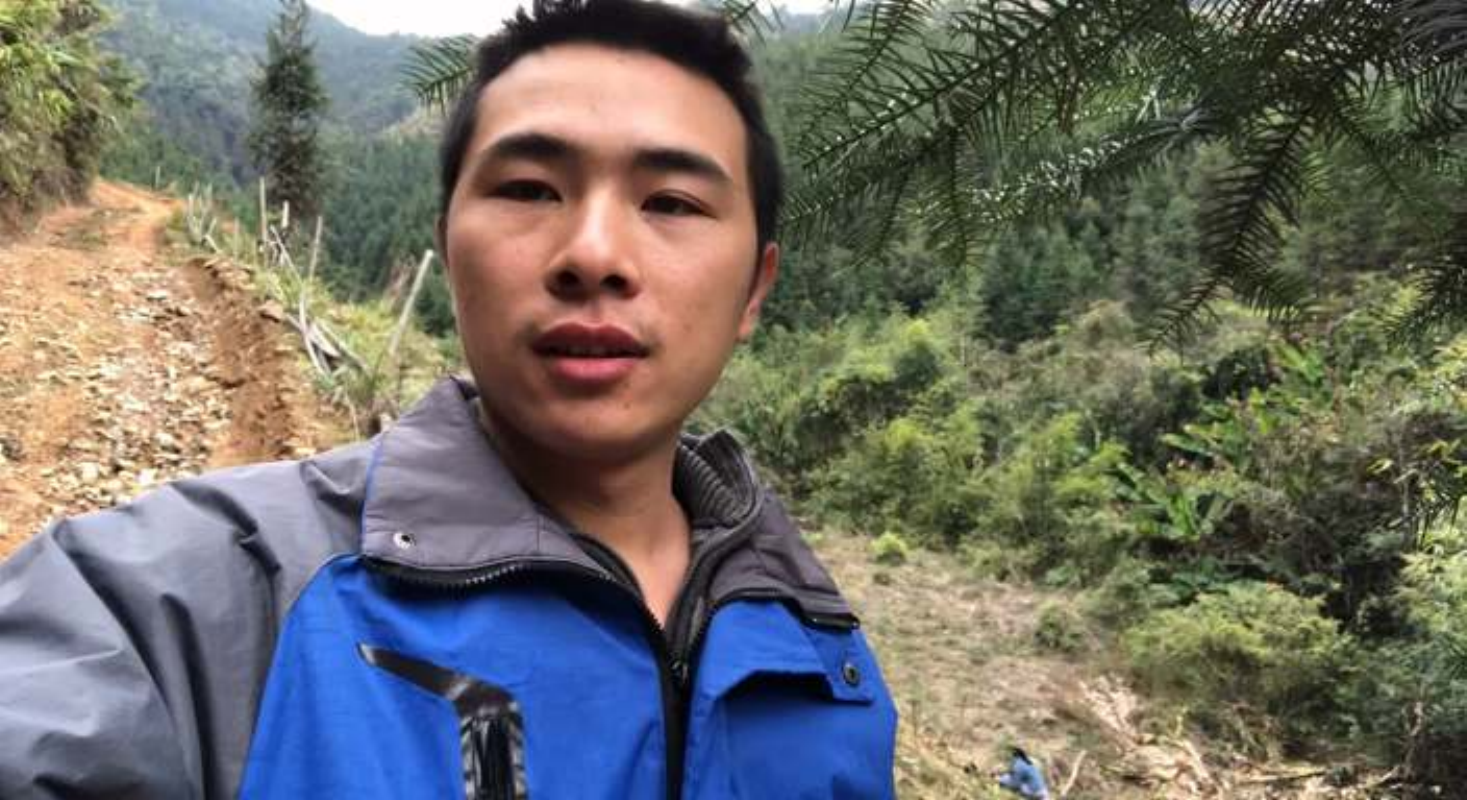} &
        \includegraphics[width=0.21\textwidth]{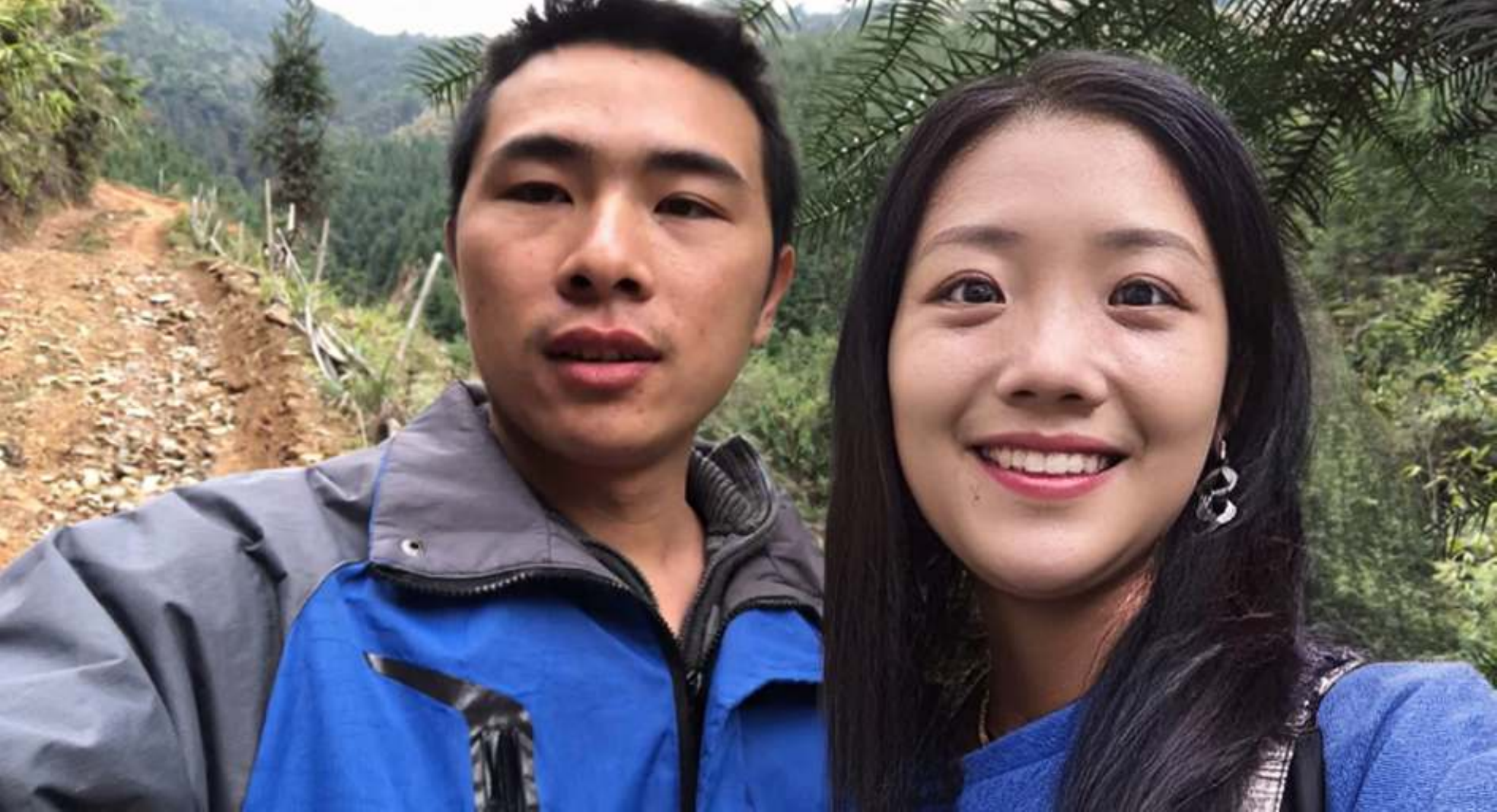} \\
  
        \multicolumn{2}{l}{Add a young Chinese woman next to the character in the image,}\\
        \multicolumn{2}{l}{with a bright smile and a pure, natural look,}\\
        \multicolumn{2}{l}{without altering the original character.} \\

        \includegraphics[width=0.21\textwidth]{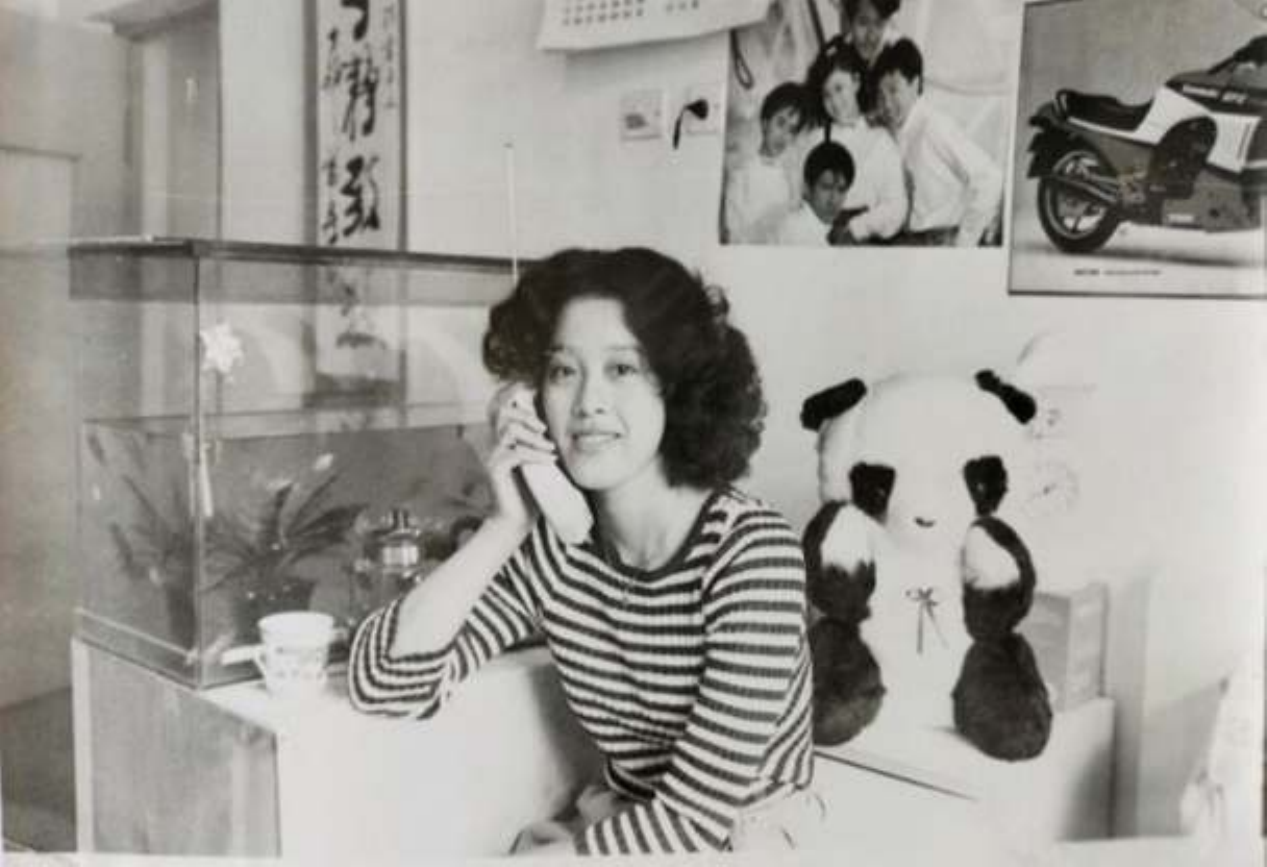} &
        \includegraphics[width=0.21\textwidth]{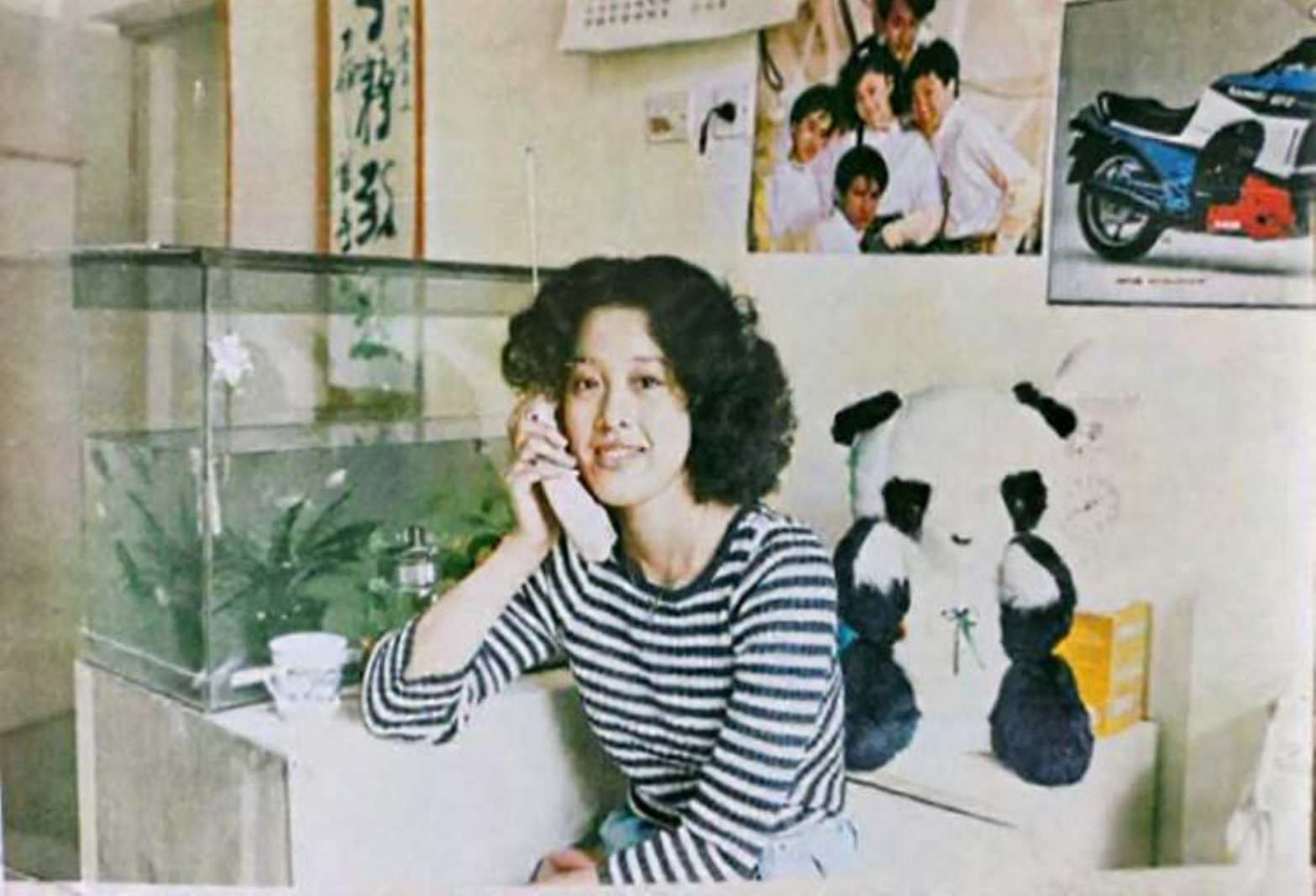} \\
  
        \multicolumn{2}{c}{Colorize the photo to make it clearer.} \\

        \includegraphics[width=0.21\textwidth]{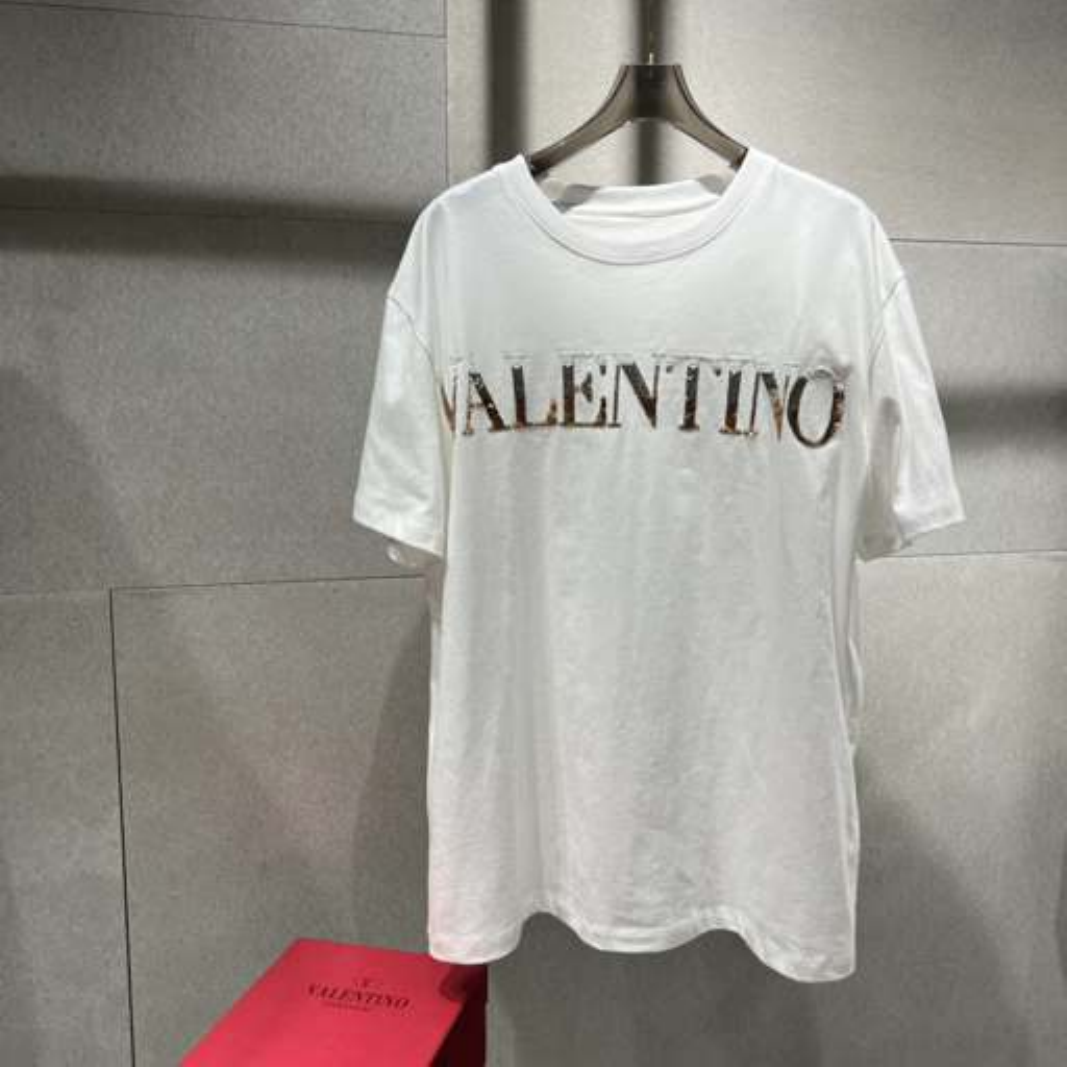} &
        \includegraphics[width=0.21\textwidth]{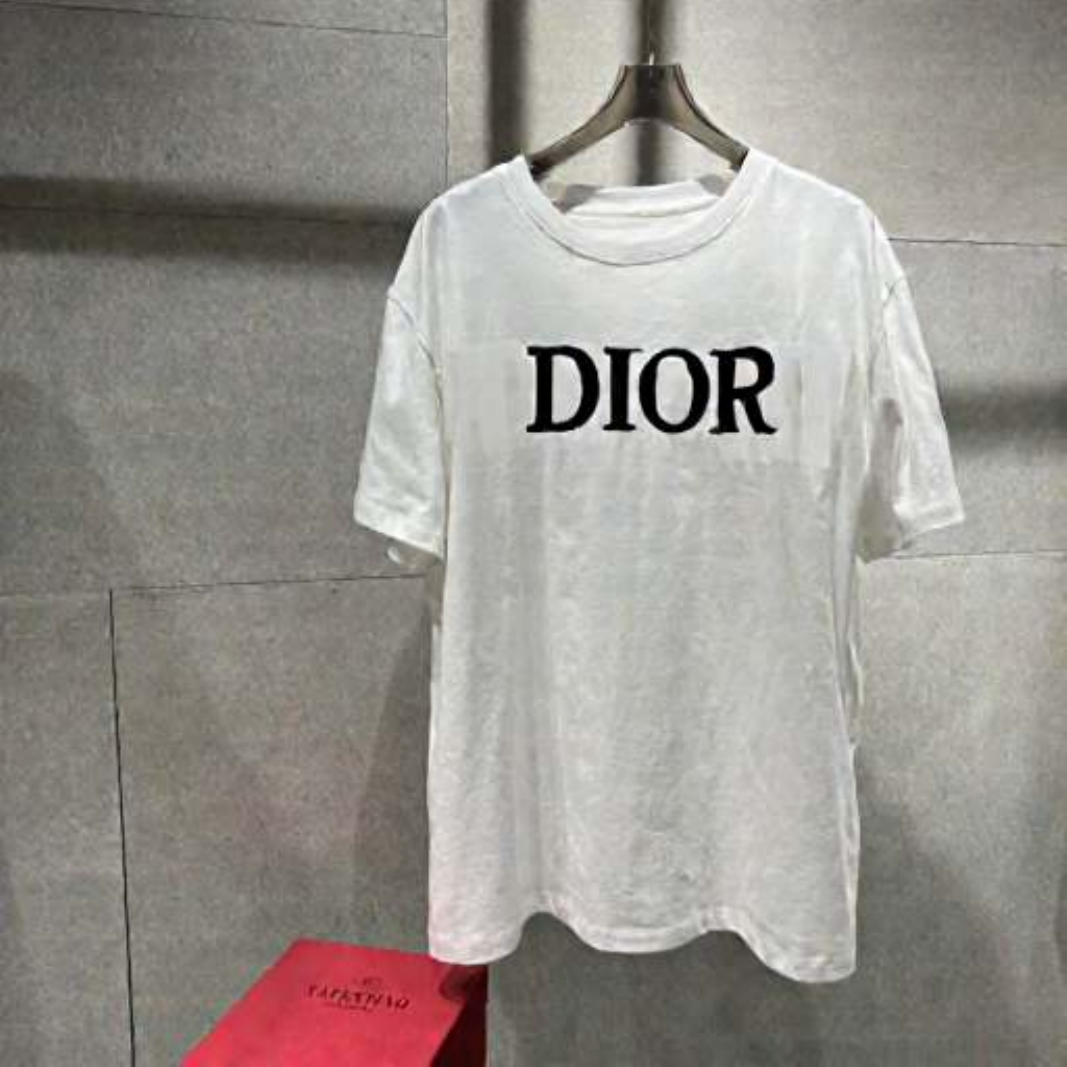} \\
     
        \multicolumn{2}{c}{I want to change the letters on this piece of clothing to "DIOR".} \\

        \includegraphics[width=0.21\textwidth]{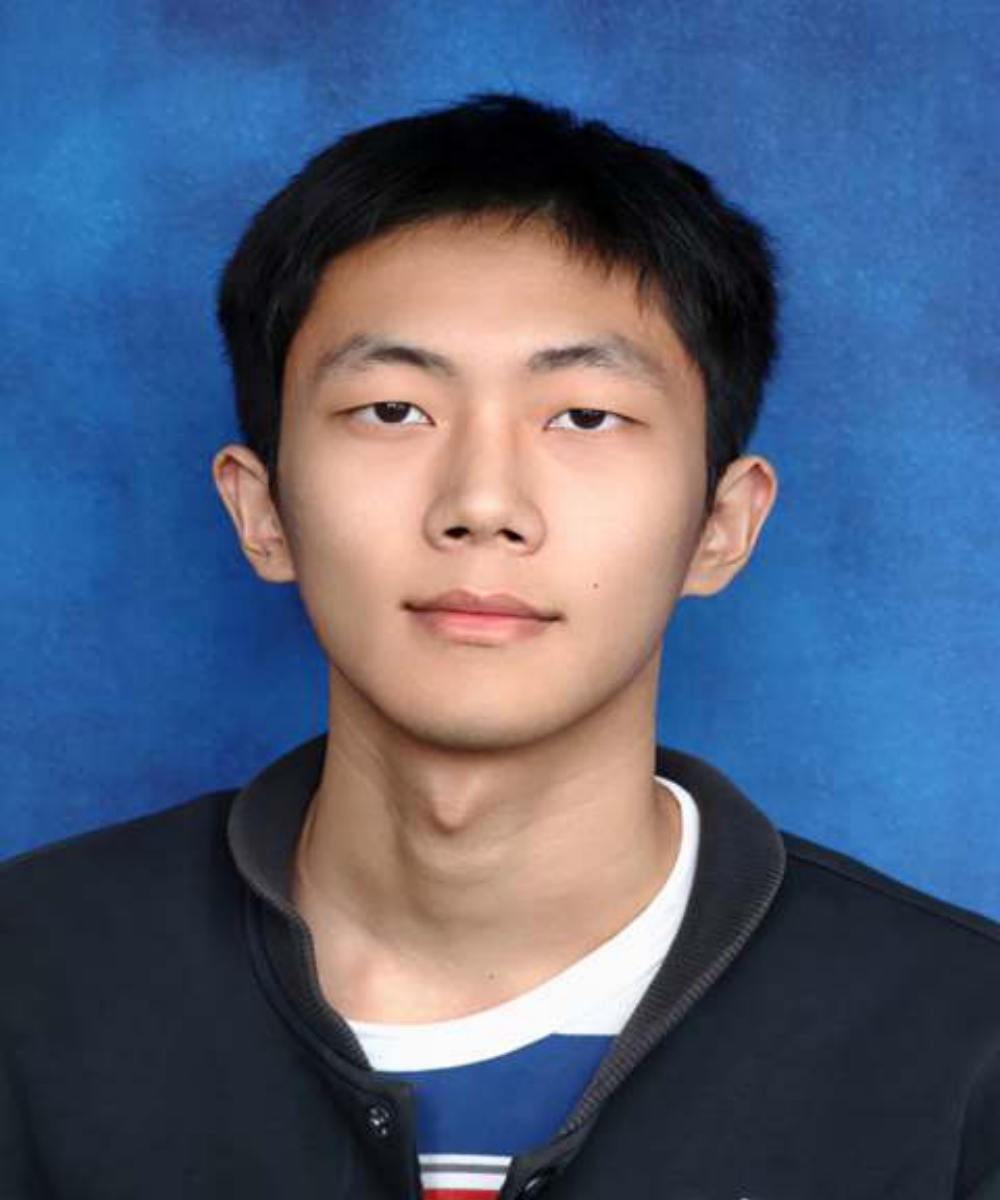} &
        \includegraphics[width=0.21\textwidth]{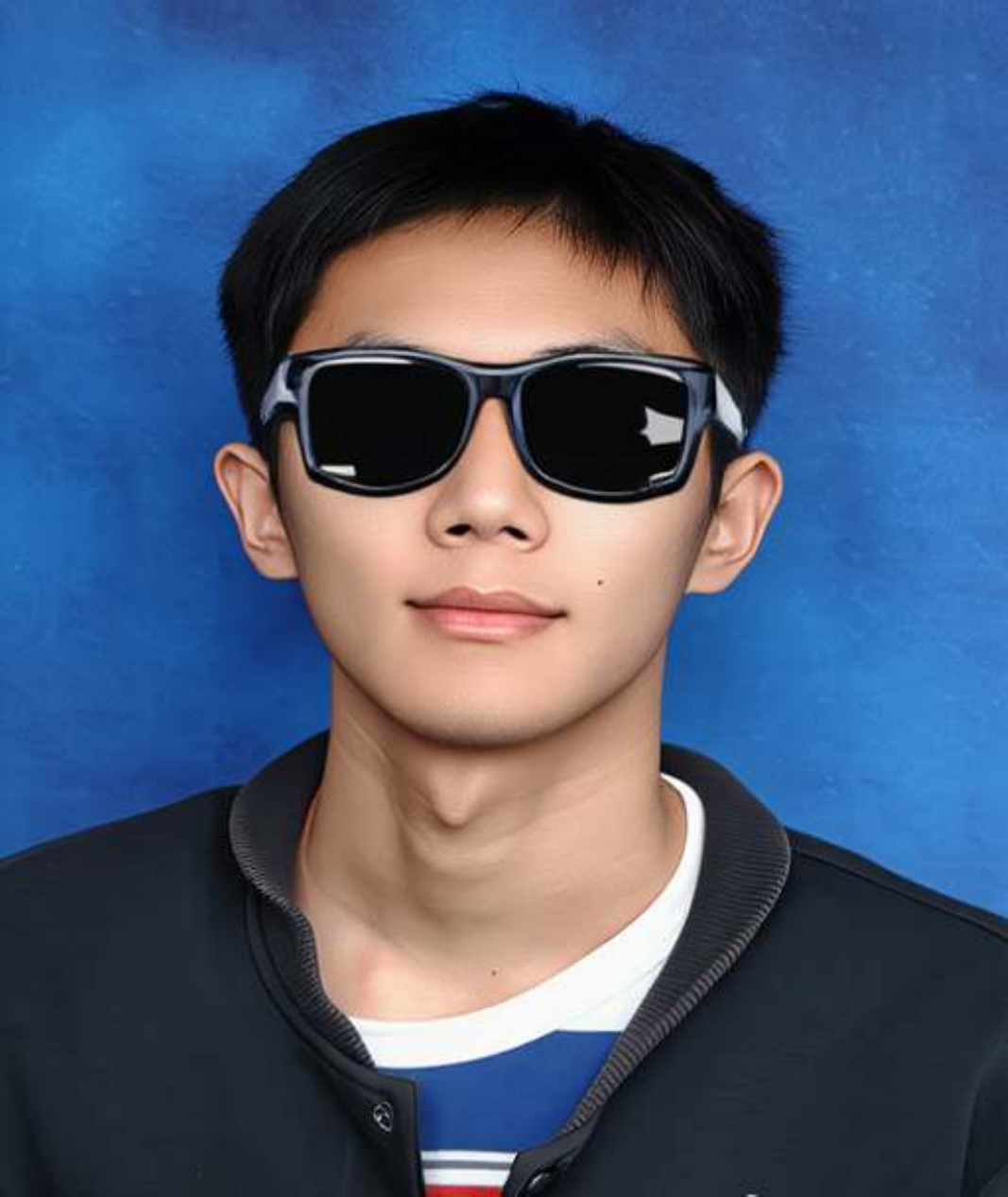} \\
 
        \multicolumn{2}{l}{Add a young Chinese woman next to the character in the image,}\\
        \multicolumn{2}{l}{with a bright smile and a pure, natural look, }\\
        \multicolumn{2}{l}{without altering the original character.} \\
        
    \end{tabular}
    }
    \caption{Demo images.}
    \label{figure-more-demos}
\end{figure}

\begin{table*}[ht]
\centering
\resizebox{0.7\textwidth}{!}{%
\begin{tabular}{c|ccccc} 
\toprule
Model & (SC, PQ)$\uparrow$ & (LSC, LPQ)$\uparrow$ & LScore$\uparrow$ & Overall$\uparrow$ \\ 
\midrule
Step1X-Edit~\citep{liu2025step1x} & (6.950, 7.260) & (5.831, 6.080) & 5.160 & 6.529 \\
FLUX.1 Kontext~\citep{labs2025flux} & (7.025, 7.000) & (5.418, 6.346) & 4.852 & 6.443 \\
BAGEL~\citep{bagel} & (7.454, 7.044) & (6.418, 5.667) & 5.375 & 6.639 \\
LGCC(ours) & (7.319, 7.159) & (6.359, 5.886) & \textbf{5.461} & \textbf{6.679} \\
\bottomrule
\end{tabular}
}
\caption{Performance scores on GEditBench and MagicCrush dataset. For each score, the high the better. All scores are evaluated by Qwen2.5-VL-72B. }
\label{table_bagel_score_gedit}
\end{table*}

\section{Demo Images}
We present additional qualitative results generated by our method, $\mmname$, in~\cref{figure-more-demos-1} and~\cref{figure-more-demos}, for the text-guided image editing task. The results demonstrate that $\mmname$ is capable of handling a diverse range of editing tasks and scenarios, consistently producing visually realistic and semantically coherent outputs.

\section{Evaluations on GEditBench and MagicBrush}

We further evaluate our method, $\mmname$, on the GeditBench and MagicBrush datasets, comparing it against existing baselines including Step1X-Edit, Flux.1, Kontext, and Bagel. The results are summarized in~\cref{table_bagel_score_gedit}. Our method achieves the highest LScore and Overall score, indicating that it not only produces globally coherent images but also preserves fine-grained details in the edited regions.

As described in the experimental section, GeditBench provides two sets of scores: $(\mathrm{SC}, \mathrm{PQ})$, which evaluate the entire image, and $(\mathrm{LSC}, \mathrm{LPQ})$, which focus exclusively on the edited regions. The edited areas are manually annotated. The Overall score is computed by aggregating both global and local quality metrics, defined as follows:

\begin{equation*}
\mathrm{Overall} = \sqrt{ \operatorname{mean}(\mathrm{SC}, \mathrm{LSC}) \cdot \operatorname{mean}(\mathrm{PQ}, \mathrm{LPQ}) }
\end{equation*}

Here, $\operatorname{mean}(\cdot, \cdot)$ denotes the arithmetic mean. This formulation ensures that both global perceptual quality and localized editing accuracy are taken into account when evaluating the overall performance of the model.

\section{More Comparison of Edited Images}
\begin{figure*}[htbp!]
    \centering
    \resizebox{\textwidth}{!}{
    \begin{tabular}{c c c c c}
    Input Image & Step1X-Edit & FLUX.1 Kontext & BAGEL & Ours\\ 
    
    \includegraphics[width=0.2\textwidth]{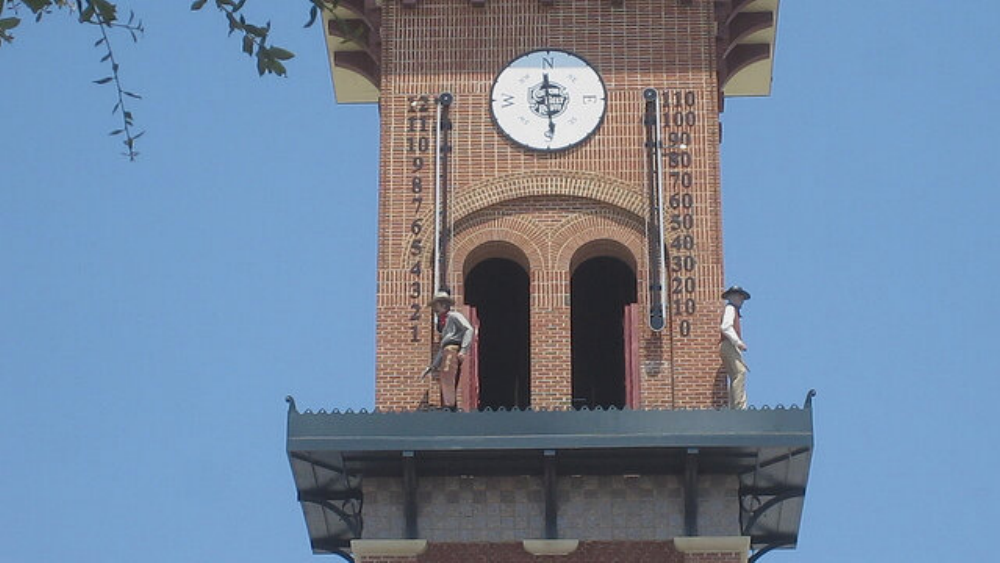} &
        \includegraphics[width=0.2\textwidth]{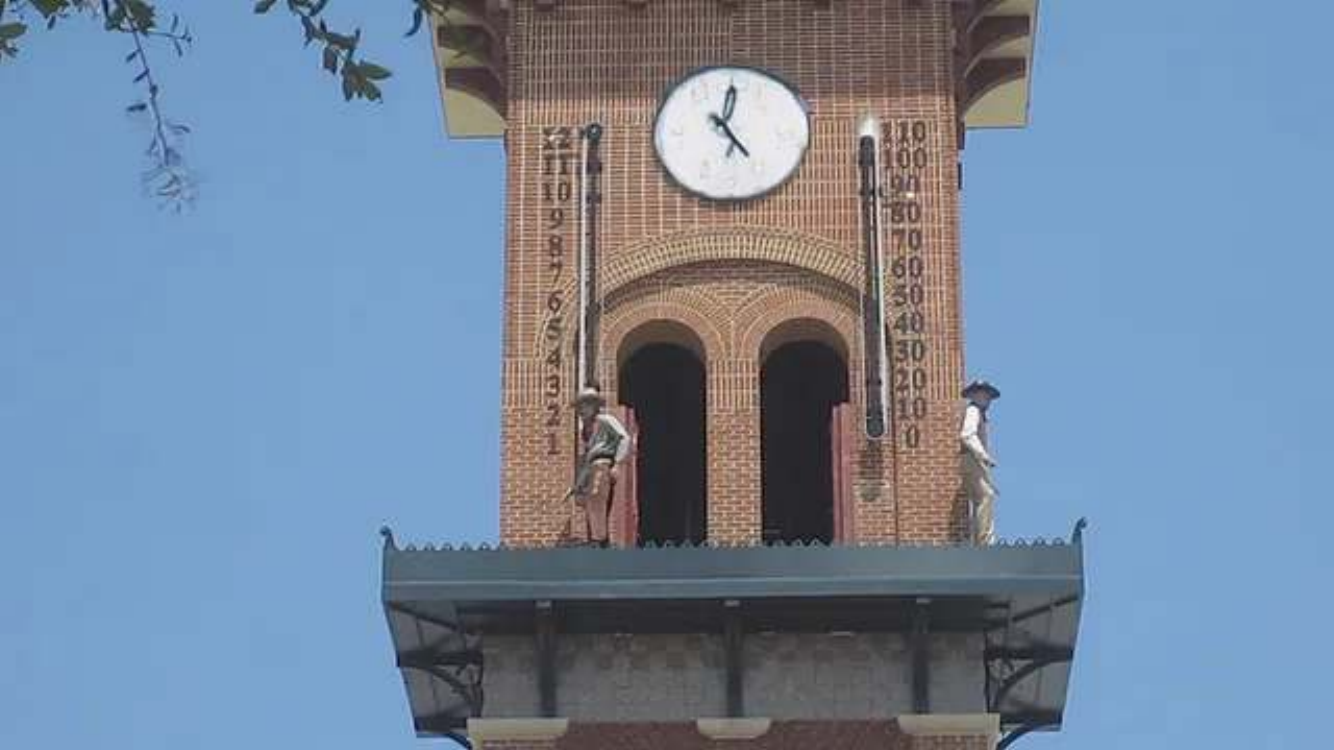} &
        \includegraphics[height=2cm, keepaspectratio]{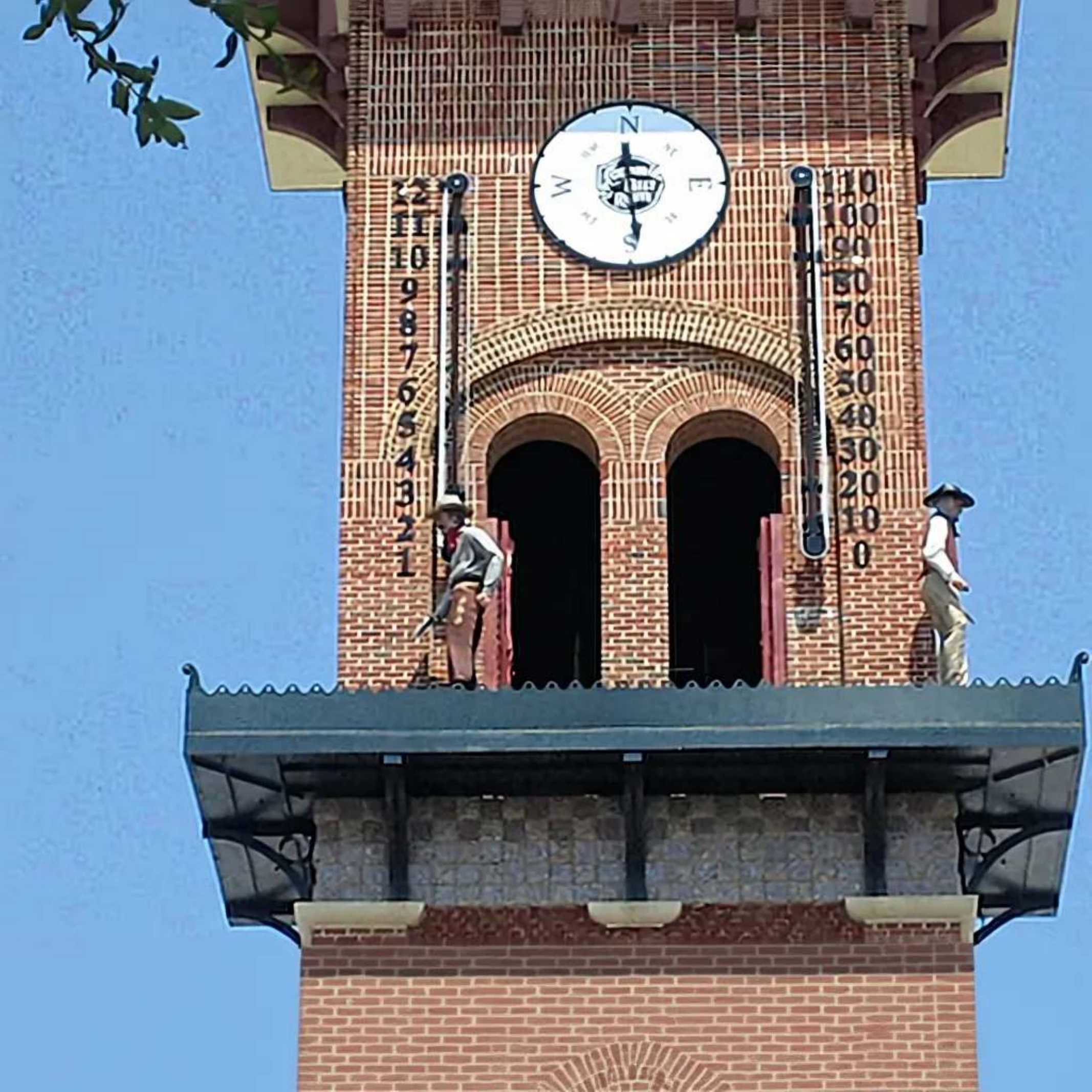} &
        \includegraphics[width=0.2\textwidth]{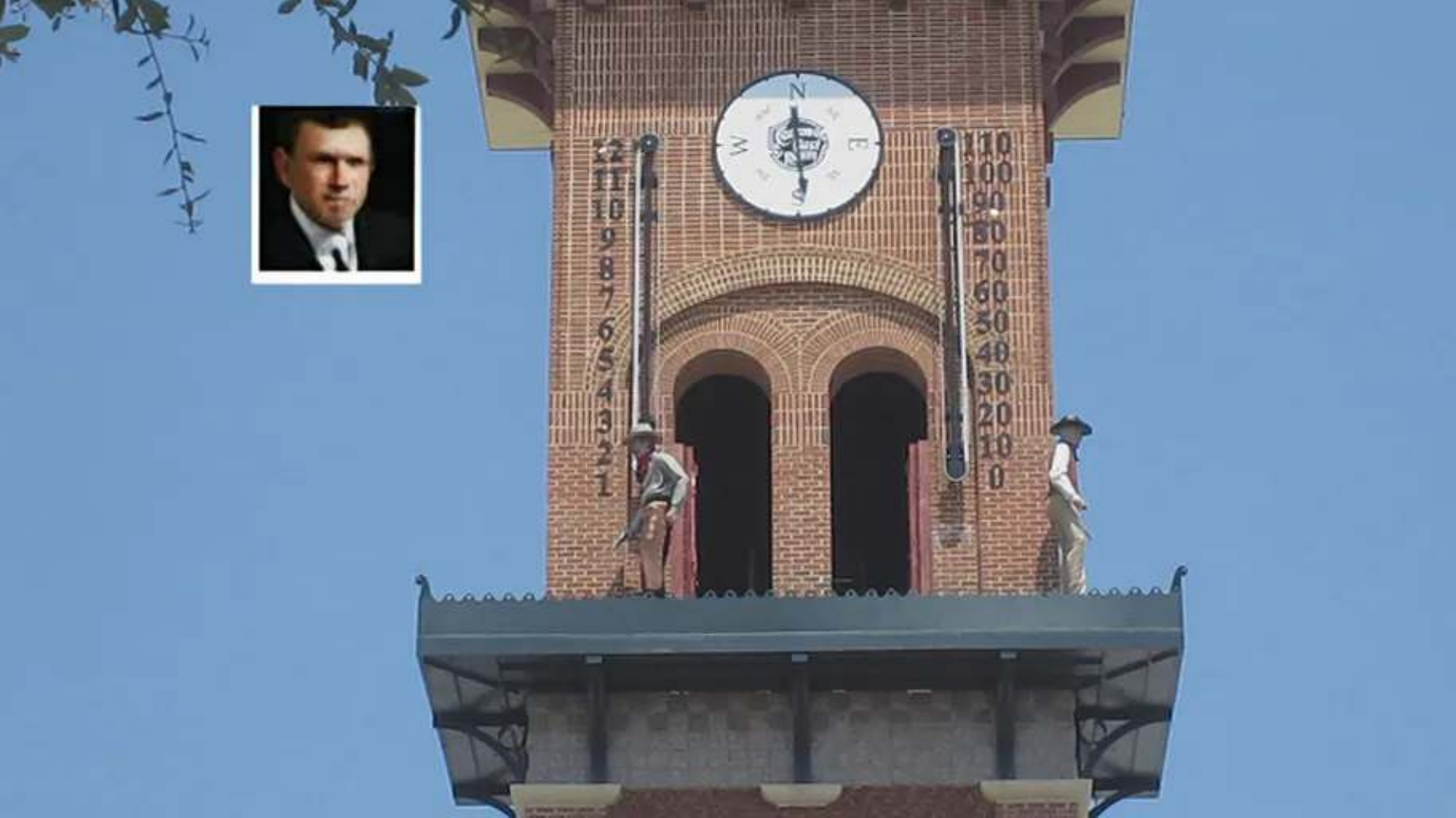} &
        \includegraphics[width=0.2\textwidth]{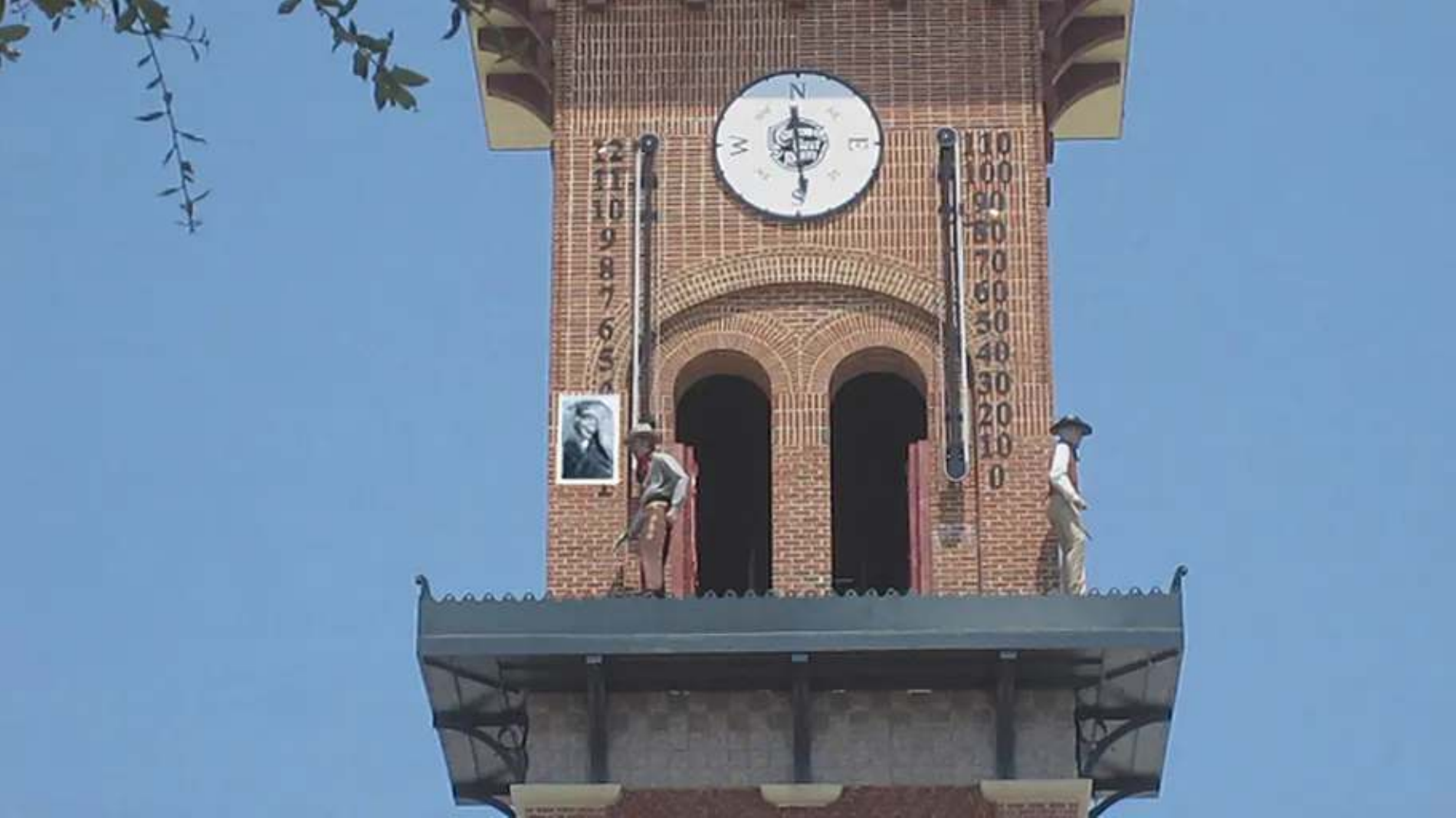} \\
        \multicolumn{5}{c}{(a) Add-a-photo-to-the-left-of-the-clock.}\\
        \multicolumn{5}{c}{}\\
        
        \includegraphics[width=0.2\textwidth]{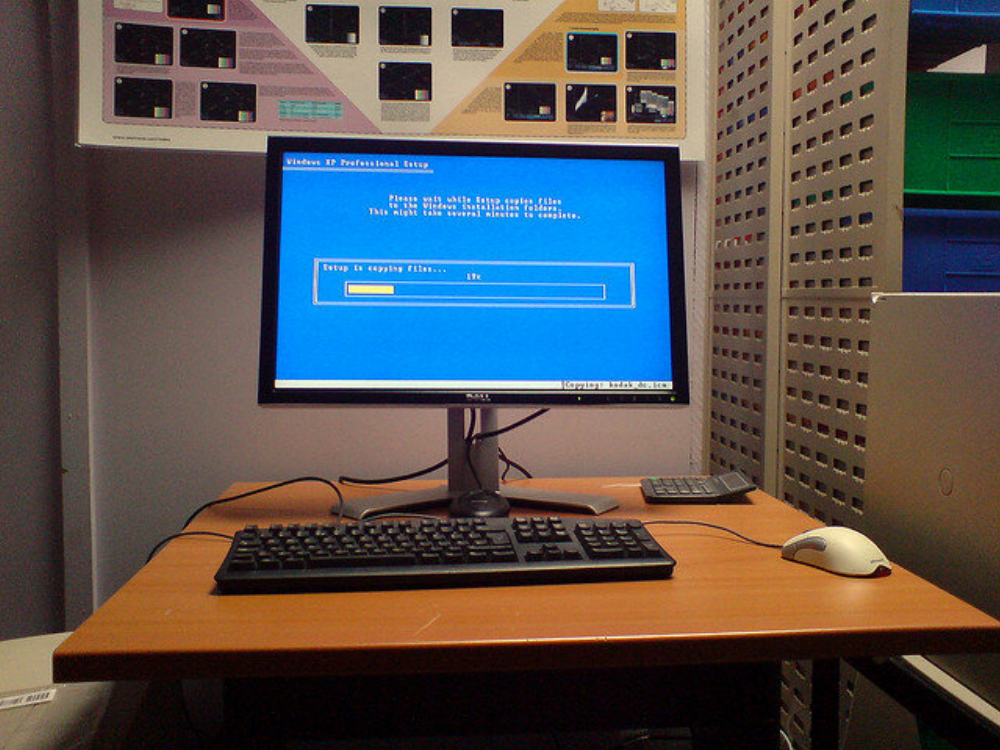} &
        \includegraphics[width=0.2\textwidth]{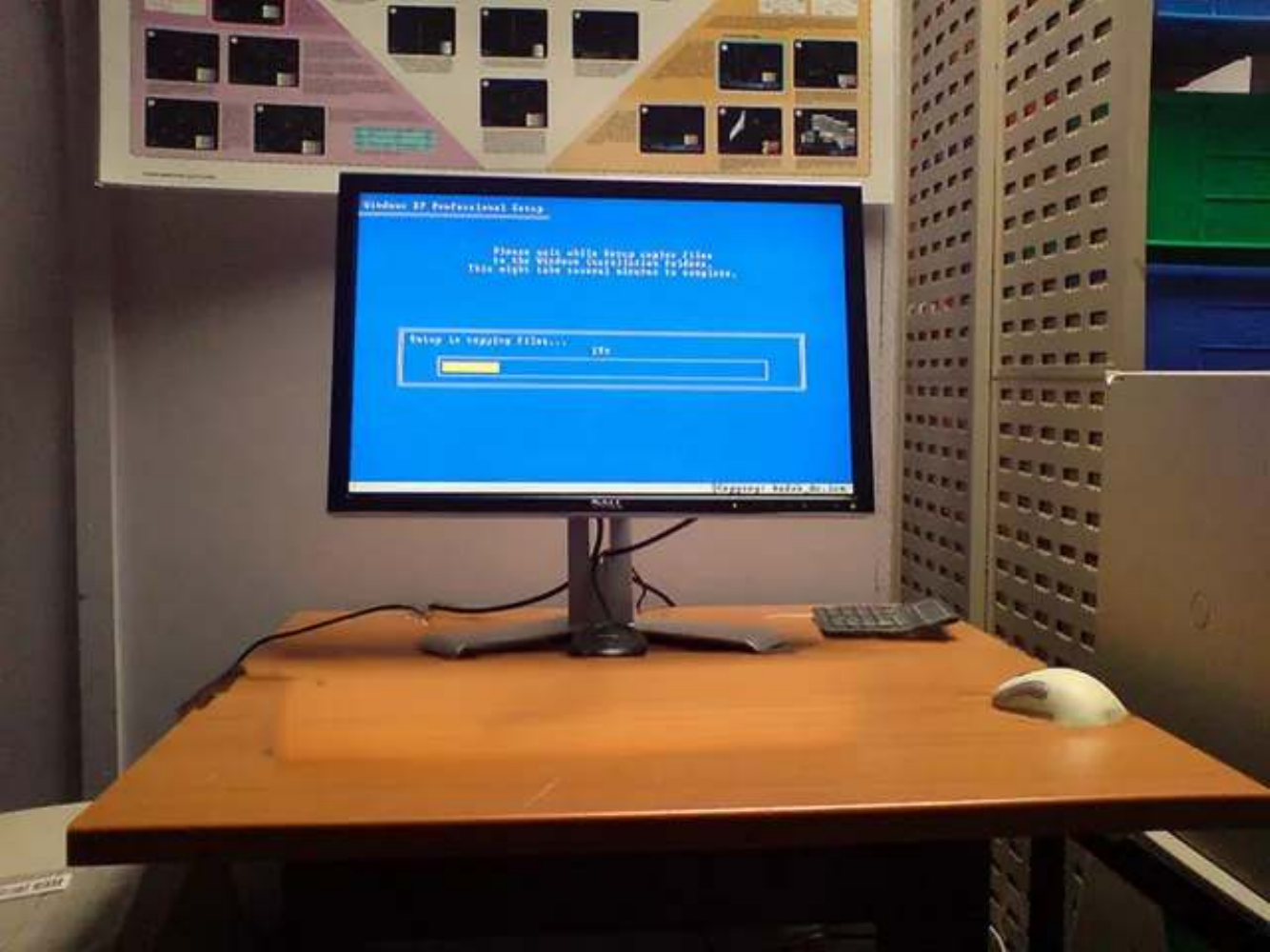} &
        \includegraphics[height=2.5cm, keepaspectratio]{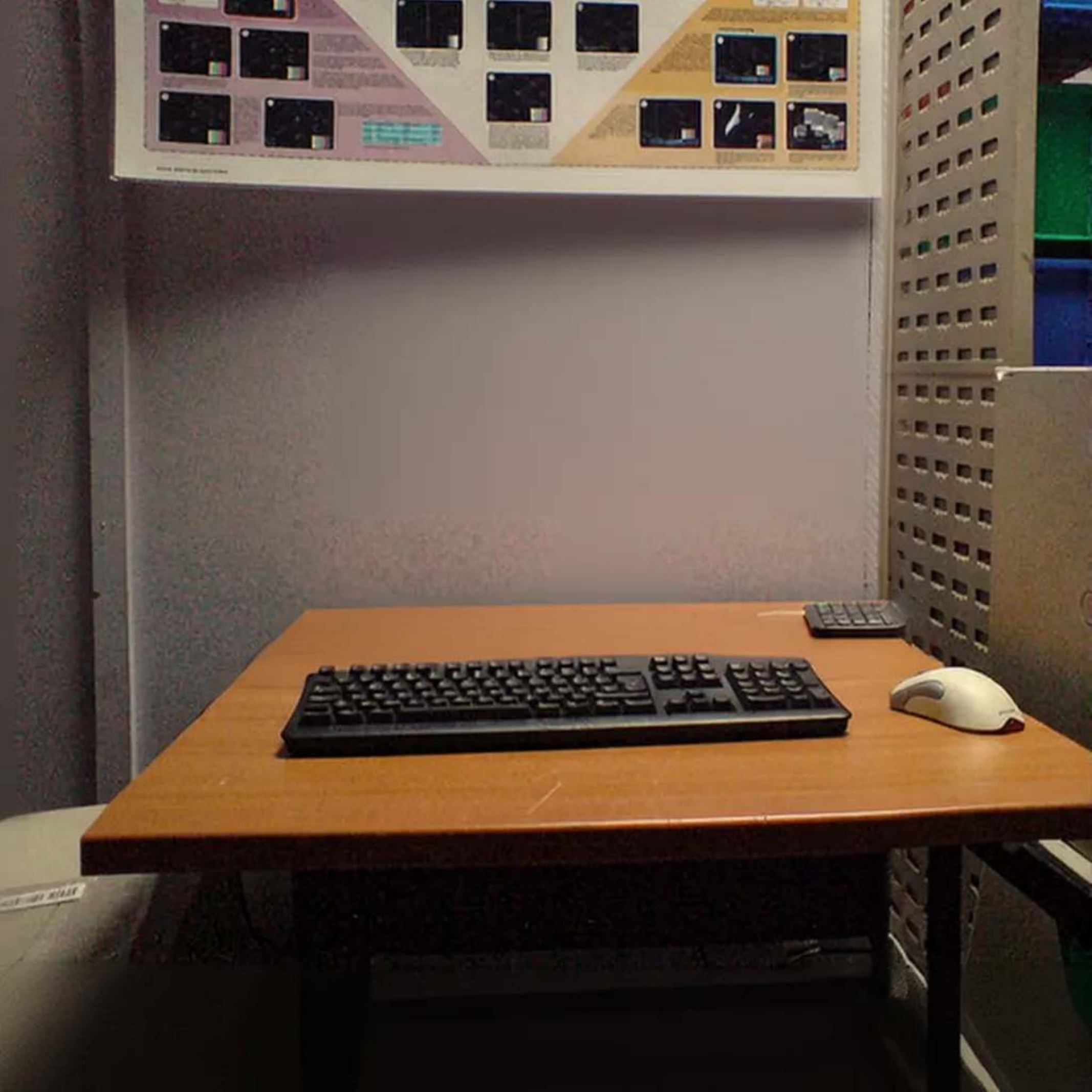} &
        \includegraphics[width=0.2\textwidth]{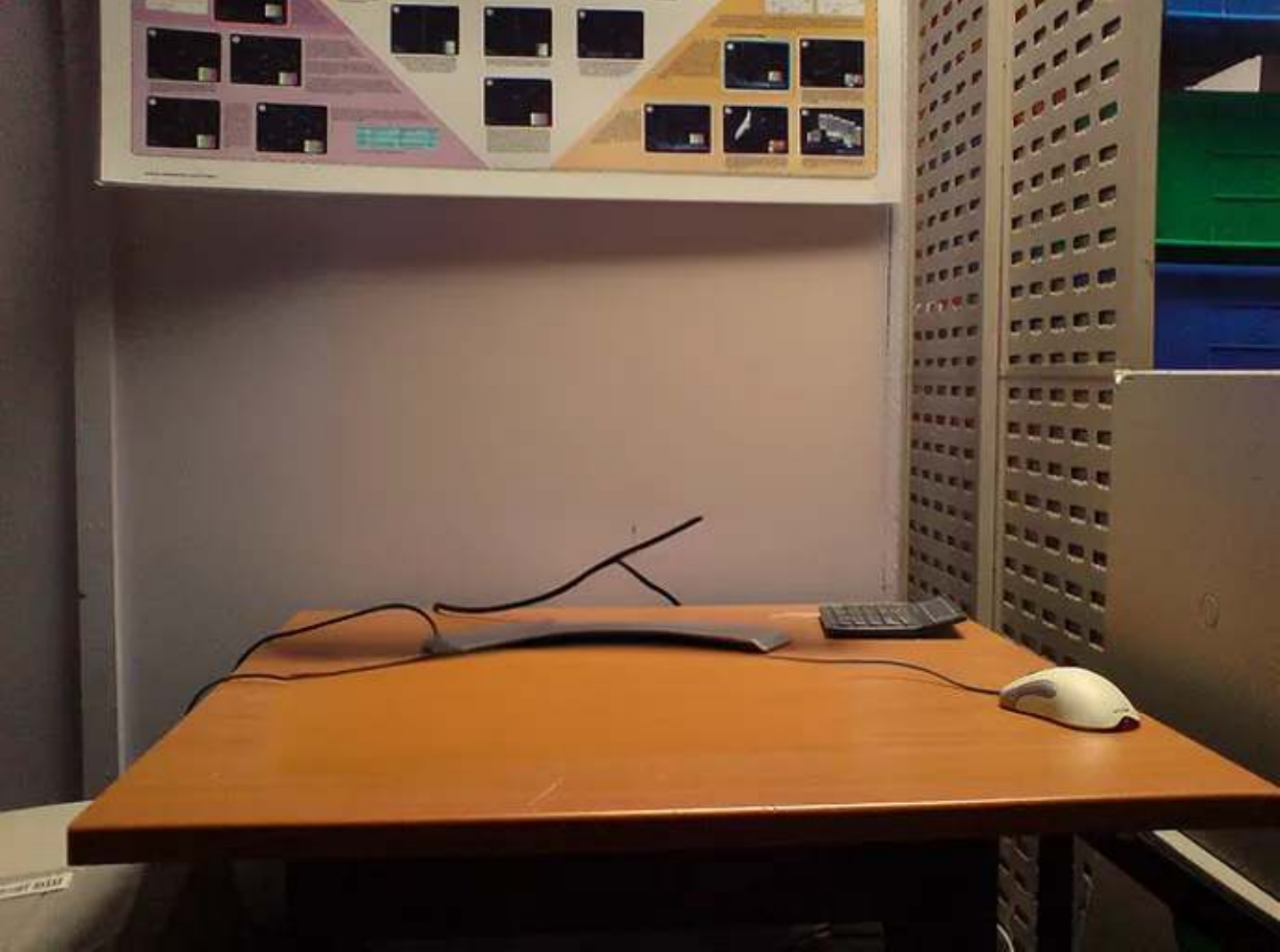} &
        \includegraphics[width=0.2\textwidth]{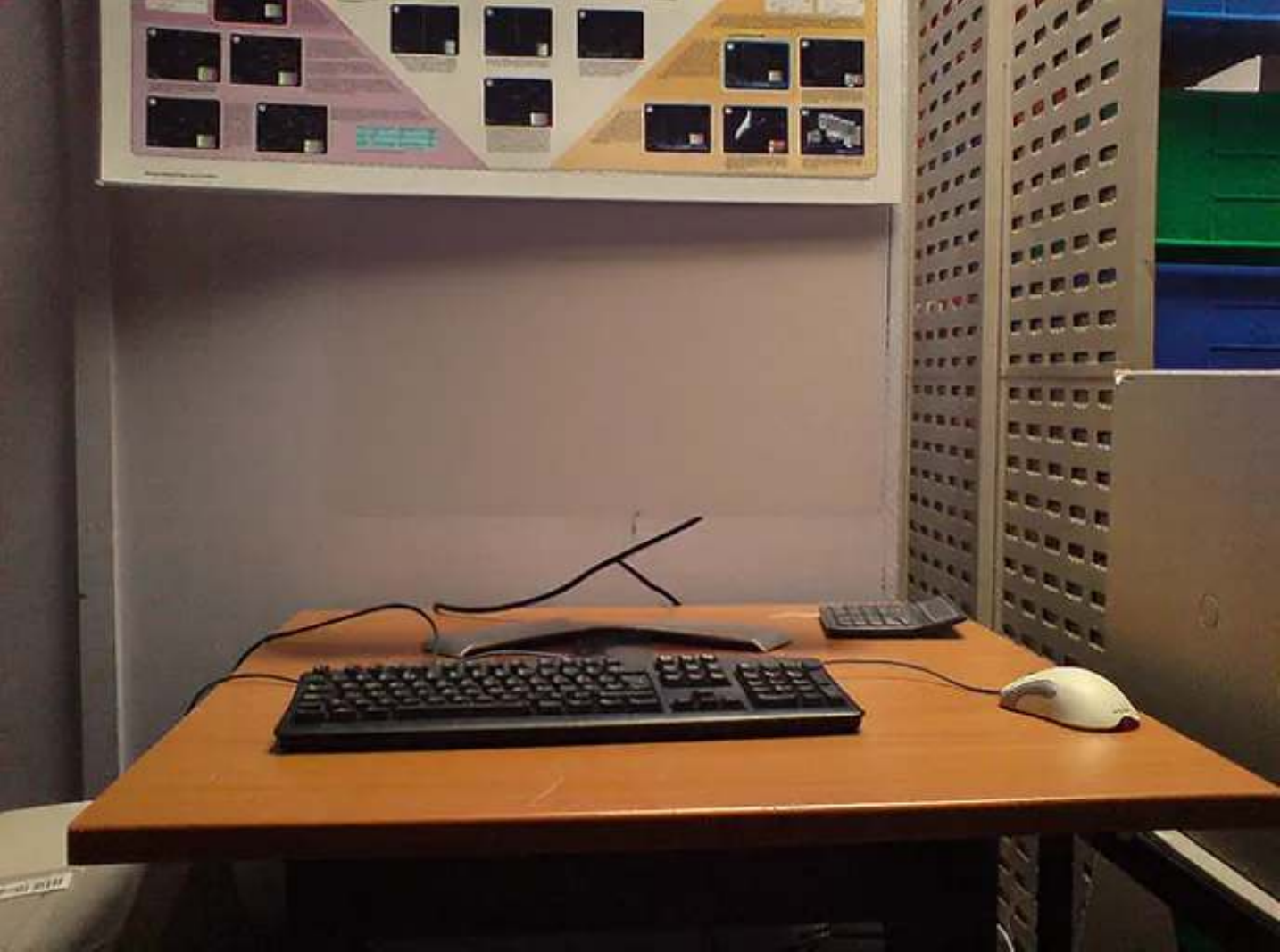} \\
        \multicolumn{5}{c}{(b) Remove-the-computer-from-the-image.}\\
        \multicolumn{5}{c}{}\\
        
        \includegraphics[width=0.2\textwidth]{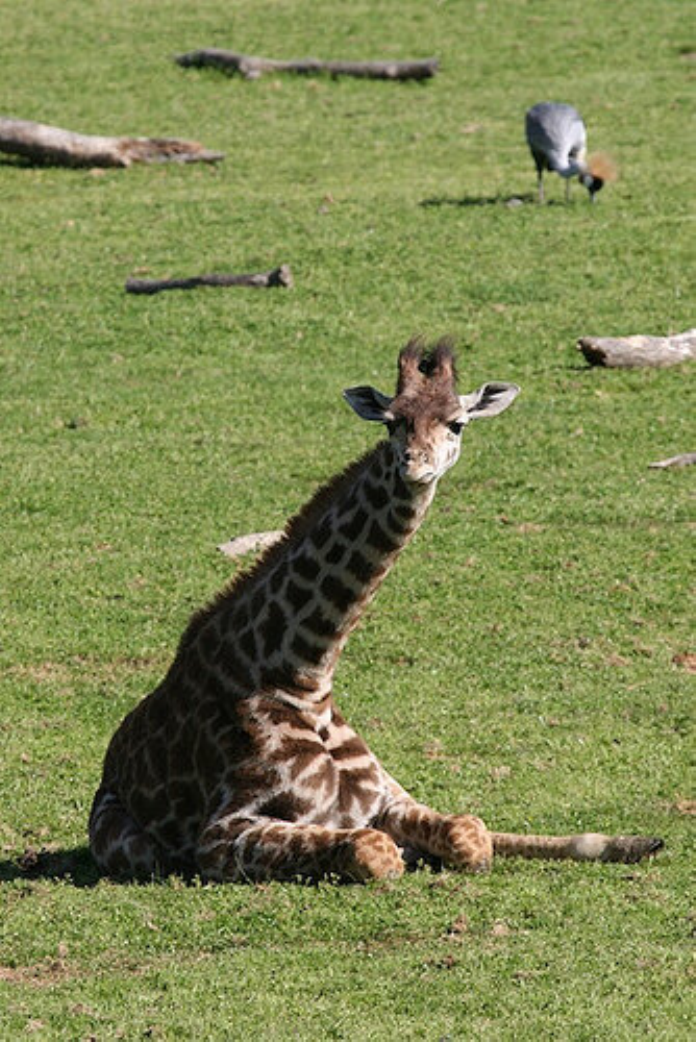} &
        \includegraphics[width=0.2\textwidth]{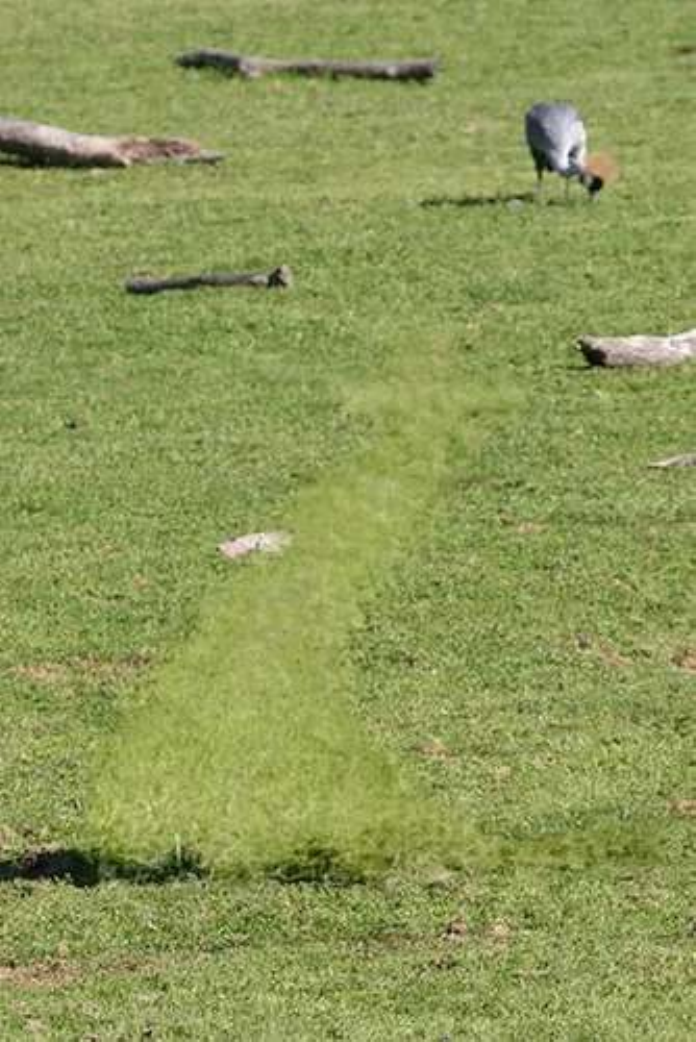} &
        \includegraphics[height=5.3cm, keepaspectratio]{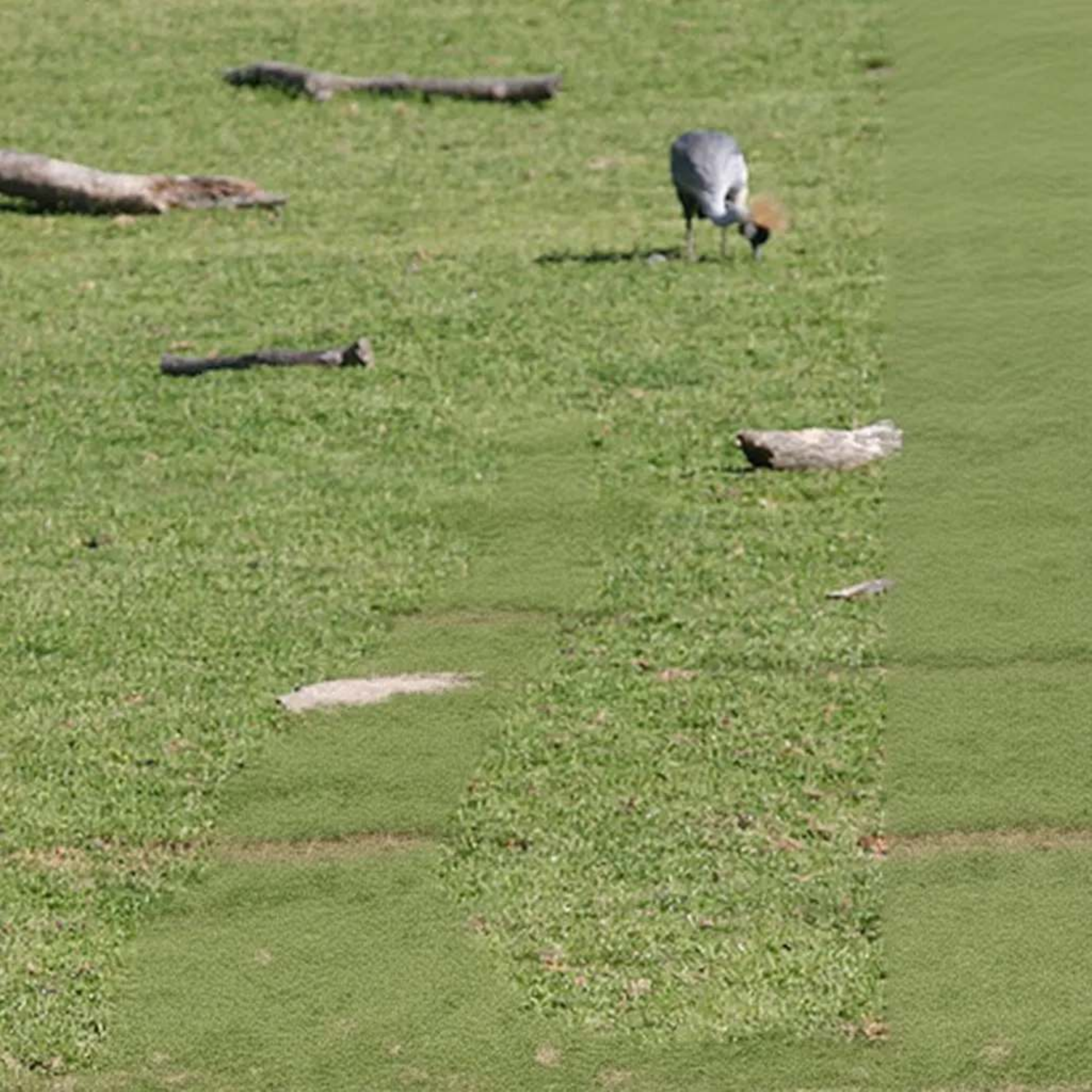} &
        \includegraphics[width=0.2\textwidth]{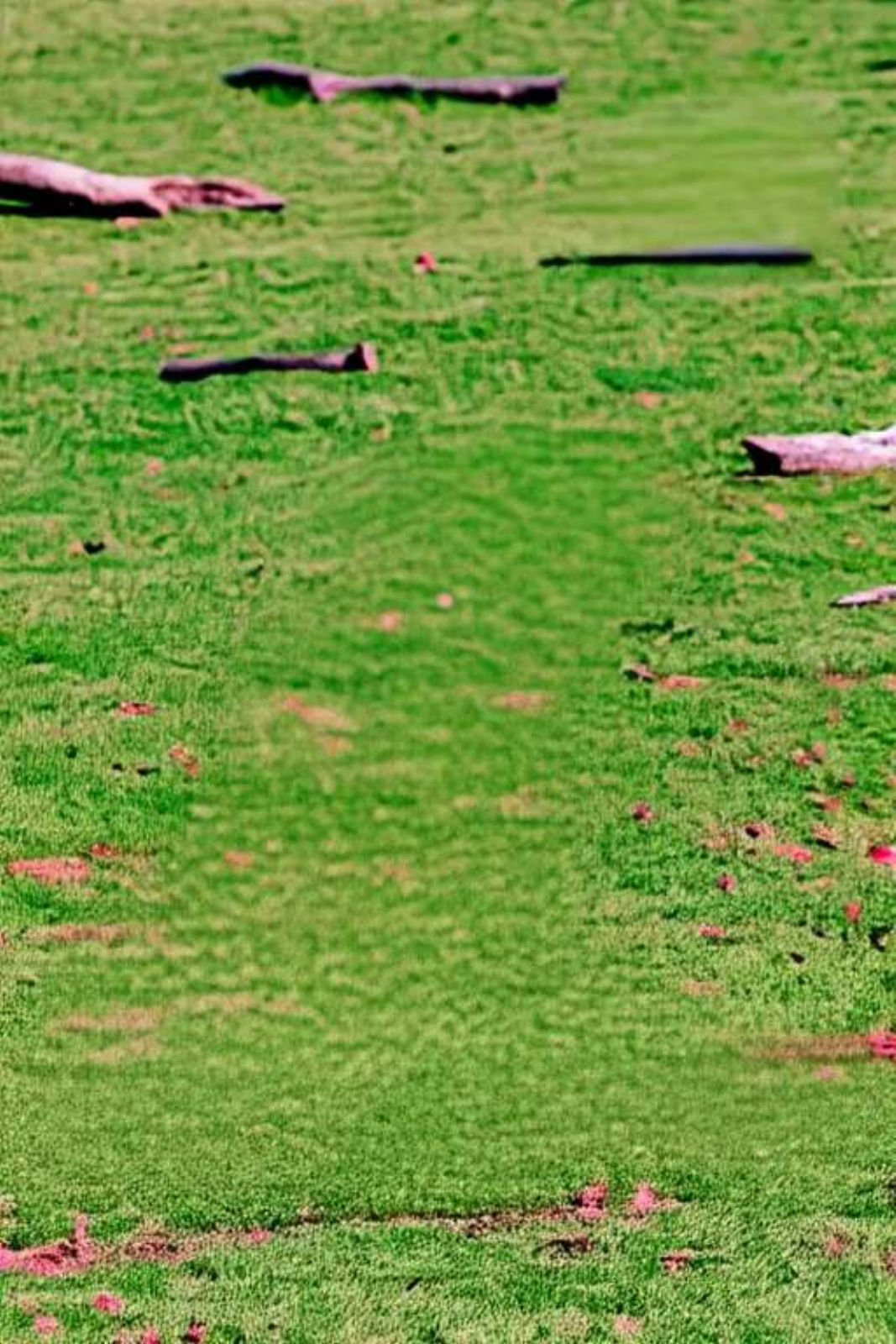} &
        \includegraphics[width=0.2\textwidth]{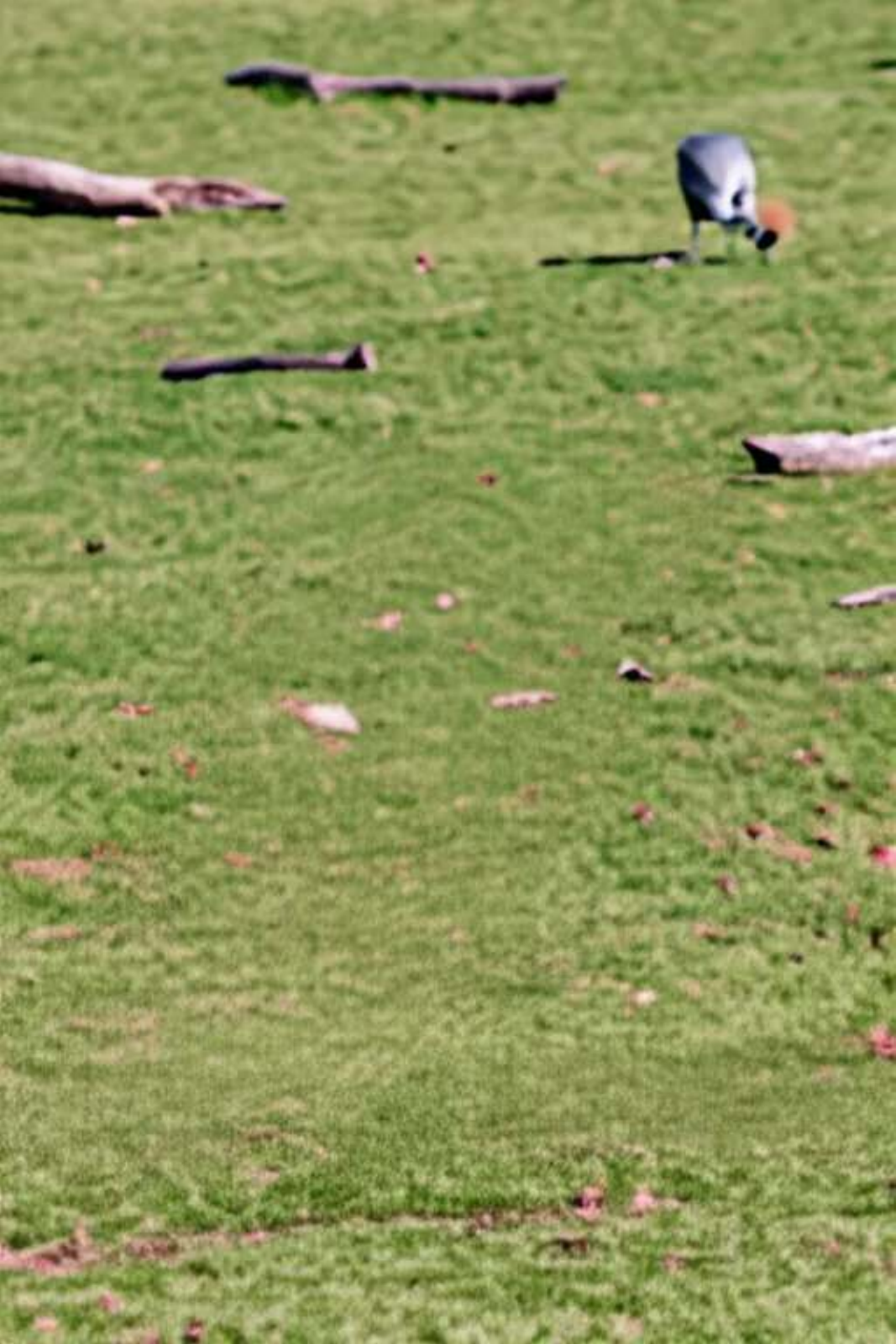} \\
        \multicolumn{5}{c}{(c) Remove-the-giraffe-from-the-image.}\\
        \multicolumn{5}{c}{}\\

        \includegraphics[width=0.2\textwidth]{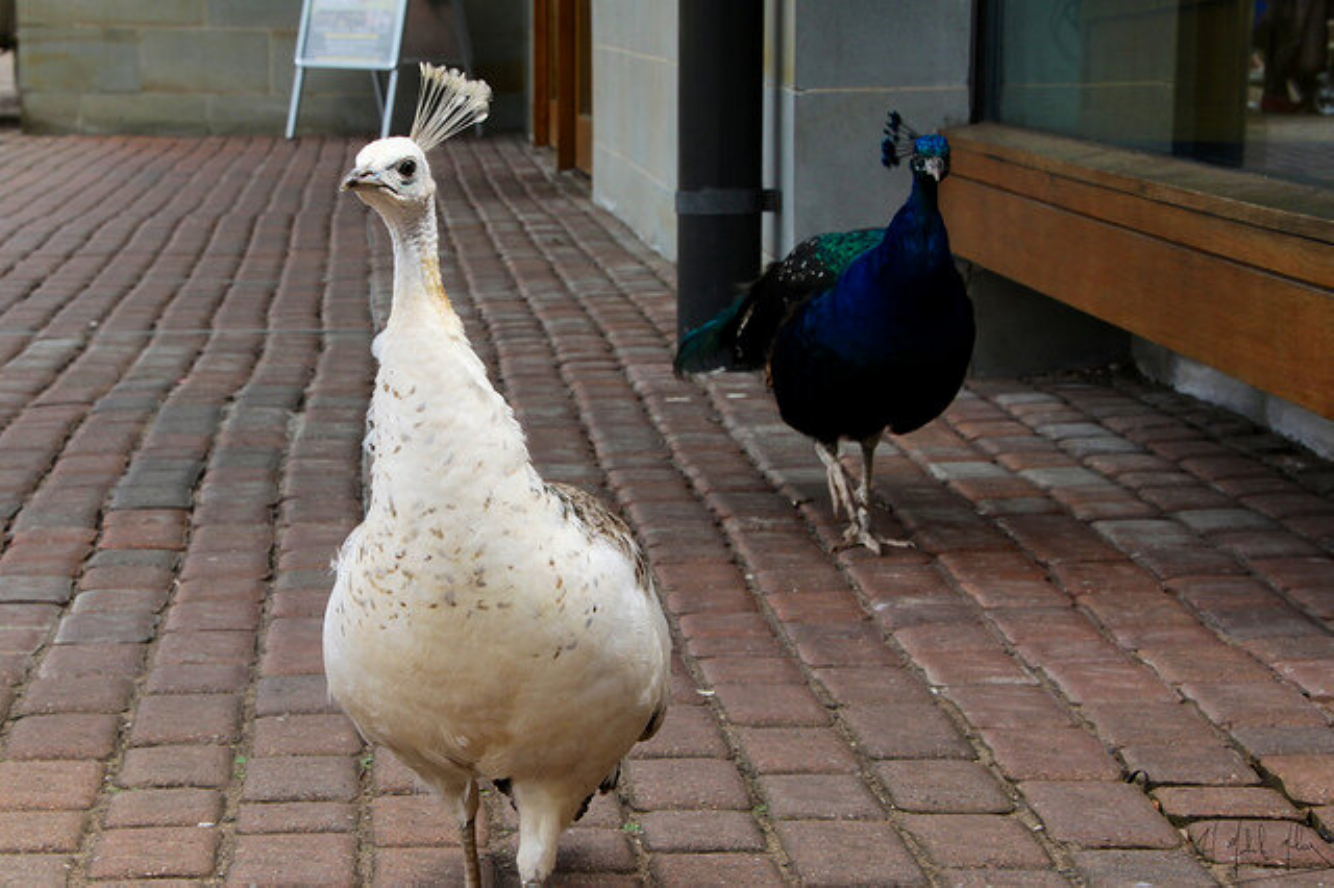} &
        \includegraphics[width=0.2\textwidth]{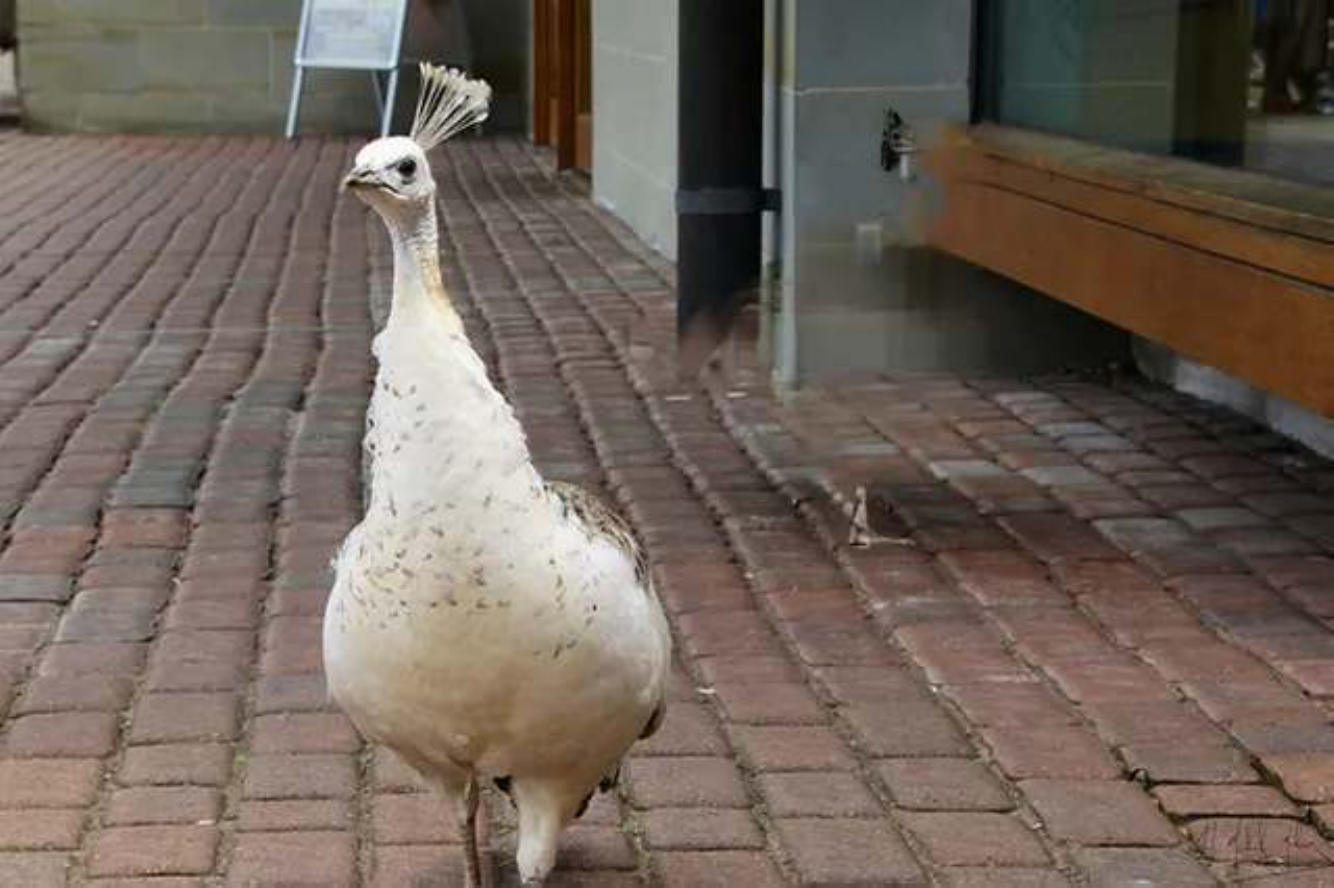} &
        \includegraphics[height=2.3cm, keepaspectratio]{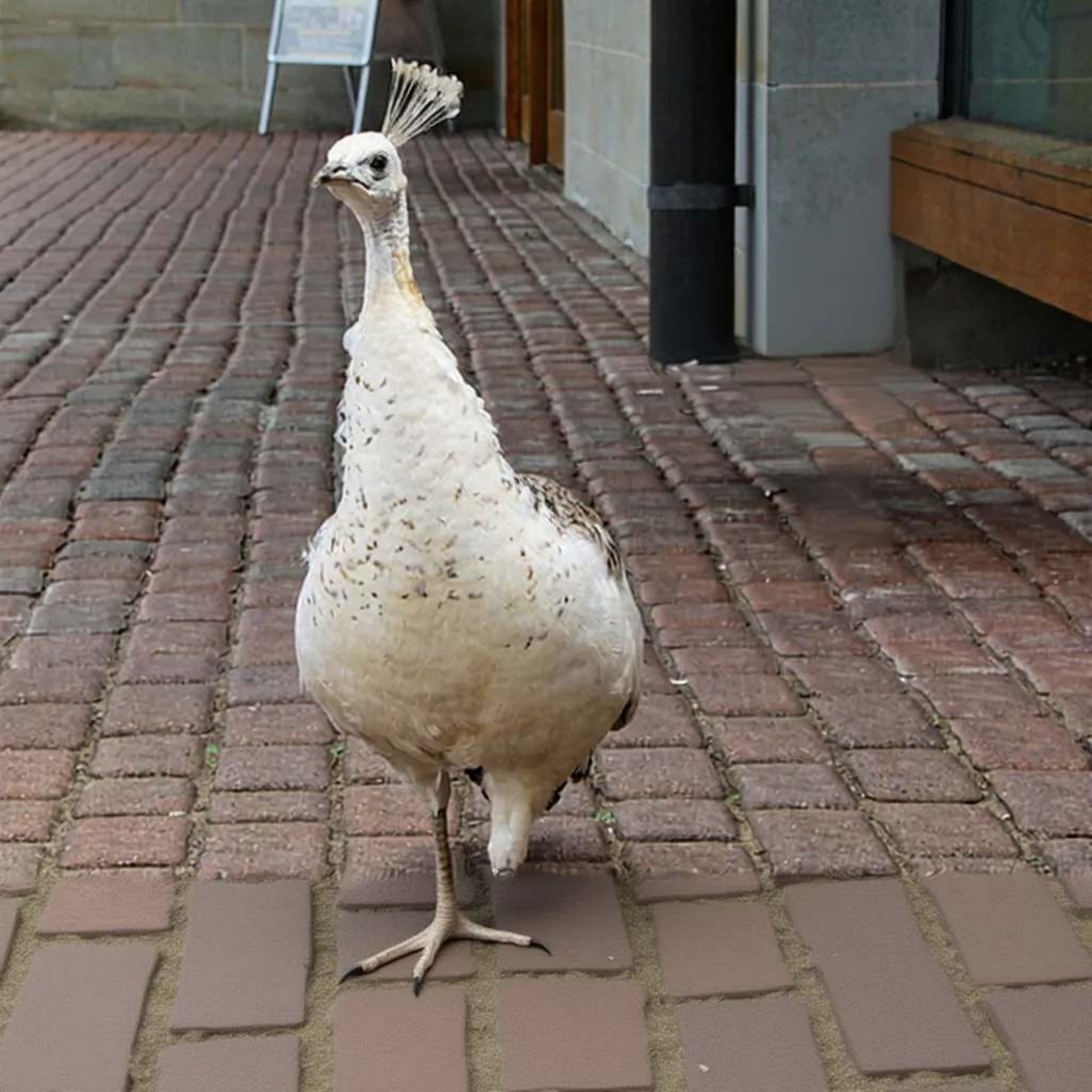} &
        \includegraphics[width=0.2\textwidth]{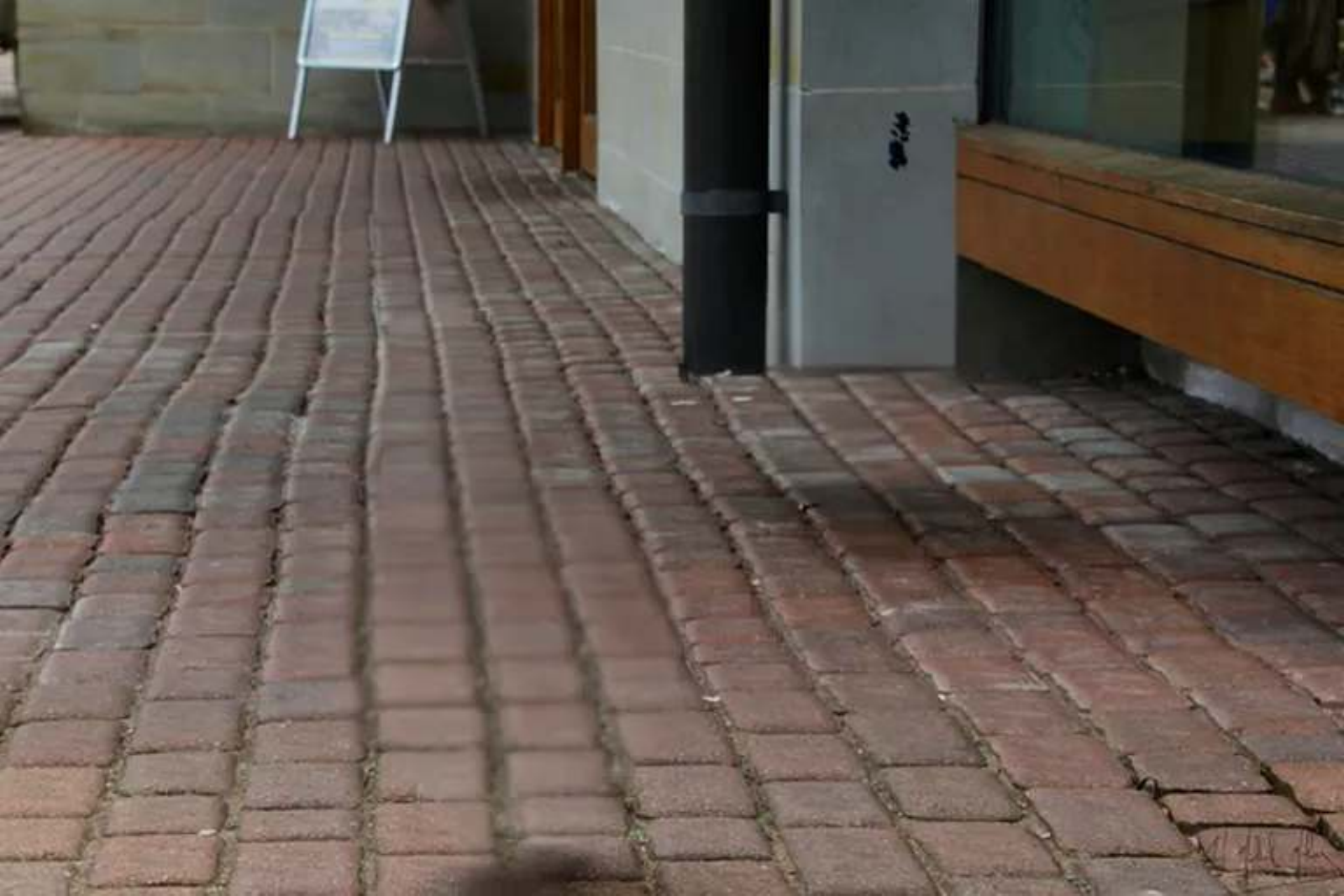} &
        \includegraphics[width=0.2\textwidth]{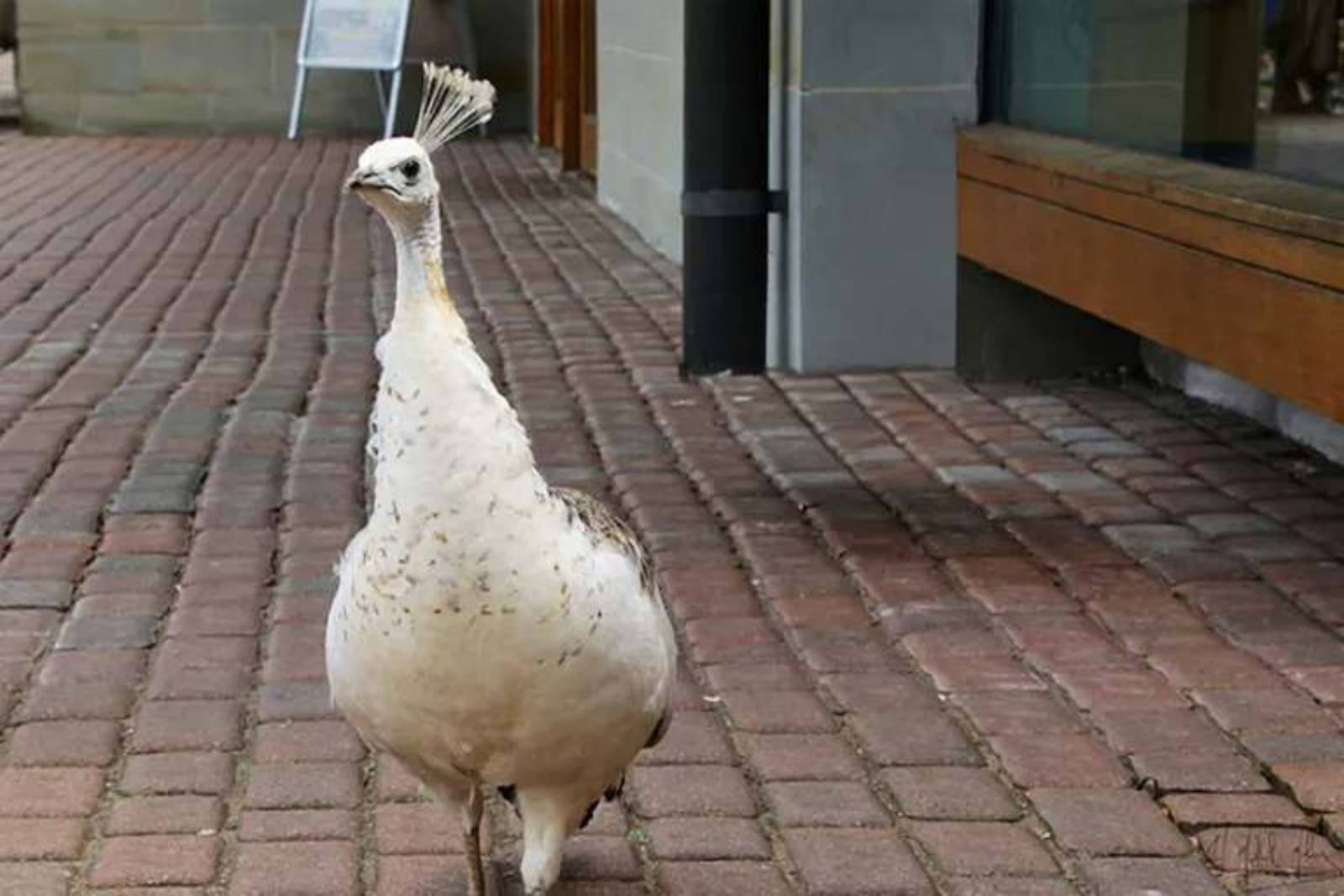} \\
        \multicolumn{5}{c}{(d) Remove-the-peacock-from-the-image.}\\
        \multicolumn{5}{c}{}\\

        \includegraphics[width=0.2\textwidth]{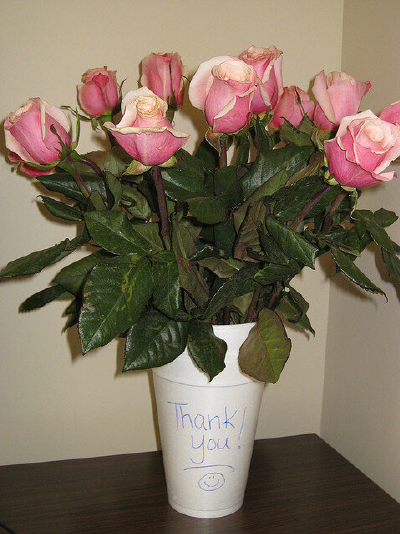} &
        \includegraphics[width=0.2\textwidth]{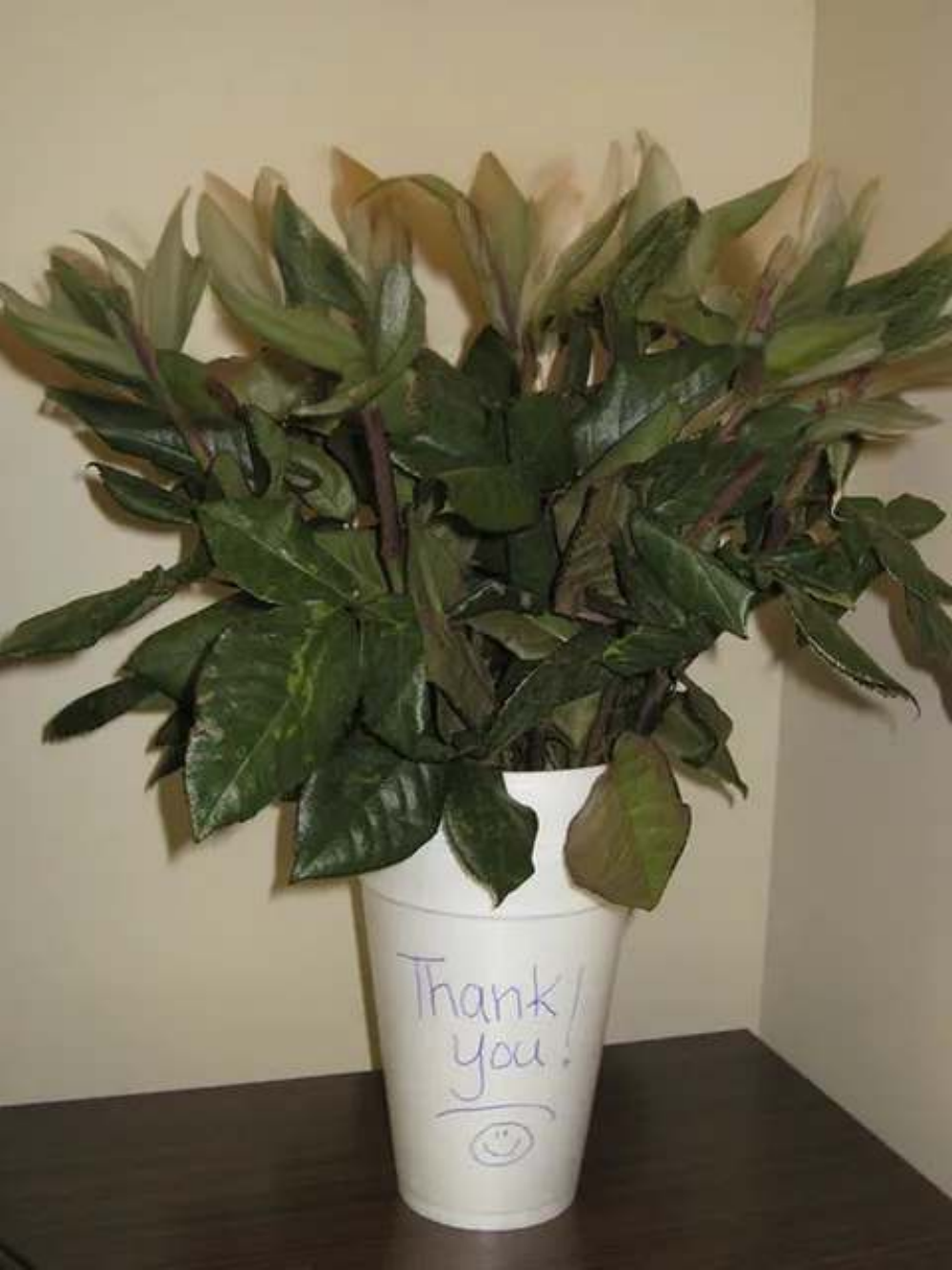} &
        \includegraphics[height=4.7cm, keepaspectratio]{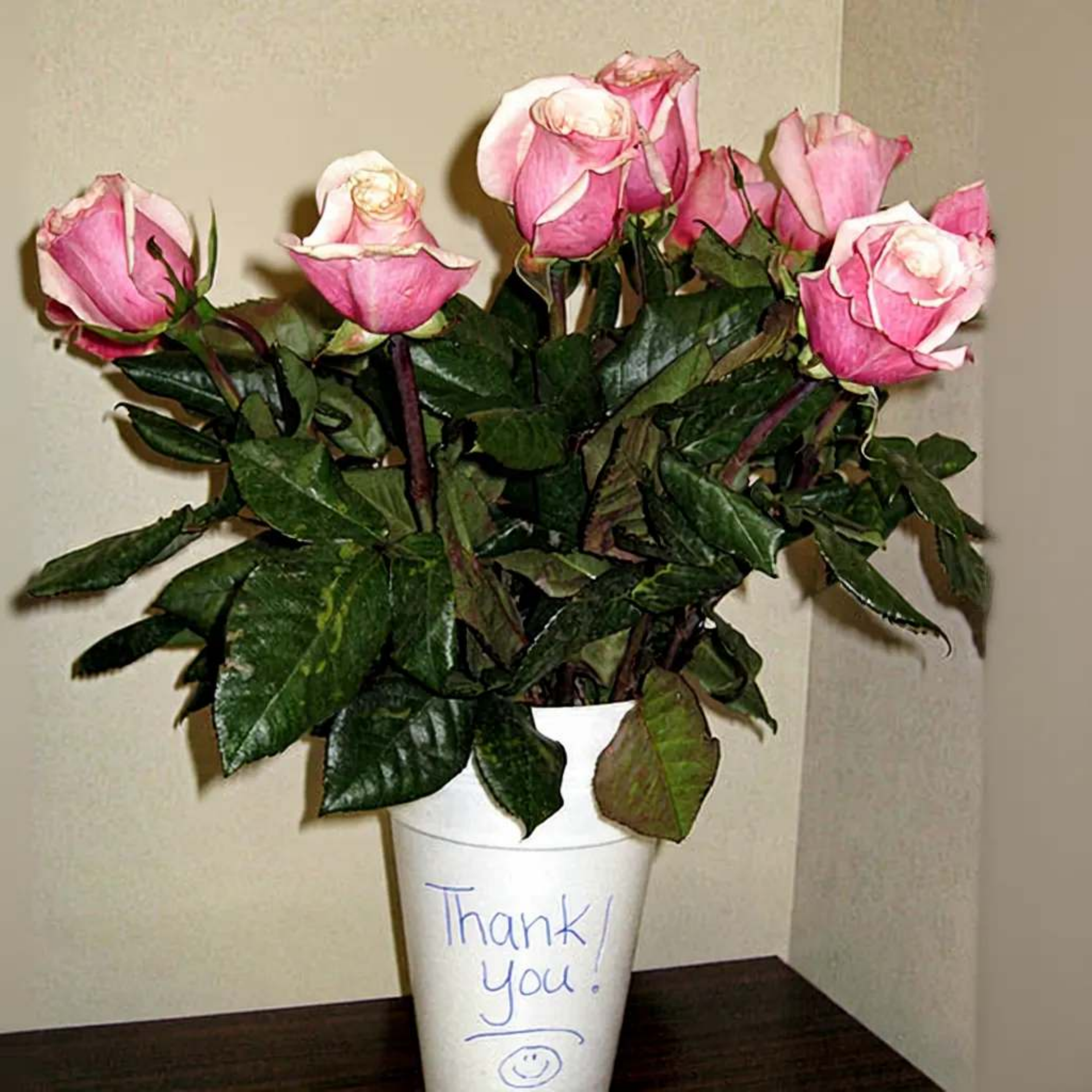} &
        \includegraphics[width=0.2\textwidth]{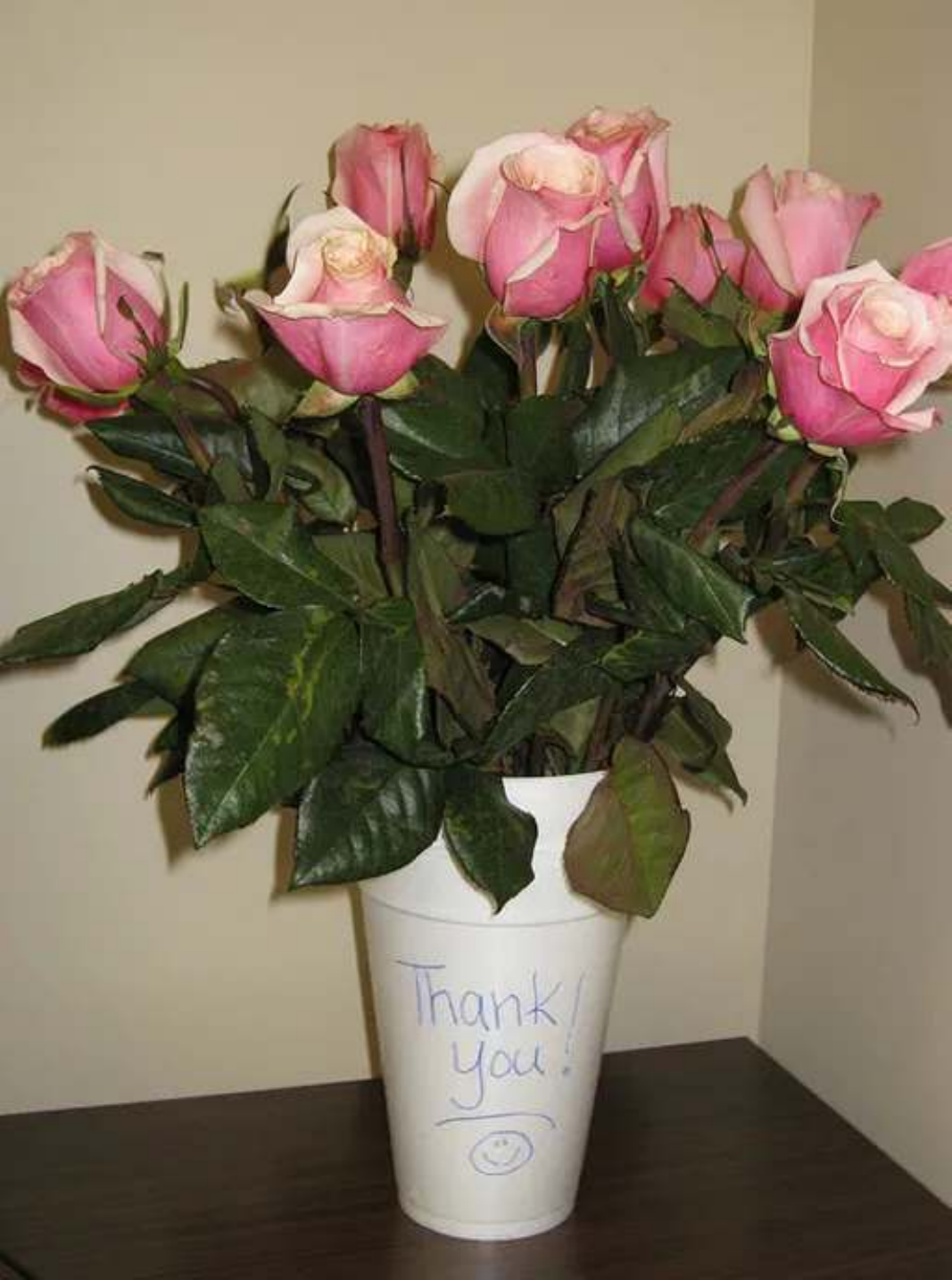} &
        \includegraphics[width=0.2\textwidth]{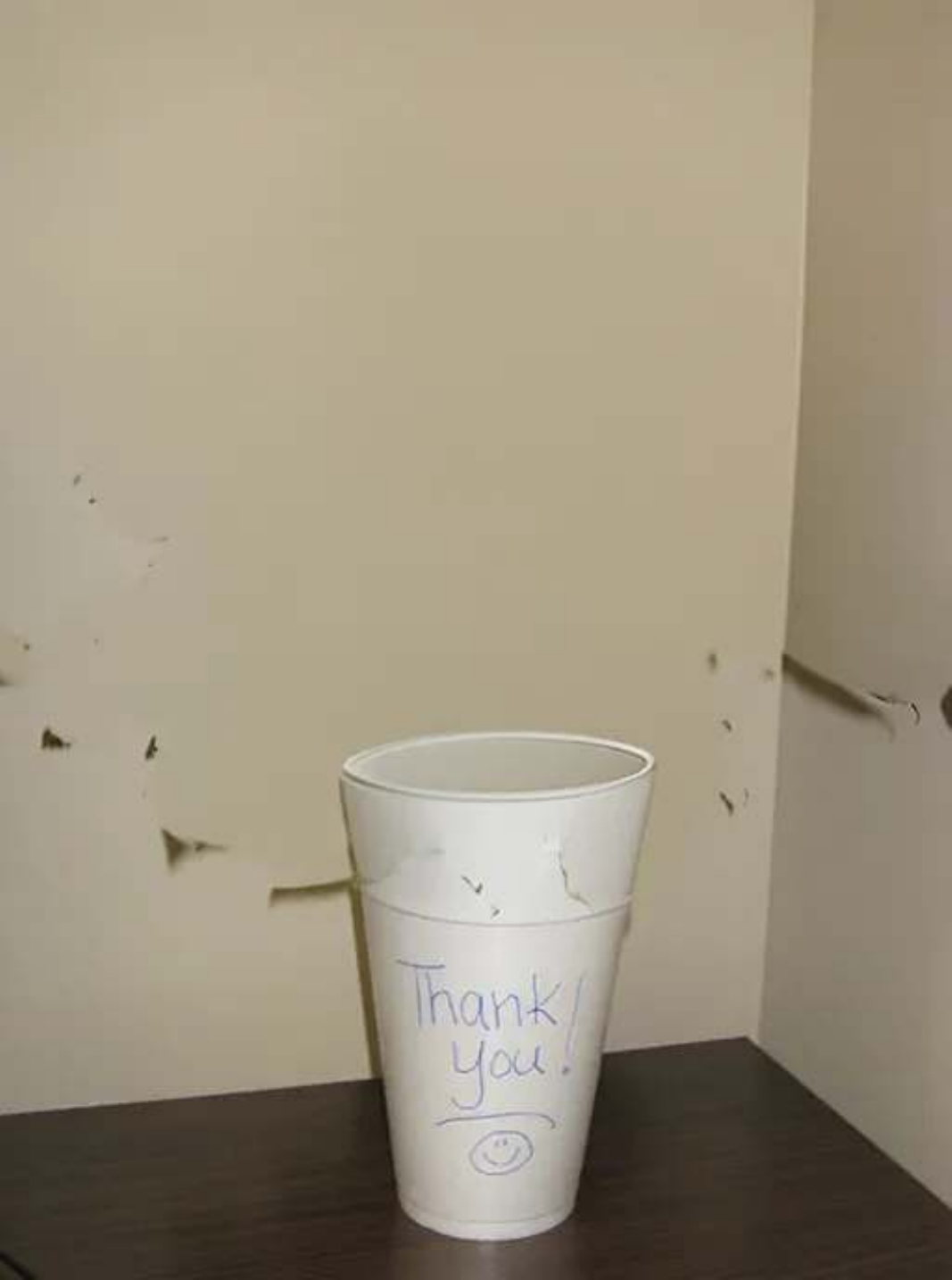} \\
        \multicolumn{5}{c}{(e) Remove-the-tulip-from-the-image.}\\
        \multicolumn{5}{c}{}\\

\includegraphics[width=0.2\textwidth]{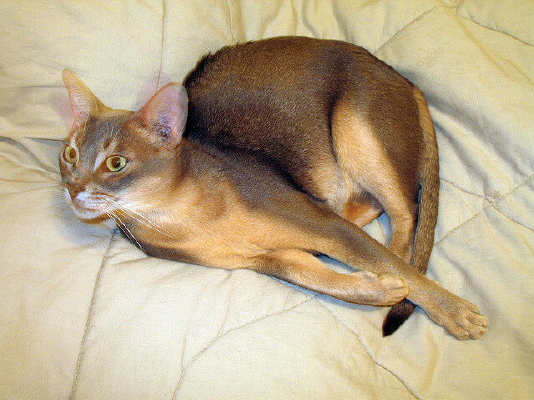} &
        \includegraphics[width=0.2\textwidth]{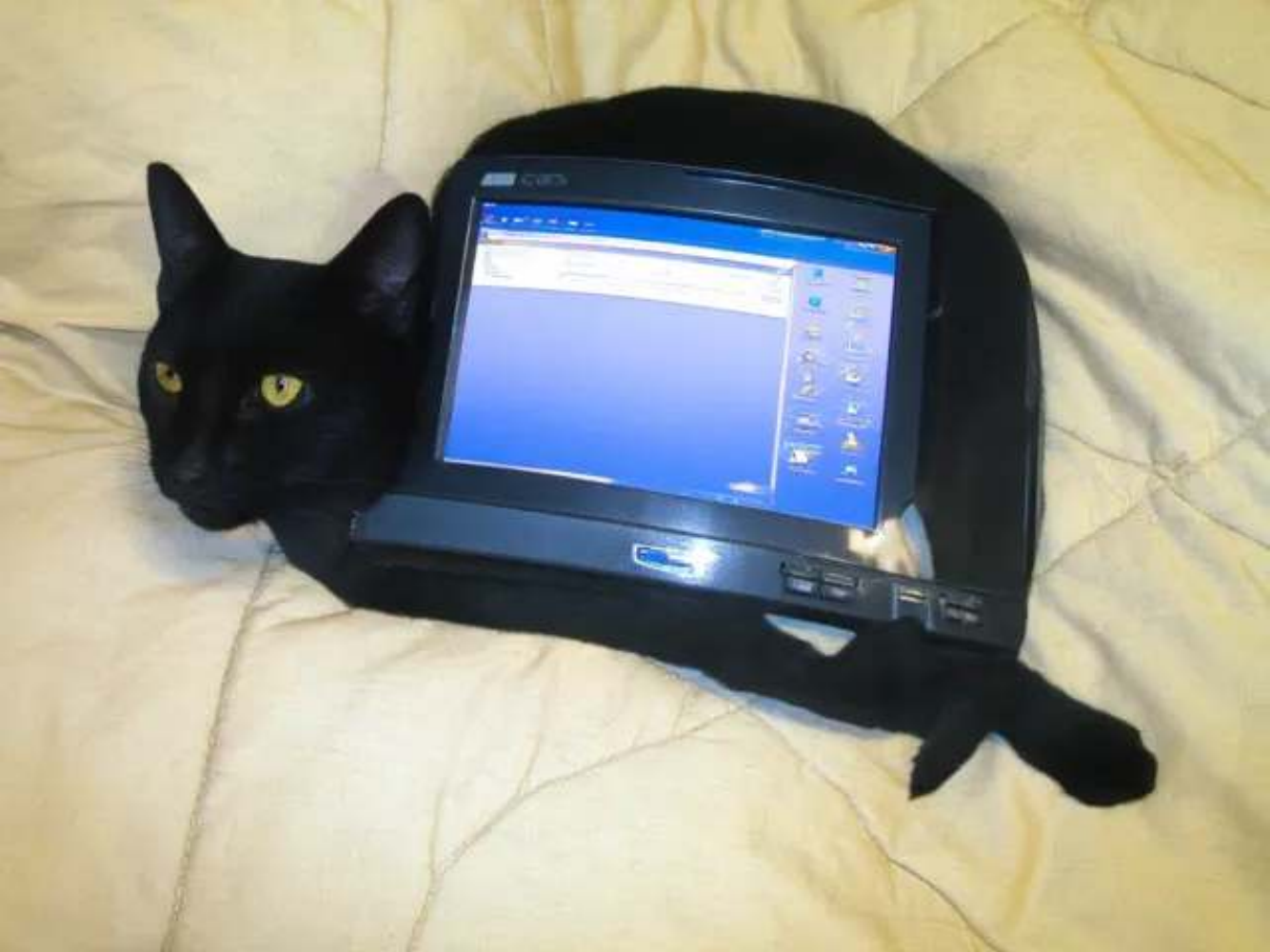} &
        \includegraphics[height=2.7cm, keepaspectratio]{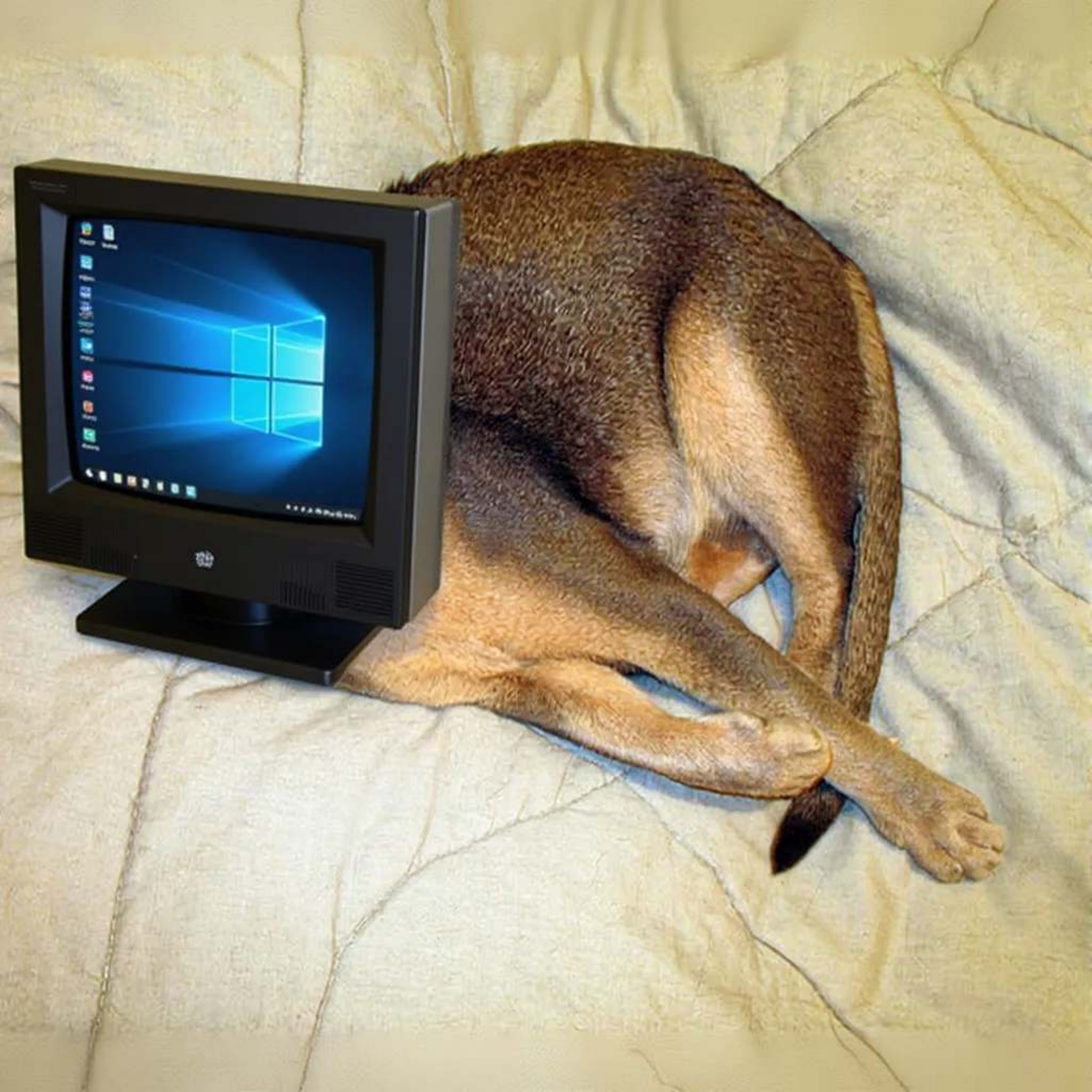} &
        \includegraphics[width=0.2\textwidth]{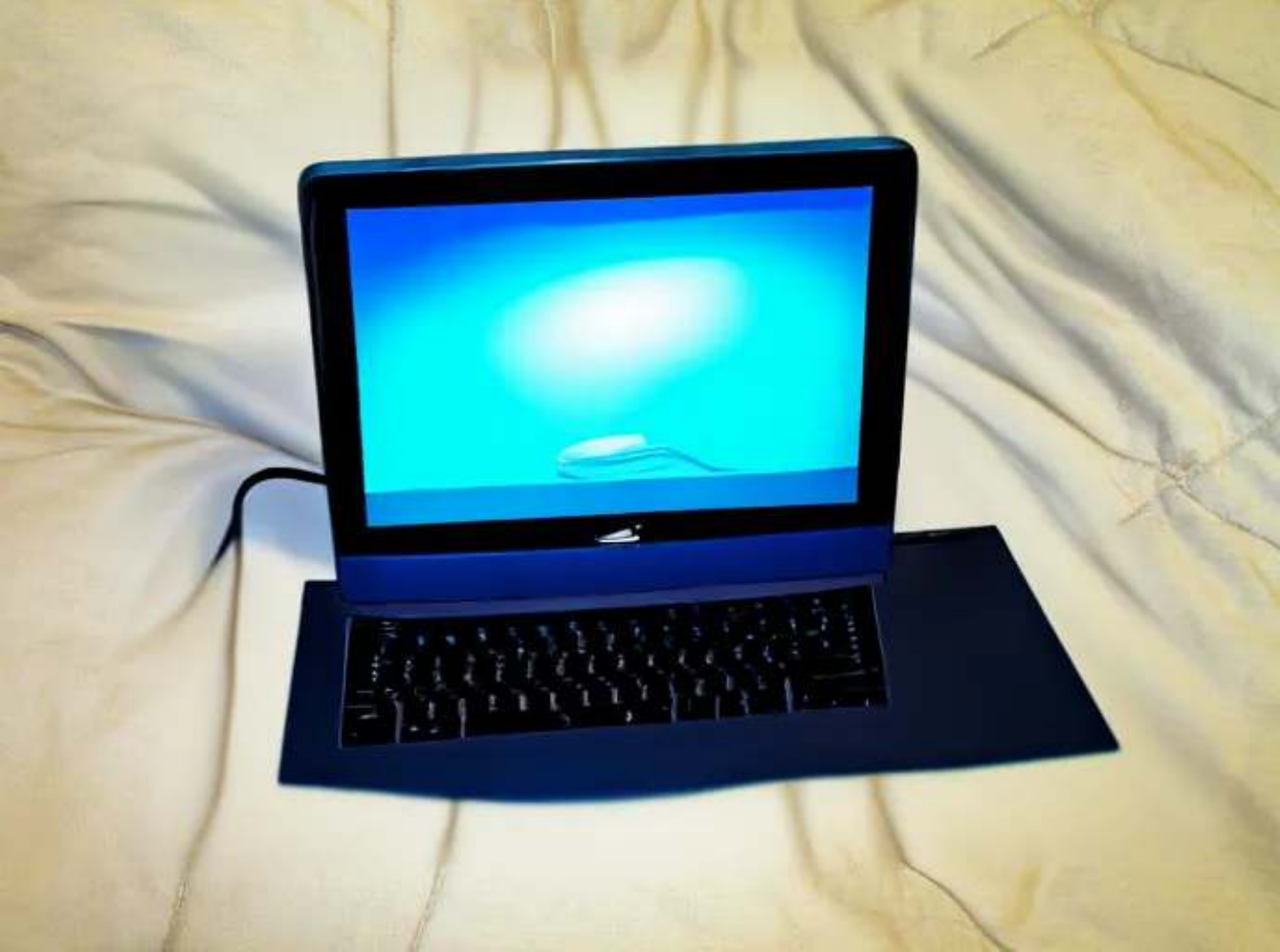} &
        \includegraphics[width=0.2\textwidth]{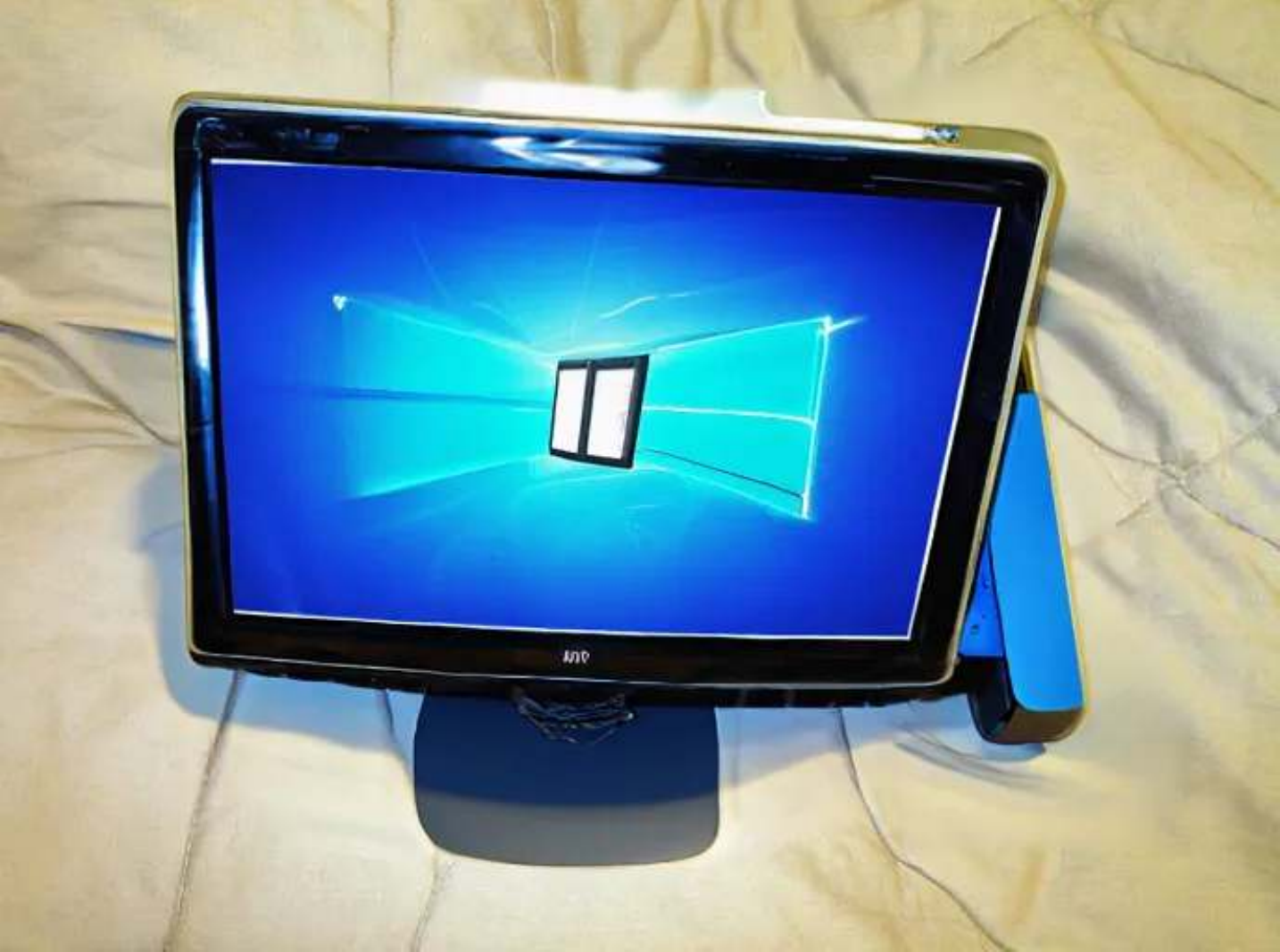} \\
        \multicolumn{5}{c}{(g) replace-cat-with-computer.}\\
        \multicolumn{5}{c}{}\\

    % \includegraphics[width=0.2\textwidth]{figs/replace-cat-with-bag/COCO_val2014_000000012443.pdf} &
    %     \includegraphics[width=0.2\textwidth]{figs/replace-cat-with-bag/COCO_val2014_000000012443-step1x.pdf} &
    %     \includegraphics[height=2.7cm, keepaspectratio]{figs/replace-cat-with-bag/COCO_val2014_000000012443-flux.pdf} &
    %     \includegraphics[width=0.2\textwidth]{figs/replace-cat-with-bag/COCO_val2014_000000012443-bagel.pdf} &
    %     \includegraphics[width=0.2\textwidth]{figs/replace-cat-with-bag/COCO_val2014_000000012443-ours.pdf} \\
    %     \multicolumn{5}{c}{(f) replace-cat-with-bag.}\\
        \multicolumn{5}{c}{}\\

    \end{tabular}
    }

    \caption{Output images of models for text-guided image editing.}
    \label{table-bagel-appendix-pic1}
\end{figure*}

We present qualitative comparisons of text-guided image editing results across different models in~\cref{table-bagel-appendix-pic1}, where the corresponding text prompts are provided in the rows beneath each row image. The first column displays the original input images, while the second to fifth columns show the outputs of Step1X-Edit, FLUX.1 Kontext, BAGEL, and $\mmname$ (ours), respectively.

Overall, $\mmname$ demonstrates superior performance in both semantic understanding of the text prompts and preservation of fine-grained visual details. For instance, in example (a), both Step1X-Edit and FLUX.1 Kontext fail to correctly interpret the instruction, with Step1X-Edit even altering the clock's pattern incorrectly. BAGEL places the photo in an unintended location, whereas \mmname\ accurately positions it on the left wall of the tower, adhering to the prompt.

Another representative case is example (c). Step1X-Edit and FLUX.1 Kontext not only remove the giraffe but also significantly alter the meadow, reflecting signs of over-editing. Similarly, BAGEL removes both the giraffe and the ostrich, again indicating excessive modification. In contrast, $\mmname$ successfully removes only the giraffe as instructed, while preserving all other elements in the scene.

\end{document}